\documentstyle[12pt]{article}

\def\cm{{\cal M}}

\def\car{{\cal R}}

\def\pp{{\mathchoice
          {%displaystyle
              \kern 1pt%
              \raise 1pt
              \vbox{\hrule width5pt height0.4pt depth0pt
                    \kern -2pt
                    \hbox{\kern 2.3pt
                          \vrule width0.4pt height6pt depth0pt
                          }
                    \kern -2pt
                    \hrule width5pt height0.4pt depth0pt}%
                    \kern 1pt
           }
            {%textstyle
              \kern 1pt%
              \raise 1pt
              \vbox{\hrule width4.3pt height0.4pt depth0pt
                    \kern -1.8pt
                    \hbox{\kern 1.95pt
                          \vrule width0.4pt height5.4pt depth0pt
                          }
                    \kern -1.8pt
                    \hrule width4.3pt height0.4pt depth0pt}%
                    \kern 1pt
            }
            {%scriptstyle
              \kern 0.5pt%
              \raise 1pt
              \vbox{\hrule width4.0pt height0.3pt depth0pt
                    \kern -1.9pt  %[e]=0.15pt
                    \hbox{\kern 1.85pt
                          \vrule width0.3pt height5.7pt depth0pt
                          }
                    \kern -1.9pt
                    \hrule width4.0pt height0.3pt depth0pt}%
                    \kern 0.5pt
            }
            {%scriptscriptstyle
              \kern 0.5pt%
              \raise 1pt
              \vbox{\hrule width3.6pt height0.3pt depth0pt
                    \kern -1.5pt
                    \hbox{\kern 1.65pt
                          \vrule width0.3pt height4.5pt depth0pt
                          }
                    \kern -1.5pt
                    \hrule width3.6pt height0.3pt depth0pt}%
                    \kern 0.5pt%}
            }
        }}

\def\mm{{\mathchoice
                       {%displaystyle
                             \kern 1pt
               \raise 1pt    \vbox{\hrule width5pt height0.4pt depth0pt
                                  \kern 2pt
                                  \hrule width5pt height0.4pt depth0pt}
                             \kern 1pt}
                       {%textstyle
                            \kern 1pt
               \raise 1pt \vbox{\hrule width4.3pt height0.4pt depth0pt
                                  \kern 1.8pt
                                  \hrule width4.3pt height0.4pt depth0pt}
                             \kern 1pt}
                       {%scriptstyle
                            \kern 0.5pt
               \raise 1pt
                            \vbox{\hrule width4.0pt height0.3pt depth0pt
                                  \kern 1.9pt
                                  \hrule width4.0pt height0.3pt depth0pt}
                            \kern 1pt}
                       {%scriptscriptstyle
                           \kern 0.5pt
             \raise 1pt  \vbox{\hrule width3.6pt height0.3pt depth0pt
                                  \kern 1.5pt
                                  \hrule width3.6pt height0.3pt depth0pt}
                           \kern 0.5pt}
                       }}

\def\ad{{\kern0.5pt
                   \alpha \kern-5.05pt \raise5.8pt\hbox{$\textstyle.$}\kern
0.5pt}}

\def\bd{{\kern0.5pt
                   \beta \kern-5.05pt \raise5.8pt\hbox{$\textstyle.$}\kern
0.5pt}}

\def\qd{{\kern0.5pt
                   q \kern-5.05pt \raise5.8pt\hbox{$\textstyle.$}\kern
0.5pt}}
\def\Dot#1{{\kern0.5pt
     {#1} \kern-5.05pt \raise5.8pt\hbox{$\textstyle.$}\kern
0.5pt}}

\def\fracm#1#2{\hbox{\large{${\frac{{#1}}{{#2}}}$}}}

\def\@magscale#1{ scaled \magstep #1}
\catcode`@=11
\def\un#1{\relax\ifmmode\@@underline#1\else
        $\@@underline{\hbox{#1}}$\relax\fi}
\catcode`@=12

\def\a{\alpha}
\def\b{\beta}

\def\d{\delta}
\def\e{\epsilon}

\def\g{\gamma}

\def\l{\lambda}
\def\m{\mu}

\def\o{\omega}

\def\q{\theta}

\def\s{\sigma}

\def\z{\zeta}
\def\D{\Delta}

\def\G{\Gamma}

\def\S{\Sigma}

\def\cf{{\cal F}}

\def\cm{{\cal M}}

\def\car{{\cal R}}

\def\cy{{\cal Y}}

\def\dslash{\not{\hbox{\kern-2pt $\partial$}}}
\def\Dslash{\not{\hbox{\kern-4pt $D$}}}
\def\pslash{\not{\hbox{\kern-2.3pt $p$}}}
 \newtoks\slashfraction
 \slashfraction={.13}
 \def\slash#1{\setbox0\hbox{$ #1 $}
 \setbox0\hbox to \the\slashfraction\wd0{\hss \box0}/\box0 }
\def\slsh{\!\bigm|} 
\font\ro=cmsy10                          % font with rope
\def\kcr{{\hbox{\ro \char'170}}}                % right-handed rope
\def\ktl{{\hbox{\ro \char'170}}}        % top end for left-handed rope
\def\ktr{{\hbox{\ro \char'170}}}        % " right
\def\kbl{{\hbox{\ro \char'170}}}        % " bottom left
\def\kbr{{\hbox{\ro \char'170}}}        % " right
\def\plpl{\raise-2pt\hbox{$\raise3pt\hbox{$_+$}\hskip-6.67pt\raise0.0pt
\hbox{$^+$}\hskip 0.01pt$}}
\def\mimi{\raise-2pt\hbox{$\raise3pt\hbox{$_-$}\hskip-6.67pt\raise0.0pt
\hbox{$^-$}\hskip 0.01pt$}} 

\def\bo{{\raise.15ex\hbox{\large$\Box$}}}               % D'Alembertian
\def\pa{\partial}                                       % curly d
\def\de{\nabla}                                         % del
\def\TH{{\raise.2ex\hbox{$\displaystyle \bigodot$}\mskip-4.7mu \llap H \;}}
\def\face{{\raise.2ex\hbox{$\displaystyle \bigodot$}\mskip-2.2mu \llap {$\ddot
        \smile$}}}                                      % happy face
\def\sp#1{{}^{#1}}                              % superscript (unaligned)
\def\sb#1{{}_{#1}}                              % sub"
\def\Tilde#1{\widetilde{#1}}                    % big tilde
\def\Hat#1{\widehat{#1}}                        % big hat
\def\Bar#1{\overline{#1}}                       % big bar
\def\leftrightarrowfill{$\mathsurround=0pt \mathord\leftarrow \mkern-6mu
        \cleaders\hbox{$\mkern-2mu \mathord- \mkern-2mu$}\hfill
        \mkern-6mu \mathord\rightarrow$}
\def\dvec#1{\vbox{\ialign{##\crcr
        \leftrightarrowfill\crcr\noalign{\kern-1pt\nointerlineskip}
        $\hfil\displaystyle{#1}\hfil$\crcr}}}           % <--> accent
\def\fracm#1#2{\hbox{\large{${\frac{{#1}}{{#2}}}$}}}
\def\frac#1#2{{\textstyle{#1\over\vphantom2\smash{\raise.20ex
        \hbox{$\scriptstyle{#2}$}}}}}                   % fraction
\def\ha{\frac12}                                        % 1/2
\def\sfrac#1#2{{\vphantom1\smash{\lower.5ex\hbox{\small$#1$}}\over
        \vphantom1\smash{\raise.4ex\hbox{\small$#2$}}}} % alternate fraction
\def\bfrac#1#2{{\vphantom1\smash{\lower.5ex\hbox{$#1$}}\over
        \vphantom1\smash{\raise.3ex\hbox{$#2$}}}}       % "
\def\afrac#1#2{{\vphantom1\smash{\lower.5ex\hbox{$#1$}}\over#2}}    % "
\newskip\humongous \humongous=0pt plus 1000pt minus 1000pt
\def\caja{\mathsurround=0pt}
\def\eqalign#1{\,\vcenter{\openup2\jot \caja
        \ialign{\strut \hfil$\displaystyle{##}$&$
        \displaystyle{{}##}$\hfil\crcr#1\crcr}}\,}
\newif\ifdtup

\def\ref#1{$\sp{#1)}$}

\parskip=\medskipamount                 % space between paragraphs (LaTeX)
\thicklines                         % thick straight lines for pictures (LaTeX)

\def\oldheadpic{                                % old UM heading
        \setlength{\unitlength}{.4mm}
        \thinlines
        \par
        \begin{picture}(349,16)
        \put(325,16){\line(1,0){4}}
        \put(330,16){\line(1,0){4}}
        \put(340,16){\line(1,0){4}}
        \put(335,0){\line(1,0){4}}
        \put(340,0){\line(1,0){4}}
        \put(345,0){\line(1,0){4}}
        \put(329,0){\line(0,1){16}}
        \put(330,0){\line(0,1){16}}
        \put(339,0){\line(0,1){16}}
        \put(340,0){\line(0,1){16}}
        \put(344,0){\line(0,1){16}}
        \put(345,0){\line(0,1){16}}
        \put(329,16){\oval(8,32)[bl]}
        \put(330,16){\oval(8,32)[br]}
        \put(339,0){\oval(8,32)[tl]}
        \put(345,0){\oval(8,32)[tr]}
        \end{picture}
        \par
        \thicklines
        \vskip.2in}
\def\oldtitle#1#2#3#4{\oldheadpic\begin{center}\vglue.5in{\large\bf #1}\\[.6in]
        {#2}\\[.1in] {\it Department of Physics and Astronomy}\\
        {\it University of Maryland, College Park, MD 20742}\\[.6in]
        Physics Publication \#{#3}\\ {#4}\\[1.5in] {\bf ABSTRACT}\\[.1in]
        \end{center} \begin{quotation}}                 % old title stuff
\def\oldTitle#1#2#3#4#5#6#7{\oldheadpic\begin{center} \vglue .4in
        {\large\bf #1}\\[.4in]
        {#2}\\[.1in] {\it Department of Physics and Astronomy}\\
        {\it University of Maryland, College Park, MD 20742}\\[.1in]
        {#3}\\[.1in] {\it {#4}}\\ {\it {#5}}\\[.4in]
        Physics Publication \#{#6}\\ {#7}\\[.5in] {\bf ABSTRACT}\\[.1in]
        \end{center} \begin{quotation}}                 % " for 2 authors
\def\border{                                            % border
        \setlength{\unitlength}{1mm}
        \newcount\xco
        \newcount\yco
        \xco=-21
        \yco=12
        \begin{picture}(140,0)
        \put(\xco,\yco){$\ktl$}
        \advance\yco by-1
        {\loop
        \put(\xco,\yco){$\kcr$}
        \advance\yco by-2
        \ifnum\yco>-240
        \repeat
        \put(\xco,\yco){$\kbl$}}
        \xco=158
        \yco=12
        \put(\xco,\yco){$\ktr$}
        \advance\yco by-1
        {\loop
        \put(\xco,\yco){$\kcr$}
        \advance\yco by-2
        \ifnum\yco>-240
        \repeat
        \put(\xco,\yco){$\kbr$}}
        \put(-20,13){\tiny University of Maryland Elementary Particle
Physics University of Maryland Elementary Particle Physics University of
Maryland Elementary Particle Physics}
        \put(-20,-241.5){\tiny University of Maryland Elementary
Particle Physics University of Maryland Elementary Particle Physics
University of Maryland Elementary Particle Physics}
        \end{picture}
        \par\vskip-8mm}
\def\bordero{                                           % alternate border
        \setlength{\unitlength}{1mm}
        \newcount\xco
        \newcount\yco
        \xco=-31
        \yco=12
        \begin{picture}(140,0)
        \put(\xco,\yco){$\ktl$}
        \advance\yco by-1
        {\loop
        \put(\xco,\yco){$\kclr}
        \advance\yco by-2
        \ifnum\yco>-240
        \repeat
        \put(\xco,\yco){$\kbl$}}
        \xco=151
        \yco=12
        \put(\xco,\yco){$\ktr$}
        \advance\yco by-1
        {\loop
        \put(\xco,\yco){$\kcr$}
        \advance\yco by-2
        \ifnum\yco>-240
        \repeat
        \put(\xco,\yco){$\kbr$}}
        \put(-20,12){\ooo bacdefghidfghghdhededbihdgdfdfhhdheidhdhebaaahjhhdahba

hgdedge
   hgfdiehhgdigicba}
        \put(-20,-241.5){\ooo ababaighefdbfghgeahgdfgafagihdidihiidhiagfedhadbfd

ecdcdfa
   gdcbhaddhbgfchbgfdacfediacbabab}
        \end{picture}
        \par\vskip-8mm}
\def\headpic{                                           % UM heading
        \indent
        \setlength{\unitlength}{.4mm}
        \thinlines
        \par
        \begin{picture}(29,16)
        \put(165,16){\line(1,0){4}}
        \put(170,16){\line(1,0){4}}
        \put(180,16){\line(1,0){4}}
        \put(175,0){\line(1,0){4}}
        \put(180,0){\line(1,0){4}}
        \put(185,0){\line(1,0){4}}
        \put(169,0){\line(0,1){16}}
        \put(170,0){\line(0,1){16}}
        \put(179,0){\line(0,1){16}}
        \put(180,0){\line(0,1){16}}
        \put(184,0){\line(0,1){16}}
        \put(185,0){\line(0,1){16}}
        \put(169,16){\oval(8,32)[bl]}
        \put(170,16){\oval(8,32)[br]}
        \put(179,0){\oval(8,32)[tl]}
        \put(185,0){\oval(8,32)[tr]}
        \end{picture}
        \par\vskip-6.5mm
        \thicklines}
\def\title#1#2#3#4{\border\headpic {\hbox to\hsize{#4 \hfill UMDEPP #3}}\par
        \begin{center} \vglue .5in {\large\bf #1}\\[.6in]
        {#2}\\[.1in] {\it Department of Physics and Astronomy}\\
        {\it University of Maryland, College Park, MD 20742}\\[1.5in]
        {\bf ABSTRACT}\\[.1in] \end{center} \begin{quotation}}  % title stuff
\def\Title#1#2#3#4#5#6#7{\border\headpic
        {\hbox to\hsize{#7 \hfill UMDEPP #6}}\par
        \begin{center} \vglue .4in {\large\bf #1}\\[.4in]
        {#2}\\[.1in] {\it Department of Physics and Astronomy}\\
        {\it University of Maryland, College Park, MD 20742}\\[.1in]
        {#3}\\[.1in] {\it {#4}}\\ {\it {#5}}\\[.5in] {\bf ABSTRACT}\\[.1in]
        \end{center} \begin{quotation}}                 % " for 2 authors
\def\endtitle{\end{quotation}\newpage}                  % end title page

\def\sect#1{\bigskip\medskip \goodbreak \noindent{\bf {#1}} \nobreak \medskip}

\begin{document}

\border\headpic {\hbox to\hsize{September 1998 \hfill UMDEPP 98-06}}\par
\begin{center}
\vglue .4in
{\large\bf ECTOPLASM HAS NO TOPOLOGY\footnote{Research supported
by NSF grant \# PHY-98-02551 and by NATO Grant CRG-93-0789} 
}\\[.2in]
S. James Gates, Jr.\footnote{gates@bouchet.physics.umd.edu} \\[.1in]
{\it Department of Physics\\
University of Maryland at College Park\\
College Park, MD 20742-4111, USA} 
\\[2.5in]

{\bf ABSTRACT}\\[.1in]
\end{center}
\begin{quotation}

In a new approach to the theory of integration over Wess-Zumino 
supermanifolds, we suggest that a fundamental principle is 
their consistency with an ``Ethereal Conjecture''  that asserts 
the topology of the supermanifold must be generated essentially from 
its bosonic submanifold.  This naturally leads to a theory of
``ectoplasmic'' integration based on super $p$-forms. One consequence 
of this approach is that the derivation of ``density projection 
operators'' becomes trivial in a number of supergravity theories. 
\endtitle

\sect{I. Introduction}

In some of the literature on supermanifolds, the bosonic sub-manifold of the 
superspace is referred to as the ``body'' of the superspace. Similarly, the 
remainder of the superspace has been called its ``soul\footnote{We take our 
``otherworldly'' sounding title in deference to this set of conventions.}.''  
This set of conventions permits us the frivolity of referring to the ``spiritual''
part of superspace as its ``ectoplasm.''  Integration theory over bosonic 
manifolds has been defined in a number of ways, i.e. Riemann-Steljes, 
Lebesque, etc. However, for a long time there was an open question regarding 
the issue of a general theory of integration over local supersymmetry 
manifolds \cite{c3}, also called Wess-Zumino superspaces. One approach 
requires the existence of a local supersymmetry measure over which the 
integration of all the superspace coordinates can be performed.  In principle, 
for a superspace with $N_B$ (sometimes denoted by the symbol D in the following) 
bosonic and $N_F$ (sometimes denoted by the symbol $N$ in the following) 
fermionic coordinates, such measures are provided by $\int d^{N_B + N_F}z 
\, [ sdet ({\rm E}_{\un A}{}^{\un M})]^{-1}$.  In practice knowing this 
does {\underline {not}} simply lead to an explicit expression in terms of 
the component fields of the supergravity multiplet and other supermultiplets 
to which it may couple. For this purpose, it is most convenient to define 
an operator known as the ``local density projector.''  The local density 
projector is the crucial ingredient in conveniently obtaining component 
results directly from superspace without need for explicit $\q$-expansions.  
Alternately, the use of the local density projector is equivalent to local 
``ectoplasmic'' integration (i.e. integration over all $\q$'s in a Wess-Zumino 
superspace).

Until quite recently, the construction of such projectors had been done on a 
case-by-case basis.  Only recently \cite{ectonorcor} has a complete theory 
that extends the definition of rigid Berezinian-type Grassmannian integration 
to the construction of density projectors in the local case of superspace manifolds 
been initiated along the two distinct lines; (a.) super-differential 
forms \cite{prelude} and (b.) normal coordinate expansion \cite{norcor}. 
In the latter work it was shown that a normal coordinate expansion technique 
is well able to compute density projectors in complete generality. The former 
approach, however, seems to be related to the deeper issue of topology of 
supersymmetry manifolds.

An outstanding question in supergravity theory (as formulated in its 
natural setting of curved Wess-Zumino superspace) may be cast in the form, 
``What is the fundamental reason why superspace supergravity theories must 
be formulated in terms of constraints on torsion and curvature supertensors?''  
A closely related question is,``How should these constraints be chosen for 
an arbitrary supergravity theory?''\footnote{We can expand these questions, 
of course, to cover all supersymmetric gauge theories.}  These types of questions 
have plagued the superspace formulation of supergravity theory for about two 
decades with no satisfactory answers to date.  In one of our earliest works
\cite{c1}, we began to grope toward answers that led to the realization of 
the importance of representation-preserving constraints, conventional constraints, 
and conformal scale-breaking constraints \cite{c2} in classifying the types of 
constraints imposed in supergravity theory.  A definite role for the 
superconformal group was noted and it was found that different versions of 
off-shell 4D, $N$ = 1 supergravity correspond to different choices of how 
superconformal symmetry is broken to become Poincar\' e supersymmetry.

This scheme represented progress but did not provide a completely satisfactory 
rationale for why the constraints exists. Nor does it lend great insight into 
how to choose {\it {complete}} sets of constraints. Another serious deficit of 
this approach is that it explains the supergravity constraints in terms of the 
local extension of certain matter field representations from the case of 
global supersymmetry to local supersymmetry.  Such an approach necessarily 
requires the existence of matter supermultiplets.  For many cases of interest, 
such matter multiplets do not exist. The need for a deeper understanding 
is clearly indicated.  We now believe this issue is closely related to the 
first class of questions that we raised in the three paragraphs above.  

One other question concerning the nature of local supersymmetry is the 
possible topological super-extensions of arbitrary manifolds \cite{c4}. 
Stated most simply, ``Is it possible to have superspaces where the topological 
properties of the superspace are substantially different from the bosonic 
sub-manifold it contains?''  In all known examples in supergravity theories 
the answer to this question is no.  

This suggests a very interesting approach to solving the puzzles described
above.  It, perhaps, is reasonable to assert that a hitherto unrecognized
guiding principle may be at work here.   As a working assumption we will assert 
that this is the case by proposing what we will refer to in the following 
discussion as the ``Ethereal Conjecture of Local Extended Manifolds'' Our 
colloquial statement of this conjecture is, 

${~~~~~~~~}$ {\it {The topology of an ``extended'' manifold {\underline 
{must}} essentially arise from \newline ${~~~~~~~~~~~~}$ its bosonic 
sub-manifold.}}  \newline \noindent
Since it is known that superspace does not possess a de Rahm cohomology, 
it may well be that the Ethereal Conjecture is precisely what is needed 
to fill this gap.  In the following, we will show evidence for the 
conjecture.   

The organization of this paper is as follows. In section two, we 
introduce our approach to the formulation of topological invariants
in superspace.  The discussion begins by introducing the ``Ectoplasmic
Integration Theorem'' for general curved Wess-Zumino superspaces.  This
is shown to be consistent with the usual ``components-by-projection''
technique for evaluating rigid superfield actions.  The generalization 
of this technique to curved superspace is found to lead to an expansion 
as first envisioned by Zumino.  For the first time to our knowledge,
a definition of a general superspace topological index is proposed
and related to the Ectoplasmic Integration Theorem. 

In sections three through five, since at present it is not possible
to construct a proof\footnote{A proof of the Ethereal Conjecture
within the context of supersymmetrical theories would {\it {first}}
\newline ${~~~~~}$ require that the off-shell constraints of {\it {all}} 
supersymmetric theories be known!}, we have gathered together a survey of
theories in 2D, 3D and 4D superspaces which all demonstrate
the previously unrecognized realization of the Ethereal Conjecture
as a universal feature.

In section six, we discuss the case of the 2D, (4,4) theory. Since we 
have yet to establish a complete understanding of this new approach, 
we use this particular example as an illustration of some difficulties 
that remain. As we shall show, complications do arise here. We believe 
that these problems are generic when $N_F > N_B$ (as is the case with 
most of the interesting theories).

In section seven, we slightly change our perspective. In the previous 
sections, our attention was directed toward the problem of defining a 
general theory of integration on local supersymmetry manifolds.  In the 
seventh section, the focus is mainly directed to the question of how 
the Ethereal Conjecture may be {\it {la}} {\it {forza}} {\it {del}} 
{\it {primo}} behind why constraints must be imposed on all supersymmetric
theories. In the example discussed, we shall see that the constraints 
may be interpreted as the topological obstructions to the 
realization of the Ethereal Conjecture.

In section eight we give our prospectives on applying these ideas to 
find new techniques with which to attack long unsolved problems of 
finding off-shell constraints for 10D supersymmetrical field theories.

\sect{II. The Ectoplasmic Integration Theorem and Topological Invariants}

~~~~The theory of integrating fermionic numbers received its beginning with
the work of Berezin \cite{bz}, who defined properties of quantities such as 
$\int d^{N_B + N_F}z$ over ``flat supermanifolds.''  This was extended 
to the case  of ``curved supermanifolds'' when Arnowitt, Nath and Zumino 
(ANZ) \cite{anz} showed how to formally (and explicitly) define the
super-determinant.   This permits one to write $\int d^{N_B + N_F}z \, 
[ sdet ({\rm E}_{\un A} {}^{\un M})]^{-1}$ as a formal way to integrate over
curved supermanifolds\footnote{See also DeWitt \cite{dw} who discussed 
related issues. Unfortunately, his discussion of integration 
\newline ${~~~~~}$ over curved manifolds is restricted to Riemannian 
supermanifolds which are not relevant to \newline ${~~~~~}$ supergravity 
theories.}.  As we pointed out above, this expression is of limited
practical value. For practical (i.e.~component) calculations it is required 
to have an equation of the form\footnote{We call this the ``Ectoplasmic 
Integration Theorem.'' It vaguely resembles the Stoke's Theorem \newline 
${~~~~~}$ of multi-variable calculus and allows us to completely 
perform the integrations over the ``soul'' \newline ${~~~~~~}$or ectoplasm 
of the superspace.},
\begin{equation} \eqalign{
\int d^{N_B + N_F} z ~ {\rm E}^{-1} {\cal L} &=~ \int d^{N_B} z ~
{\rm e}^{-1} \, [~ {\cal D}^{N_F} {\cal L} \slsh ~] ~~~,  \cr
\to ~ {\cal D}^{N_F}  &\equiv ~{\rm e} \,\int d^{N_F} z ~ {\rm E}^{-1} ~~~,}
\end{equation} 
for an arbitrary superfunction ${\cal L}$.  Here ${\rm E}_{\un A}{}^{\un M}$ 
denotes the inverse supervielbein for the entire superspace and ${\rm e}_{
\un a}{}^{\un m}$ denotes the inverse vielbein on the body of the superspace.  
(Also we have used the notations $[ sdet ({\rm E}_{\un A} {}^{\un M} )]^{-1} 
\equiv {\rm E}^{-1}$ and $[ det ({\rm e}_{{\un a}}{}^{\un m})]^{-1} \equiv 
{\rm e}^{-1}$.) The symbols $ \int d^{N_B + N_F} z$ and $\int d^{N_B} z$ 
denote integrations over the full superspace and the body of the superspace, 
respectively.  In the equation above, ${\cal D}^{N_F}$ denotes a particular ${
N_F}$-order differential operator constructed from the supergravity covariant 
derivative $\nabla_{\a}$.  Furthermore, ${\cal D}^{N_F} {\cal L} \slsh$ 
corresponds to the action of first applying the operator ${\cal D}^{N_F}$ 
to ${\cal L}$ and afterward in that result setting all the fermionic 
coordinates to zero.  The range of integration on the right hand side of
the equation above is evaluated on a sub-space of the range of integration
on the left hand side.  The {\it a} {\it {priori}} derivation 
of ${\cal D}^{N_F}$ has until recently been an unsolved problem that will 
be our main concern in the following.   We call the quantity ${\rm e}^{-1} 
[ {\cal D}^{N_F} {\cal L} \slsh \, ]$ the ``local density projector'' 
acting on ${\cal L}$.

Let us point out that it is {\underline {not}} our goal to derive equation 
(1). This is the starting point of the works in \cite{ectonorcor, norcor}.
There it is shown that a normal coordinate expansion can be applied to
the left hand side of (1) and used to rigorously derive the Ectoplasmic 
Integration Theorem or E.~I.~T.   Instead we wish to take the right hand 
side of the equation as a starting point and attempt to formulate a 
logically consistent formalism that can be used to {\underline {derive}} 
the operator ${\cal D}^{N_F}$ from some principle and with no reference
to the superdeterminant.  As we will see in the following there is no need
to introduce the notion of a superdeterminant. Our operatorial oriented 
approach is an entirely different way to view this problem and based upon
the fact that super forms have a well defined meaning.  We have previously 
given a short introduction to this alternate approach \cite{prelude}.

The operator ${\cal D}^{N_F}$ must be the local extension of a
well known result from rigid supersymmetry.  Within rigid supersymmetry 
theories, it is permissible to write as an extension of Berezin's original 
definition\footnote{This differs from Berezin's original definition only 
by total derivative terms.}
\begin{equation}
\int d^{N_B + N_F} z \, {\cal L} ~\equiv ~ \int d^{N_B} z \, \Big[ \,
(D \, \cdot\,  \, \cdot\, \, \cdot\, D)^{N_F}  {\cal L} | ~ \Big] ~~~,
\end{equation}
where $D$ is a symbolic representation of all of the spinorial
derivatives of the superspace with $N_F$ Grassmann coordinates.
For 4D, $N$ = 1 superspace as an example, we write,
\begin{equation}
\int d^{4} x \, d^2 \q \, d^2 {\bar \q} ~ {\cal L} ~\equiv~ \fracm 12
\Big\{ \int d^{4} x ~ [\, D^2 \, {\Bar D}{}^2  \,  {\cal L} \, | ~] 
~+~{\rm {h. \, c.}} ~ \Big\} ~~~.
\end{equation}
In the local case, there must exist the extension
\begin{equation}
\eqalign{
\int d^{N_B + N_F} z ~{\rm E}^{-1} \, {\cal L} ~=~ &\int d^{N_B} z ~{\rm 
e}^{-1} \, \Big[ ~ \sum_{i = 0}^{N_F} \, c_{(N_F - i)} \, (\nabla \, 
\cdot\,  \, \cdot\, \, \cdot\, \nabla)^{N_F - i}  \, {\cal L} | ~ \Big]
\cr
~\equiv~ &\int d^{N_B} z ~{\rm e}^{-1} \,  \Big[ ~ {\cal D}^{N_F} {\cal 
L} \, | ~ \Big] ~~~, }
\end{equation}
where $\nabla$ represents the {\it {supergravity}} spinorial derivatives 
that describe the corresponding curved Wess-Zumino superspace.  The 
coefficients $c_{(N_F - i)}$ in this expansion are not constants. In 
general they depend on the supergravity component fields (both physical 
and auxiliary).  The challenged raised by Zumino \cite{c3} (which has now 
been definitively answered \cite{ectonorcor,norcor}) was to find a 
method by which these coefficients might be calculated from some principle.  
This problem had remained unsolved (and usually unrecognized) in most of 
the years since it was first noted.

For a long time it has been known that the leading coefficient,
$c_{(N_F)}$, can be set equal to a constant. In the rigid limit, where all
supergravity component fields vanish, this permits the result of
(4) to agree with (2).  Similarly, it is known that the remaining
coefficients represent an expansion involving the gravitino and
other fields of the supergravity multiplet.  For example, in many 
(but not all) theories we find
\begin{equation}
c_{(N_F - 1)} ~\propto~ i \psi_{\un a} {}^{\b} (\g^{\un a})_{\b} {}^{\a}
~~~.
\end{equation}
It is also clear that the volume of the full superspace is given
by 
\begin{equation}
\int d^{N_B + N_F} z ~{\rm E}^{-1}  ~=~ \int d^{N_B} z ~{\rm 
e}^{-1} ~  c_{(0)} ~~~~,
\end{equation}
so that the volume vanishes whenever $ c_{(0)} = 0$.

Since to our knowledge, the whole topic of topology of supermanifolds is not 
well developed mathematically, we will proceed as cautiously as possible
recognizing that there may not be rigorous mathematical definitions for 
all of the steps we define operationally and heuristically.  Let us set up 
the general situation.  We use ${\cal L}(\tilde I)$ to represent some superspace 
topological invariant (i.e. we have in mind some index associated with 
a field theory).  If it is a topological invariant then we demand that by 
definition the following condition be satisfied. 
\begin{eqnarray}
\int d^{N_B} z ~{\rm e}^{-1} [~ {\cal D}^{N_F} {\cal L} (\tilde I)  \slsh ~] 
&=&~ \int d^{N_B} z ~{\rm e}^{-1} \Big[~ {\rm e}_{\un a} {}^{\un m} \pa_{\un m} 
(~{J_{(I)}}^{\un a} ~+~ {J_{(E)}}^{\un a}  ~) ~\Big]  ~~~, \nonumber\\
&=&~ \int d^{N_B} z ~ \Big[~ \pa_{\un m} (\, {J_{(I)}}^{\un m} ~+~ {J_{(E
)}}^{\un m}  ~) ~ \Big] ~~~.
\end{eqnarray}
for some quantities ${J_{(I)}}^{\un m} \equiv {J_{(I)}}^{\un a} {\rm 
e}_{\un a}{}^{\un m}$ and ${J_{(E)}}^{\un m} \equiv {J_{(E)}}^{\un a} 
e_{\un a} {}^{\un m}$.  As can been clearly seen, these are integrals of 
total divergences and hence their values can only depend on the values 
of fields at the boundary of the integration.  The first integral on 
the right hand side is, in fact, the corresponding topological invariant 
on the body of the superspace. The integrand of the second  must correspond 
to an exact globally well defined quantity.  Roughly speaking the above 
equations suggest that in all local supersymmetry theories the following 
equation is true 
\begin{equation}
{\cal H} ({\rm {sM}}_{N_B + N_F} ) ~\approx~ {\cal H} ({\rm {M}}_{N_B} ) ~~~,
\end{equation}
where ${\cal H} ({\rm {sM}}_{N_B + N_F} ) $ denotes any homotopy group
(or element thereof) of the supermanifold with $N_B$ bosonic coordinates 
and $N_F$ fermionic  coordinates.  A similar interpretation of the symbol 
${\cal H} ({\rm  {M}}_{N_B}) $ is to be understood for the purely bosonic 
sub-manifold of dimension ${N_B}$.  

Another question that arises is, ``Which topological invariants are to 
be used in calculating the local density projector?''  Our answer to 
this is that all topological invariants which can be constructed in the 
class of field theories over a given manifold should satisfy this
condition if that manifold is regarded as the body of a supermanifold.  
This statement implies that any topological invariant that can be written 
is a candidate to use for the derivation of the local density projector.  
Furthermore, given that the local density projector has been calculated 
using one topological invariant, our previous statement implies that the 
same answer will be  obtained from the use of any other topological invariant 
that exists over the body of the superspace. 

With the idea that topology lies at the heart of the problem of finding 
the $c_{(N_F - i)}$ coefficients, a role for super p-forms can be discerned.  
The 4D, $N$ = 1 superspace formulation of irreducible super p-forms \cite{pfm} 
can, in principle, be extended to all values\footnote{Discussions of this nature
can be seen in some of the recent works on D-p-branes and type IIB \newline 
${~~~~~}$ supergravity \cite{IIB}.} of D and $N$.  In a superspace of $N_B$ 
bosonic coordinates, a special role is accorded to super $N_B$-forms. A 
topological index $\D$ results from an integral of a closed but not exact 
$N_B$-form.

For any closed super $N_B$-form, by a choice of Wess-Zumino gauge\footnote{
We first derived this result for the case of the 2-form within 4D, $N$ = 4
supergravity theory \cite{N4D4}.  \newline ${~~~~~}$ That derivation can 
easily be extended to all values $p$, $N_B$ and $N_F$.} it follows that in 
the presence of supergravity, the component of the super $N_B$-form that 
possesses only bosonic indices satisfies\footnote{We emphasize that this 
equation is valid {\it {independent}} of the constraints to which $F_{{\un 
A}_1 \cdots {\un A}_{N_B}}$ is \newline ${~~~~~}$ subject.},
\begin{equation}
\eqalign{ {~~~~}
\Big( \, F_{{\un a}_1 \cdots {\un a}_{N_B} } | \, \Big) ~=~ \Big[~
&{\Tilde f}_{{\un a}_1 \cdots {\un a}_{N_B} } ~+~ \l^{(N_B , 1)} \psi_{[ 
{\un a}_1 |}{}^{\a_1} \Big( \, F_{\a_1 | {\un a}_2 \cdots {\un a}_{N_B} 
] } | \, \Big)   \cr
&~+~ \l^{(N_B , 2)} \psi_{[ {\un a}_1|}{}^{\a_1} \psi_{ |{\un a}_2|}
{}^{\a_2} \Big( \, F_{\a_1 \a_2 | {\un a}_3 \cdots {\un a}_{N_B} ] } | 
\, \Big) \cdots \cr
&~+~ \l^{(N_B , N_B)} [\, \psi_{ {\un a}_1}{}^{\a_1} \psi_{ {\un a}_2}
{}^{\a_2} \cdots \psi_{ {\un a}_{N_B}}{}^{\a_{N_B}} \,]
\Big( \, F_{\a_1 \a_2 \cdots  \a_{N_B} } | \, \Big) ~\Big]  ~~~, }
\end{equation}
where $\l^{(N_B , i)}$ are some normalization constants that are easy to 
calculate and $\psi_{ {\un a}}{}^{\a}$ denotes the gravitino. The explicit
value of these constants depend on $N_B$ and $N_F$.  In fact, a normal
coordinate expansion should offer the simplest way to derive this equation.

Above ${\Tilde f}_{{\un a}_1 \cdots {\un a}_{N_B} }$ is a non-supersymmetric 
bosonic $N_B$-form component field.  If ${\Tilde f}_{{\un a}_1 \cdots {\un 
a}_{N_B} }$ is closed (which implies that $F_{{\un A}_1 \cdots {\un A}_{N_B}}$ 
is super closed), it follows that a topological index (${\Tilde \D}$) is 
defined through the equation,
\begin{equation}
\Tilde \D \, \equiv \, ({N_B}!)^{-1} \, 
\int d^{N_B} z ~{\rm e}^{-1} \, \e^{{\un a}_1 \cdots {\un a}_{N_B} }
{\Tilde f}_{{\un a}_1 \cdots {\un a}_{N_B} } ~~~.
\end{equation}
Now using (9) we see that $\Hat \D = {\Tilde \D}$ where
\begin{equation}
\eqalign{
{\Hat \D} ~\equiv~ \int d^{N_B} z ~{\rm e}^{-1} \, \e^{{\un a}_1 \cdots 
{\un a}_{N_B} } \, \Big[ & ~({N_B}!)^{-1} \Big( \, F_{{\un a}_1 \cdots 
{\un a}_{N_B} } | \, \Big) \,-\, \l^{(N_B , 1)} \psi_{ {\un a}_1}{}^{\a_1} 
\Big( \, F_{\a_1  {\un a}_2  \cdots {\un a}_{N_B}  } | \, \Big)   \cr
&-\, \l^{(N_B , 2)} \psi_{ {\un a}_1}{}^{\a_1} \psi_{ {\un a}_2}{}^{\a_2} 
\Big( \, F_{\a_1 \a_2  {\un a}_3 \cdots {\un a}_{N_B}  } | \, \Big) 
\cdots \cr
&-\, \l^{(N_B , N_B)}({N_B}!)^{-1} [\,\psi_{ {\un a}_1}{}^{\a_1} \cdots \psi_{ 
{\un a}_{N_B}}{}^{\a_{N_B}} \,] \Big( \, F_{\a_1 \a_2 \cdots \a_{N_B} } | 
\, \Big) ~\Big] 
 }
\end{equation}
is used as a definition.  One might not expect the equation $\Hat \D = {\Tilde 
\D}$ (which may be regarded as an expansion in terms of the gravitino) to 
contain any information.  It seems to be a simple tautology.  However, in the 
presence of constraints (required to define irreducible p-forms) and via the 
solution to the Bianchi identities on $F_{{\un A}_1 \cdots {\un A}_{N_B}
}$, this equation can in many cases be used to derive the operator ${\cal D}^{N_F}$.   What emerges from this approach is that the gravitino expansion
in (11) often directly produces the coefficients $c_{(N_F - i)}$ of the 
local density projector.   Let us further note that we may write
\begin{equation}
d^{N_B} z ~{\rm e}^{-1} \, \e^{{\un a}_1 \, \cdots \, {\un a}_{N_B} } ~
= ~ d z^{{\un m}_1}  \, \wedge \, \cdots  \, \wedge \, d z^{{\un m}_{N_B}} 
\, {\rm e}_{{\un m}_1}{}^{{\un a}_1} \, \cdots \, 
{\rm e}_{{\un m}_{N_B}}{}^{ {\un a}_{N_B} }
\, \equiv ~ d \o^{{\un a}_1 \, \cdots  \, {\un a}_{N_B} } 
~~~.  
\end{equation}

In the paragraph above, we mentioned that in order to be irreducible,
the super $p$-form $F_{{\un A}_1 \cdots {\un A}_{N_B} }$ must be subject to
a set of constraints. It is our belief that even in those cases where 
the equation (11) does not lead to a complete determination of the density 
projector\footnote{We define such density projectors as stage-II
density projectors. An example of such a system \newline ${~~~~~}$ will be 
discussed in a later section.}, it is still of importance because it seems 
to completely determine the gravitino field dependence of the density 
projector.  We conjecture\footnote{With the application of the normal 
coordinate expansion technique \cite{ectonorcor,norcor}, it should be possible
\newline ${~~~~~}$ to investigate this.} that it is always the case that
\begin{equation}
\eqalign{ {~~~}
&{\cal D}^{N_F}  ~=~ \e^{{\un a}_1 \cdots {\un a}_{N_B} } \, 
 \Big[ \, \, ( {N_B}!)^{-1}  {\cal D}_{{\un a}_1 \cdots 
{\un a}_{N_B} }   ~-~ \l^{(N_B , 1)} \psi_{ {\un a}_1}{}^{\a_1} 
 {\cal D}_{\a_1 {\un a}_2  \cdots {\un a}_{N_B}  }    \cr
&{~~~~~~~~~~~~~~~~~~~~~~~~~~~}-~ \l^{(N_B , 2)} \psi_{ {\un a}_1}
{}^{\a_1} \psi_{ {\un a}_2}{}^{\a_2} {\cal D}_{\a_1 \a_2  {\un a}_3 
\cdots {\un a}_{N_B}  }  \cdots \cr
&{~~~~~~~~~~~~~~~~~~~~~~~~~~~}-~ \l^{(N_B , N_B)}({N_B}!)^{-1} [\,\psi_{ 
{\un a}_1}{}^{\a_1} \cdots \psi_{ {\un a}_{N_B}}{}^{\a_{N_B}} \,]  
\, {\cal D}_{\a_1 \a_2 \cdots \a_{N_B} }  ~\Big] ~~~,}
\end{equation}
where the operators ${\cal D}_{{\un a}_1 \cdots {\un a}_{N_B} 
}$,..., ${\cal D}_{{\a}_1 \cdots {\a}_{N_B} }$ are independent of the
gravitino field and spacetime derivatives.  Dimensional analysis implies 
that ${\cal D}_{{\un a}_1 \cdots {\un a}_{N_B} }$ must be of order $N_F$ 
in $\nabla_{\a}$, ${\cal D}_{{\a}_1 {\un a}_{2} \cdots {\un a}_{N_B} }$ 
must be of order $N_F - 1$ in $\nabla_{\a}$, and so forth until one gets
to ${\cal D}_{\a_1 \a_2 \cdots \a_{N_B} }$ which must be of order
$N_F - N_B$ in $\nabla_{\a}$. The coefficients of the operators ${\cal D
}_{{\un a}_1 \cdots {\un a}_{N_B} }$,..., ${\cal D}_{{\a}_1 \cdots 
{\a}_{N_B} }$ can only depend on the superspace torsions and their 
spinorial derivatives. A notable implication of (13) is that the gravitino 
has a maximum power to which it is raised in the density projector, i.e. 
$N_B$ in the final term above.  This property is not at all obvious from 
the ANZ local measure.

We note that any closed super $N_B$-form can be used to play the role 
of $F_{{\un A}_1 \cdots {\un A}_{N_B} }$.  Thus, even in a 
theory with no matter superfields, topological indices may be constructed 
directly from the supergravity multiplet. In this way, we may say that 
the topology of the bosonic sub-manifold determines the local theory 
of superspace integration. 

Let us discuss a bit about some additional notation. The index $\Tilde \D$
is defined for ${\Tilde f}_{{\un a}_1 \cdots {\un a}_{N_B} }$, the leading 
term in (9).  Since this quantity is closed, it is possible to add an
exact $N_B$-form to it without changing the fact that ${\Tilde f}_{{\un a}_1 
\cdots {\un a}_{N_B} }$ is closed.  In many classes of interest,
this form can be thus separated where the exact form is constructed
from fermionic fields or other ``matter'' fields.  This is
responsible for the structure of (7).  We may thus introduce another
non-supersymmetric index (denoted by $\D$) by taking the limit 
of $f_{{\un a}_1 \cdots {\un a}_{N_B} }$ where all matter fields
are set to zero.

Our considerations also suggest some further interesting observations.
For example, we can consider a closed super $p$-form (with $p < N_B$
in a superspace of $N_B$ bosonic and $N_F$ fermionic coordinates), for
which it follows there exists ${\Tilde f}^{(p)}_{{\un a}_1 \cdots 
{\un a}_p }$ such that
\begin{equation}
\eqalign{
{\Tilde f}^{(p)}_{{\un a}_1 \cdots 
{\un a}_p } ~\equiv~  \Big[ & ~ \Big( \, F_{{\un a}_1 \cdots {\un 
a}_{p} } | \, \Big) \,-\, \l^{(p , 1)} \psi_{ [ \, {\un a}_1 | }{}^{\a_1} 
\Big( \, F_{\a_1  | \, {\un a}_2  \cdots {\un a}_{p} \, ] } | \, 
\Big)   \cr 
&-\, \l^{(p , 2)} \psi_{ [\, {\un a}_1 |}{}^{\a_1} \psi_{ |{\un a}_2
| }{}^{\a_2} \Big( \, F_{ \a_1 \a_2  | \,{\un a}_3 \cdots {\un a}_{p} 
\, ] } | \, \Big) \cdots \cr
&-\, \l^{(p , p)}\, [\,\psi_{ {\un a}_1}{}^{\a_1} \cdots \psi_{ 
{\un a}_{p}}{}^{\a_{p}} \,] \Big( \, F_{\a_1 \a_2 \cdots \a_{p} } | 
\, \Big) ~\Big] ~~~.
 }
\end{equation}
If there are defects or surfaces of interest characterized by differentials
$d \o^{{\un a}_1 \cdots {\un a}_{p} }$, these may be coupled to a super
$p$-form $F_{{\un A}_1 \cdots {\un A}_{p} }$ by use of (14) and
\begin{equation}
{\cal S} (\o) ~\equiv ~ (p!)^{-1} \, \int ~ d \o^{{\un a}_1 \cdots {\un 
a}_{p} } \, {\Tilde f}^{(p)}_{{\un a}_1 \cdots {\un a}_p } ~~~.
\end{equation}
Since defects or surfaces are used to define ${\cal S} (\o) $, this may 
(partially) break supersymmetry. This observation may be of use in discussing 
the coupling to branes. Equations (14,15) constitute the definition of a 
theory of integration for super $p$-forms.

The crucial features of the arguments in this section are the existence of
the formulae (9) and (14).  In turn the most critical features of these
formulae are the coefficients $\l^{(N_B , \ell)}$, and $\l^{(p , \ell)}$
where $\ell$ is a dummy index that ranges appropriately for each of these.
It turns out that their determination follows by a simple observation
\cite{ectonorcor}.  We may return to (12) and consider its existence in
a purely bosonic space where we can use it to show
\begin{equation}
\eqalign{ {~~~~~}
d \o^{{\un a}_1 \, \cdots  \, {\un a}_{N_B} } \, f_{{\un a}_1 \,
\cdots  \, {\un a}_{N_B} } &=~ d z^{{\un m}_1}  \, \wedge \, \cdots  
\, \wedge \, d z^{{\un m}_{N_B}} \, {\rm e}_{{\un m}_1}{}^{{\un a}_1} 
\, \cdots \, {\rm e}_{{\un m}_{N_B}}{}^{ {\un a}_{N_B} } \, f_{{\un 
a}_1 \, \cdots  \, {\un a}_{N_B} } \cr
&=~ d z^{{\un m}_1}  \, \wedge \, \cdots \, \wedge \, d z^{{\un 
m}_{N_B}} \, f_{{\un m}_1 \, \cdots  \, {\un m}_{N_B} } 
 ~~~. }
\end{equation}
In a purely bosonic spacetime we also have
\begin{equation}
f_{{\un m}_1 \, \cdots  \, {\un m}_{N_B} } ~=~ \,{\rm e}_{{\un m}_1}
{}^{{\un a}_1} \, \cdots \, {\rm e}_{{\un m}_{N_B}}{}^{ {\un a}_{N_B} 
} \, f_{{\un a}_1 \, \cdots  \, {\un a}_{N_B} }  ~~~.
\end{equation}
In a superspace, however, this equation must be generalized to
\begin{equation}
f_{{\un m}_1 \, \cdots \, {\un m}_{N_B}} ~=~ (-1)^{[\fracm{N_B}{2}]} \,
{\rm E}_{{\un m}_{N_B}}{}^{ {\un A}_{N_B} }  \, \cdots \,{\rm E}_{{\un 
m}_1}{}^{{\un A}_1} \, F_{{\un A}_1 \, \cdots  \, {\un A}_{N_B} }  ~~~,
\end{equation}
where $[\fracm{N_B}{2}]$ denotes the greatest integer in $\fracm{N_B}{2}$
and
\begin{equation}
{\rm E}_{{\un m}}{}^{{\un A}} \, \equiv \, (\, - \psi_{{\un m}}
{}^{{\a}}(x) , \, \, {\rm e}_{{\un m}}{}^{{\un a}}(x)~) ~~~.
\end{equation}
When these components for the supervielbein are substituted into (18) and 
it is expanded over the super indices ${\un A}_1 ,\, \dots \, ,{\un A}_{
N_B} $, the required coefficients ($\l^{(N_B , \ell)}$) appear. This same 
argument applies to the other set of coefficients in (14)\footnote{At least
one prior effort \cite{zup} has appeared, in which some discussion of 
superspace integration \newline ${~~~~~}$ and super $p$-forms was given.}.  
The remarkable feature of the argument in this paragraph is that the 
supervielbein that appears in (19) is a superfield whereas the fields 
that appear on the right hand side are simple component fields. Although 
seemingly contradictory, this identification is correct within the context 
it is used. 

\sect{III. 2D Local Ectoplasmic Integration}

We begin with the simplest possible 2D superspace supergravity theory that 
exists (i.e., (1,0) supergravity). The superspace description of this
theory has been known for some time \cite{c6}. Its superspace supergravity
covariant derivative ($\nabla_A \equiv (\nabla_+ , \nabla_{\pp} , 
\nabla_{\mm} )$ satisfies 
\begin{equation}
[ \nabla_+ , \nabla_+ \} = i 2 \nabla_{\pp} ~~ , ~~ [ \nabla_+ ,
\nabla_{\pp} \}  = 0 ~~ , ~~ \nabla_+ {\Sigma}^+ = \fracm 12 
{\cal R} \quad , 
\end{equation}
\begin{equation}
[ \nabla_{+} , \nabla_{\mm} \} = - i 2 {\Sigma}^+ \cm  ~~ , ~~ 
[ \nabla_{\pp} , \nabla_{\mm} \}  = - ( ~ {\Sigma}^+ \nabla_+ + { \cal R }
\cm ) ~ \quad . 
\end{equation}
The quantities ${\Sigma}^+ $ and ${ \cal R }$ are superfield field
strengths and ${\cal M}$ denotes the generator of SO(1,1), the 2D Lorentz
group.  On defining ${\Sigma}^+ \slsh$ as the limit of  $ {\Sigma}^+ (z^{
\hat M})$ as $\zeta^+ ~\to 0$ and similarly for ${\cal R } \slsh$, we find 
\begin{equation}
{\Sigma}^+ \slsh ~=~ - {\psi}_{\pp, ~ \mm} {}^+ \,=\, - [~ {\rm e}_{\pp}
{\psi}_{\mm}{}^+ -  {\rm e}_{\mm}  {\psi}_{\pp}{}^+ - c_{\pp,~\mm}{}^{\pp}
{\psi}_{\pp}{}^+ - c_{\pp,~\mm}{}^{\mm} {\psi}_{\mm}{}^+ ~] ~ , 
\end{equation}
\begin{equation}
{r}_{\pp,~\mm}( \o )  ~=~ - [~ {\rm e}_{\pp} {\o}_{\mm} -  {\rm e}_{\mm}
{\o}_{\pp} - c_{\pp,~\mm}{}^{\pp} {\o}_{\pp} - c_{\pp,~\mm}{}^{\mm}
{\o}_{\mm} ~] ~ ,  
\end{equation}
\begin{equation}
{\nabla}_+ {\Sigma}^+ \slsh  ~=~ - \fracm 12 [ ~ {r}_{\pp,~\mm}( \o ) + i 2 {
\psi}_{\pp} {}^+ { \psi}_{\pp, ~\mm} {}^+ ~ ] \quad ,
\end{equation}
\begin{equation}
\nabla_{\pp}  \equiv {\rm e}_{\pp} + {\o}_{\pp} {\cal M} \qquad , \qquad 
\nabla_{\mm} \equiv {\rm e}_{\mm} + {\o}_{\mm} {\cal M} \qquad , \qquad  
{\o}_{\pp} \,=\, c_{\pp,~\mm}{}^{\mm} \quad , 
\end{equation}
\begin{equation} 
~~{\o}_{\mm} \,=\, c_{\pp,~\mm}{}^{\pp} + i 2 {\psi}_{\pp}{}^+{\psi}_{\mm}{}^+ 
~~, ~~ {\rm e}_{a} \equiv {\rm e}_a {}^m {\pa}_m ~~ , ~~ [ {\rm e}_a , 
{\rm e}_b ] = c_{a, b} {}^c {\rm e}_c \quad . 
\end{equation}
The Lorentz generator ${\cal M}$ above is defined to act according to the
rules; $[ {\cal M} , \psi_+ ] = \frac 12 \psi_+$, $[ {\cal M} , \psi_- ] = 
- \frac 12 \psi_-$, $[ {\cal M} , {\rm e}_{\pp} ] = {\rm e}_{\pp}$ and 
$[ {\cal M} , {\rm e}_{\mm} ] = - {\rm e}_{\mm}$.

From equation (22), we see that $\psi_{\pp, \mm}{}^+ = - \S^+$ and upon 
substituting into (24) we obtain
\begin{equation}
- \frac 12 \, r_{\pp, \mm} ~=~ (\, \nabla_+ ~-~ i \psi_{\pp}{}^+
\, ) \S^+ \,|  ~\equiv ~ \Big[\,  {\cal D}_+  \S^+ \,|  \, \Big] ~~~.
\end{equation}
Further multiplying this equation by ${\rm e}^{-1}$ and integrating
over the 2-manifold we find
\begin{equation}
{\Tilde \D} ~=~ - \frac 12 \int d^2 \s ~ {\rm e}^{-1} ~ r_{\pp, 
\mm} ~=~  \int d^2 \s \, {\rm e}^{-1} \, \Big[\, {\cal D}_+  
\S^+ \,|  \, \Big] ~\equiv ~ {\Hat \D} ~~~.
\end{equation}
According to the Ectoplasmic Integration Theorem
\begin{equation}
 \int d^2 \s \, d \z^- ~ {\rm E}^{-1} \, \S^+ ~\equiv ~
 \int d^2 \s \, {\rm e}^{-1} \, \Big[\, {\cal D}_+  \S^+ \,|  
\, \Big] ~~~, 
\end{equation}
and $\S^+$ may be replaced by any general (1,0) superfield Lagrangian 
${\cal L}_-$.  Notice that (28) realizes the Ethereal Conjecture precisely 
in the form of (1) and (7), since the the spin-connection (defined 
in (25) and (26)) contains a contribution from the gravitino bilinear.  As 
shown previously \cite{c6} the density projector in (27) leads to all 
of the (1,0) superspace results appropriate for describing the heterotic 
string.  The index ${\Tilde \D}$ can be related to ${\D}$ (the index of the 
non-supersymmetric theory) by the following arguments.

In the non-supersymmetric theory, the curvature tensor is defined as in 
(23). However, the connection in the non-supersymmetric theory does not 
contain the fermionic terms in $\o_{\mm}$ as given in (26).  This means 
that the curvature tensor in the non-supersymmetric theory (denoted by 
$r_{\pp, \mm}^B$) is related to $r_{\pp, \mm}$ via
\begin{equation}
r_{\pp, \mm}  ~=~ r_{\pp, \mm}^B ~+~ i 2 \Big[ \,
\nabla_{\pp}(e)  ( \, \psi_{\pp}{}^+ \psi_{\mm}{}^+ \,) \, \Big]
~~~,
\end{equation}
so that integrating both sides of this equation yields
\begin{equation}
{\Tilde \D} ~=~ {\D} ~-~ i \int d^2 \s \, \Big\{ ~ \pa_m [\, 
{\rm e}_{\pp}{}^m (\,  \psi_{\pp}{}^+ \psi_{\mm}{}^+
\,) \,] ~ \Big\} ~~~. 
\end{equation}
In equation (30) $\nabla_{\pp}(e)$ refers to the 2D gravitational 
covariant derivative constructed solely from the zweibein fields. 
Equations (30,31) provide a concrete example of our general discussion
surrounding equation (7). In all examples known to us, a topological
index calculated from superfields possesses the structure of (7).
So even when adding surface terms to supersymmetrical theories, such 
terms must have more than purely bosonic parts if they are to be 
consistent with a superfield formulation. This feature has often
been ignored in the literature in various discussions at the component
level of anomalies, boundary terms, duality transformations, etc.

Along the same lines, one can look at 2D, $N$ = 1 supergravity. The
solution to the Bianchi identities are given as,
\begin{equation}
[ ~ \nabla_{\a} ~,~  \nabla_{\b} ~ \} ~= ~ i 2 (\g^a)_{\a \b}
\nabla_{a} ~+~  2 (\g^3)_{\a \b} {\rm R} {\cal M} ~~, 
{~~~~~~~~~~~~~~~~~~~~~~~}
\end{equation}
\begin{equation}
[ ~ \nabla_{\a} ~,~  \nabla_{b} ~ \} ~= ~ i  [ ~ \ha {\rm R}
(\g_b)_{\a}{}^{\b} \nabla_{\b} ~+~ (\g^3 \g_b)_{\a}{}^{\b} 
( \nabla_{\b}{\rm R} ) {\cal M} ~ ] ~~, {~~~~~~~~~~}
\end{equation}
\begin{equation}
[ ~ \nabla_{a} ~,~  \nabla_{b} ~ \} ~= ~ - \e_{a b} ~[ ~ \ha
( \nabla^{\a} {\rm R} ) (\g^3)_{\a}{}^{\b} \nabla_{\b} ~-~ 
(\nabla^2 {\rm R} ~-~  {\rm R}^2 ) {\cal M} ~ ] ~~.
\end{equation}
(We alert the reader that our conventions are such that we write $
\nabla^2 \equiv \ha \nabla^{\a} \nabla_{\a}$.)  In writing these results, 
we have also simplified them by replacing the usual Lorentz generator 
according to:
\begin{equation}
{\cal M}_{ b c} \to \e_{b c } {\cal M} ~~, 
\end{equation}
so that when acting on a spinor $\psi_{\a}$ or a vector $v_a$ we have
\begin{equation}
{\cal M} \psi_{\a} =  \fracm 12 (\g^3)_{\a}{}^{\b} \psi_{\b}
~~~,~~~{\cal M} v_a =  \e_{a}{}^{b} v_{b}   ~~~~. 
\end{equation}
 
The component fields of the supergravity multiplet are the graviton (${\rm 
e}_{a}{}^{\rm m}$), gravitino ($\psi_{ a}{}^{\a}$) and auxiliary field 
(${\rm B}$).  These enter the superfield ${\rm R}$ as
follows 
\begin{equation}
{\rm R} | ~=~ {\rm B} ~~, ~~~~
{~~~~~~~~~~~~~~~~~~~~~~~~~~~~~~~~~~~~~~~~~~~~~~~~~~~~~~~~} \end{equation} 
\begin{equation}
\nabla_{\a} {\rm R} | ~=~ (\g^3)_{\a \b} \e^{a b} \Psi_{a b} {}^{\b} ~+~ i
{\rm B} (\g^b)_{\a \b} \psi_b {}^{\b} ~~,  {~~~~~~~~~~~~~~~~~~~~~~~~~~~~~}
\end{equation} 
\begin{equation}
\nabla^2 {\rm R} | ~=~ - \ha \e^{a b} \, r_{a b} (\o) ~-~ i 2 \psi^a
{}^{\a} (\g^b)_{\a \b} \Psi_{a b} {}^{\b} ~+~ {\rm B} \psi^a {}^{\a}
\psi_a {}_{\a} ~+~ {\rm B}^2 ~~~,
\end{equation}
here $ \Psi_{a b} {}^{\a} $ denotes the ``curl'' of the gravitino and $
{\cal R}_{a b} $ denotes the two-dimensional curvature in terms of
a spin-connection defined by 
\begin{equation}
\o^a ~=~ - \ha \e^{b c} C_{b c}{}^a ~-~ i \e^{b c} \psi_b {}^{\a}
(\g^a)_{\a \b} \psi_c {}^{\b} ~~.
\end{equation}
Upon multiplying (38) by $\frac 12 ( \g^3)^{\a \b} \e_{a \, b}$ 
and using the resultant to eliminate $\psi_{a \, b}{}^{\g}$ from
(39), we find
\begin{equation} {~~~~}
\eqalign{ - \frac 12 \e^{a \, b} \, r_{a \, b} &=~ \Big\{ ~
\Big[ ~ \nabla^2 \,-\, i \psi^{a \, \b} (\g_a)_{\b}{}^{\g}
\nabla_{\g} \, + \, \e^{a \, b} \psi_a {}^{\a} (\g^3)_{\a \b}
\psi_b {}^{\b}  \, - \, {\rm R}  ~\Big] \, {\rm R} \, | ~ \Big\} \cr
&\equiv ~ \Big\{ {\cal D}^2 \, {\rm R} \, | \, \Big\} ~~~.
}
\end{equation}
Thus, we may define
\begin{equation}
{\Tilde \D} ~\equiv~ - \frac 12 \int d^2 \s ~ {\rm e}^{-1}
\, \e^{a \, b} \, r_{a \, b} ~=~ \int d^2 \s ~ {\rm e}^{-1}
\, \Big\{ \, {\cal D}^2 \, {\rm R} \, | ~ \Big\} ~\equiv~ {\Hat \D}
~~~,
\end{equation}
\begin{equation}
~\to ~ {\Hat \D} ~-~ \D ~=~ \int d^2 \s ~ {\pa}_m \Big[ \, \e^{c \, d}
\psi_c {}^{\g} (\g^3 \g^a)_{\g \d} \psi_d {}^{\d} {\rm e}_a {}^m
\, \Big] ~~~~,
\end{equation}
or via the Ectoplasmic Integration Theorem,
\begin{equation}
\int d^2 \s \, d^2 \theta ~ {\rm E}^{-1} {\cal L} ~=~ \int d^2 
\s  ~ {\rm e}^{-1} \, \Big[ \, {\cal D}^2 \, {\cal L} \, | ~ \Big]
~~~. \end{equation}

At this point, we have shown that both 2D (1,0) and (1,1) supergravity
theories do indeed realize the Ethereal Conjecture.  We next turn to
the the case 2D, $N$ = 2 theories.

There are two minimal irreducible off-shell formulations of 2D, $N$ = 2
supergravity which we call the $U(1)$ \cite{GLO} and $U_A(1)$ \cite{n2d2sg} 
theories, respectively.  There is also a reducible formulation \cite{GrWe} 
which we refer to as the $U(1) \otimes U_A(1)$ theory.  In the following, we
will restrict our consideration solely to the irreducible theories.
However, one can show that exactly the same arguments apply to the 
$U(1) \otimes U_A(1)$ theory.  The differences in the various theories
has to do with the structure of the holonomy group of the superspace
supergravity covariant derivative.  In 2D, $N$ = 2 superspace, the
form of this operator (for the three respective theories mentioned 
above) is
\begin{equation}
\eqalign{
\nabla_A  &\equiv ~ {\rm E}_A {}^M D_M ~+~ \o_A {\cal M} ~+~ 
\G_A {\cal Y} ~~~, \cr
\nabla_A  &\equiv ~ {\rm E}_A {}^M D_M ~+~ \o_A {\cal M} ~+~ 
{\G'}_A {\cal Y}' ~~~, \cr
\nabla_A  &\equiv ~ {\rm E}_A {}^M D_M ~+~ \o_A {\cal M} ~+~ 
\G_A {\cal Y} ~+~  {\G'}_A {\cal Y}' ~~~,
}
\end{equation}
where the $U(1)$ generator ${\cal Y}$ and $U_A(1)$ generator ${\cal Y}'$
are defined according to
\begin{equation}
\eqalign{
[ {\cal Y}\, , \,{\nabla}_{\pm} ] &=~ i \frac 12 \nabla_{\pm} ~~~,~~~ [ {\cal 
Y}  \, , \, {\Bar \nabla}_{\pm} ] ~=~ - i \frac 12 {\Bar \nabla}_{\pm} ~~~, \cr
[ {\cal Y}' \, , \, \nabla_{\pm} ] &=~ \pm i \frac 12 \nabla_{\pm} ~~~,~~~ 
[ {\cal Y}' \, , \, {\Bar \nabla}_{\pm} ] ~=~ \mp i \frac 12 {\Bar 
\nabla}_{\pm} ~~~. }
\end{equation}

Since we are now considering an extended superspace, in addition to the
measure over the full superspace, $\int d^2 \s d \z^+  d \z^- d {\bar \z}{}^+ 
d {\bar \z}{}^- {\rm E}^{-1}$, there must also be measures over the chiral 
sub-spaces of this superspace.  In the discussion to follow we use $\int d^2 
\s d \z^+ d \z^- {\cal E}^{-1}$ to denote the chiral measure.

Although our previous discussion \cite{GLO} of the irreducible 2D, $N$ =  
2 theories utilized ``covariant spinor notation,'' it is also possible to 
formulate these theories using ``light-cone spinor notation'' as
was done with the (1,0) theory earlier in this section. Using such
notation, the supergravity commutator algebra for the $U(1)$ theory
takes the form below.
$$ 
[ \nabla_+ ~,~ \nabla_+ \} ~=~ 0 ~~~,~~~ [ \nabla_- ~,~ \nabla_- \} 
~=~ 0 
~~~, ~~~ [ \nabla_+ ~,~ \nabla_- \} ~=~ 0  
$$
$$ 
[ \nabla_+ ~,~ {\Bar \nabla}_- \} ~=~ - 2 {\Bar {\cal H}} (~ {\cal M}
~+~ i {\cal Y} ~)  ~~~, ~~~ \nabla_+ {\Bar {\cal H}} ~=~ 
{\Bar \nabla}_- {\Bar {\cal H}} ~=~ 0 ~~~, 
$$
$$ 
[ \nabla_+ ~,~ {\Bar \nabla}_+ \} ~=~ i 2 \nabla_{\pp} ~~~,~~~ 
[ \nabla_- ~,~ {\Bar \nabla}_- \} ~=~ i 2 \nabla_{\mm} ~~~ , 
$$
$$ 
[ \nabla_+ ~,~ \nabla_{\pp} \} ~=~ 0 ~~~,~~~ [ \nabla_- ~,~ \nabla_{\mm} 
\} ~=~ 0 ~~~,   
$$
$$ 
[ \nabla_+ ~,~ \nabla_{\mm} \} ~=~ - i [~{\Bar {\cal H}} \nabla_- ~+~
 (\nabla_- {\Bar {\cal H}}) (~ {\cal M} ~+~ i {\cal Y} ~)
 ~] ~~~,
$$
$$ 
[ \nabla_- ~,~ \nabla_{\pp} \} ~=~ - i [~{{\cal H}} \nabla_+ ~-~
 (\nabla_+ { {\cal H}}) (~ {\cal M} ~-~ i {\cal Y} ~)
 ~] ~~~,
$$
$$ 
[ \nabla_{\pp} ~,~ \nabla_{\mm} \} ~=~ \frac 12 [~ (\nabla_+ { {\cal H}})
{\Bar \nabla}_- ~+~ ({\Bar \nabla}_- {\cal H}) \nabla_+ ~] {~~~~~~~~~~~~} 
$$
$$ 
{~~~~~~~~~~~~~~~} - \frac 12 [~ ({\Bar \nabla}_+ {\Bar {\cal H}} ) \nabla_-
~+~ (\nabla_- {\Bar {\cal H}}) {\Bar \nabla}_+ ~] {~~~~~~~~} 
$$
$$ 
{~~~~~~~~~~~~~~~~~~~} + \frac 12 [~ \nabla_+ {\Bar \nabla}_- { {\cal H}}
 ~-~ {\Bar \nabla}_+ \nabla_- {\Bar {\cal H}} ~-~ 4   { {\cal H}}
{\Bar {\cal H}}~] {\cal M} 
$$
\begin{equation} {~~~~~~~~~~~~~~} - i \frac 12 [~ \nabla_+ {\Bar \nabla}_- { 
{\cal H}} ~+~ {\Bar \nabla}_+ \nabla_-  {\Bar {\cal H}} ~] {\cal Y} ~~~. {~~~~}
\end{equation}

For the $U_A(1)$ theory using a similar notation, we find
$$ 
[ \nabla_+ ~,~ \nabla_+ \} ~=~ 0 ~~~,~~~ [ \nabla_- ~,~ \nabla_- \} ~=~ 0 
~~~, ~~~ [ \nabla_+ ~,~ {\Bar \nabla_-} \} ~=~ 0  
$$
$$ 
[ \nabla_+ ~,~ { \nabla}_- \} ~=~ - 2 {\Bar B} (~ {\cal M}
~+~ i {\cal Y}' ~)  ~~~, ~~~ {\Bar \nabla}_+ B ~=~ 
{\Bar \nabla}_- B ~=~ 0 ~~~, 
$$
$$ 
[ \nabla_+ ~,~ {\Bar \nabla}_+ \} ~=~ i 2 \nabla_{\pp} ~~~,~~~ 
[ \nabla_- ~,~ {\Bar \nabla}_- \} ~=~ i 2 \nabla_{\mm} ~~~ , 
$$
$$ 
[ \nabla_+ ~,~ \nabla_{\pp} \} ~=~ 0 ~~~,~~~ [ \nabla_- ~,~ \nabla_{\mm} 
\} ~=~ 0 
~~~,   
$$
$$ 
[ \nabla_+ ~,~ \nabla_{\mm} \} ~=~ - i [~{\Bar B} {\Bar \nabla}_- ~+~
 ({\Bar \nabla}_- {\Bar B}) (~ {\cal M} ~+~ i {\cal Y}' ~)
 ~] ~~~,
$$
$$ 
[ \nabla_- ~,~ \nabla_{\pp} \} ~=~  i [~{\Bar B} {\Bar \nabla}_+ ~-~
 ({\Bar \nabla}_+ {\Bar B}) (~ {\cal M} ~+~ i {\cal Y}' ~)
 ~] ~~~,
$$
$$ 
[ \nabla_{\pp} ~,~ \nabla_{\mm} \} ~=~ \frac 12 [~ (\nabla_+ B)
{\nabla}_- ~+~ ({ \nabla}_- B) \nabla_+ ~] {~~~~~~~~~~~~} 
$$
$$ 
{~~~~~~~~~~~~~~~} - \frac 12 [~ ({\Bar \nabla}_+ {\Bar B} ) {\Bar 
\nabla}_- ~+~ ({\Bar \nabla}_- {\Bar B}) {\Bar \nabla}_+ ~] {~~~~~~~~} 
$$
$$ 
{~~~~~~~~~~~~~~~~~~~} + \frac 12 [~ \nabla_+ { \nabla}_- B
 ~-~ {\Bar \nabla}_+ {\Bar \nabla}_- {\Bar B} ~-~ 4 B
{\Bar B}~] {\cal M} 
$$
\begin{equation} {~~~~~~~~~~~~~~} - i \frac 12 [~ \nabla_+  \nabla_-  B
 ~+~ {\Bar \nabla}_+{\Bar \nabla}_- {\Bar B} ~] {\cal Y}' ~~~. {~~~~}
\end{equation}

Since the $U_A(1)$ theory more closely resembles the case of 4D, $N$ = 1
theory (reducing the 4D, $N$ = 1 theory to 2D gives the $U_A(1)$
theory), we will first turn our attention to that case.  By looking
at the coefficients of the ${\cal M}$ and ${\cal Y}'$ generators in
(47), we can note that the 2D, $N$ = 2 superfield curvature 
($ {\cal R}_{\pp , \, \mm}$) and $U_A(1)$ field strength ($ {\cal 
F}_{\pp , \, \mm}$) must be given respectively by
\begin{equation}
\eqalign{
{\cal R}_{\pp , \, \mm} &=~  \frac 12 [~ \nabla_+ { \nabla}_- B
 ~-~ {\Bar \nabla}_+ {\Bar \nabla}_- {\Bar B} ~-~ 4 B
{\Bar B}~] ~~~, \cr
{\cal F}_{\pp , \, \mm} &=~  - i \frac 12 [~ \nabla_+  \nabla_-  B
 ~+~ {\Bar \nabla}_+{\Bar \nabla}_- {\Bar B} ~] ~~~.}
\end{equation}
Accordingly it follows that a complex quantity ${\widehat {\cal R}}{}_{\pp , 
\, \mm}$ can be defined to satisfy
\begin{equation}
{\widehat {\cal R}}{}_{\pp , \, \mm} \, \equiv \, {\cal R}_{\pp , \, \mm}  
\, + \, i {\cal F}_{\pp , \, \mm}
~=~  [~ \nabla_+ { \nabla}_- B ~-~ 2 B {\Bar B}~] ~~~.
\end{equation}
Taking the $\q \to 0$ limit, multiplying by ${\rm e}^{-1}$ and integrating
over the two bosonic coordinates yields a complex index.
\begin{equation}
\int d^2 \s ~{\rm e}^{-1} \, [~ {\widehat {\cal R}}{}_{\pp , \, \mm}
\, | ~ ] ~=~  \int d^2 \s ~{\rm e}^{-1} \,  \Big[ ~ [\, \nabla_+ {\nabla}_- 
B ~-~ 2 B {\Bar B}~] \, | ~ \Big] ~~~. 
\end{equation}
It remains only for us to evaluate ${\widehat {\cal R}}{}_{\pp , \, \mm}
\, |$ which is done by the standard method of {\it {Superspace}} to
yield
\begin{equation}
{\widehat {\cal R}}{}_{\pp , \, \mm} \, | ~=~ {\widehat {r}}{}_{\pp , 
\, \mm} ~-~ i 2 {\Bar \psi}_{\mm}{}^- \nabla_+ B | ~+~ i 2 {\Bar 
\psi}_{\pp}{}^+ \nabla_- B | ~+~ 4 ( \, {\Bar \psi}_{\pp}{}^+ {\Bar 
\psi}_{\mm}{}^- \, - \, {\Bar \psi}_{\mm}{}^+ {\Bar \psi}_{\pp}{}^- 
\, ) \, B |  ~~~.
\end{equation}
This gives us the $\l$-coefficients of (9) and upon substitution of 
this into (51), we learn that the chiral density projector may be 
defined by
$$
\int d^2 \z \,{\cal E}^{-1} ~ {\cal L}_c ~=~ {\rm e}^{-1}\, \Big[ \,
[~ \nabla_+ \nabla_- ~+~ i 2 {\Bar \psi}_{\mm}{}^- \nabla_+ ~-~ i 2 
{\Bar \psi}_{\pp}{}^+ \nabla_- ~-~ 2{\Bar B} 
$$
$$
{~~~~~~~~~~~~~~~~~~~~~}  ~-~ 4 \, (~ {\Bar \psi}_{\pp}{}^+ {\Bar \psi
}_{\mm} {}^- ~-~ {\Bar \psi}_{\mm}{}^+ {\Bar \psi}_{\pp}{}^- ~) ~] \,  
{\cal L}_c \, | ~\Big]   ~~~, 
$$
\begin{equation}
{~~~}  ~\equiv~ {\rm e}^{-1} \, {\cal D}^2 {\cal L}_c \, | 
~~~, {~~~~~~~~~~~~~~~~~~~~~~~~~~~~~~~~}
\end{equation}
where ${\Bar {\nabla}}_+ {\cal L}_c \, = \, {\Bar {\nabla}}_- {\cal L}_c \, 
= \, 0 $.
This result is in complete agreement with that derived by other means
in reference \cite{GrWe}.

The 2D (2,2) density projector in (53) depended on our examination
of an index (51) associated with the supergravity covariant derivative
in (48).  We now wish to derive this result by considering an
index that is associated with a 2D (2,2) matter system. Now let us 
add a twisted 2D, $N$ = 2 vector multiplet to the supergravity
covariant derivative so that the commutator algebra becomes,
(where $t$ is a U(1) generator)
$$ 
[ \nabla_+ ~,~ \nabla_+ \} ~=~ 0 ~~~,~~~ [ \nabla_- ~,~ \nabla_- \} 
~=~ 0 ~~~, ~~~ [ \nabla_+ ~,~ {\Bar \nabla_-} \} ~=~ 0  
$$
$$ 
[ \nabla_+ ~,~ { \nabla}_- \} ~=~ - 2 {\Bar B} (~ {\cal M}
~+~ i {\cal Y}' ~)   ~-~ i 4 g' {\Bar {\cal P}} t
~~~, ~~~ {\Bar \nabla}_+ B ~=~ {\Bar \nabla}_- B ~=~ 0 ~~~, 
$$
$$ 
[ \nabla_+ ~,~ {\Bar \nabla}_+ \} ~=~ i 2 \nabla_{\pp} ~~~,~~~ 
[ \nabla_- ~,~ {\Bar \nabla}_- \} ~=~ i 2 \nabla_{\mm} ~~~ , 
$$
$$ 
[ \nabla_+ ~,~ \nabla_{\pp} \} ~=~ 0 ~~~,~~~ [ \nabla_- ~,~ \nabla_{\mm} 
\} ~=~ 0 ~~~,   
$$
$$ 
[ \nabla_+ ~,~ \nabla_{\mm} \} ~=~ - i [~{\Bar B} {\Bar \nabla}_- ~+~
 ({\Bar \nabla}_- {\Bar B}) (~ {\cal M} ~+~ i {\cal Y}' ~) ~] 
 ~+~ 2 g' ({\Bar \nabla}_- {\Bar {\cal P}} ) t
~~~,
$$
$$ 
[ \nabla_- ~,~ \nabla_{\pp} \} ~=~  i [~{\Bar B} {\Bar \nabla}_+ ~-~
 ({\Bar \nabla}_+ {\Bar B}) (~ {\cal M} ~+~ i {\cal Y}' ~)
 ~]  ~+~ 2 g' ({\Bar \nabla}_+ {\Bar {\cal P}} ) t ~~~,
$$
\begin{equation}
[ \nabla_{\pp} ~,~ \nabla_{\mm} \} ~=~ ...~+~ i g' [~ \nabla_+
\nabla_- {\cal P} ~-~ {\Bar \nabla}_+ {\Bar \nabla}_- {\Bar {\cal P}}
~-~ 2 (\, B {\Bar {\cal P}} ~+~ {\Bar B} {\cal P} \,) ~] t
 ~~~. {~~~~}\end{equation}
On the last line above$...$ denotes the supergravity terms that were
present prior to the introduction of the twisted 2D, $N$ = 2 vector
multiplet. 

The final equation above implies,
\begin{equation}
{\cal F}_{\pp , \, \mm}^{U(1)} ~=~ \nabla_+
\nabla_- {\cal P} ~-~ {\Bar \nabla}_+ {\Bar \nabla}_- {\Bar {\cal P}}
~-~ 2 (\, B {\Bar {\cal P}} ~+~ {\Bar B} {\cal P} \,) ~~~,
\end{equation}
and by use of a {\it {Superspace}} technique we find,
\begin{equation}
\eqalign{
{\cal F}_{\pp , \, \mm}^{U(1)} \, |  ~=~ &f_{\pp , \, \mm}^{U(1)}
 ~ - ~ i 2 {\Bar \psi}_{\mm}{}^- \nabla_+ {\cal P}\, |  ~ + ~  i 2 
{\Bar \psi}_{\pp}{}^+ \nabla_- {\cal P} \, | \cr
 & - ~ i 2 {\psi}_{\mm}{}^- {\Bar {\nabla}}_+ {\Bar {\cal P}} \, |  
~ + ~  i 2 {\psi}_{\pp}{}^+ {\Bar {\nabla}}_- {\Bar {\cal P}} \, | \cr
& + ~ 4 \, (\, {\Bar \psi}_{\pp}{}^+ {\Bar \psi}_{\mm}{}^-  ~-~ 
{\Bar \psi}_{\mm}{}^+ {\Bar \psi}_{\pp}{}^- \, ) \, {\cal P} \, | \cr
& - ~ 4 \, (\, {\psi}_{\pp}{}^+ {\psi}_{\mm}{}^-  ~-~ 
{\psi}_{\mm}{}^+ {\psi}_{\pp}{}^- \,) \, {\Bar {\cal P}} \, | 
~~~. }
\end{equation}
Thus, the following result is valid,
\begin{equation}
\int d^2 \s ~ {\rm e}^{-1} \, f_{\pp , \, \mm}^{U(1)} ~=~ \int d^2 \s ~ 
{\rm e}^{-1} \, [~ {\cal D}^2 {\cal P} \, | ~+~ {\Bar {\cal D}}{}^2
{\Bar {\cal P}}\, | ~]  ~~~,
\end{equation}
where ${\cal D}^2$ is precisely the operator in (53).  The matter 
superfield ${\cal P}$ is chiral (i.e. ${\Bar \nabla}_{\pm}
{\cal P} = 0$). In the presence of $U_A(1)$ supergravity, any chiral
superfield satisfies,

A previous investigation of 2D, (2,2) supermeasures \cite{GrWe} has 
also revealed another previously unknown feature. Namely, in addition 
to the measures associated with the full and chiral superspaces, there 
can also exist other local measures! In particular, these exist for twisted 
chiral measures. In order to derives this, once more we go back to the 
pure supergravity commutator algebra and add an ordinary 2D, $N$ = 2 
vector multiplet. Under these circumstances the commutators take the 
forms (below we use ${\Bar \nabla}_+ \Psi ~=~ {\Bar \nabla}_- {\Bar 
\Psi} ~=~ 0 $ as is appropriate for a twisted chiral superfield),
$$ 
[ \nabla_+ ~,~ \nabla_+ \} ~=~ 0 ~~~,~~~ [ \nabla_- ~,~ \nabla_- \} 
~=~ 0 ~~~, ~~~ [ \nabla_+ ~,~ {\Bar \nabla_-} \} ~=~ - i 4 g {\Bar 
\Psi} t  
$$
$$ [ \nabla_+ ~,~ { \nabla}_- \} ~=~ - 2 {\Bar B} (~ {\cal M}
~+~ i {\cal Y}' ~) 
~~~,  $$
$$ 
[ \nabla_+ ~,~ {\Bar \nabla}_+ \} ~=~ i 2 \nabla_{\pp} ~~~,~~~ 
[ \nabla_- ~,~ {\Bar \nabla}_- \} ~=~ i 2 \nabla_{\mm} ~~~ , 
$$
$$ 
[ \nabla_+ ~,~ \nabla_{\pp} \} ~=~ 0 ~~~,~~~ [ \nabla_- ~,~ 
\nabla_{\mm} \} ~=~ 0 ~~~,   
$$
$$ 
[ \nabla_+ ~,~ \nabla_{\mm} \} ~=~ - i [~{\Bar B} {\Bar \nabla}_- ~+~
 ({\Bar \nabla}_- {\Bar B}) (~ {\cal M} ~+~ i {\cal Y}' ~) ~] 
 ~+~ 2  g ({\nabla}_- {\Bar {\Psi}} ) t
~~~,
$$
$$ 
[ \nabla_- ~,~ \nabla_{\pp} \} ~=~  i [~{\Bar B} {\Bar \nabla}_+ ~-~
 ({\Bar \nabla}_+ {\Bar B}) (~ {\cal M} ~+~ i {\cal Y}' ~)
 ~]  ~-~ 2 g ({\nabla}_+ { {\Psi}} ) t ~~~,
$$
\begin{equation}
 [ \nabla_{\pp} ~,~ \nabla_{\mm} \} ~=~ ...~+~ i g [~ \nabla_+
{\Bar \nabla}_- {\Psi} ~-~ {\Bar \nabla}_+ {\nabla}_- {\Bar {\Psi}}  
~] t  ~~~. {~~~~}
\end{equation}
On the last line above $...$ once again denotes the supergravity terms 
that were present prior to the introduction of the 2D, $N$ = 2 vector multiplet.
Also once again, the final equation in (58) above implies,
\begin{equation}
{\widehat {\cal F}}_{\pp , \, \mm}^{U(1)} ~=~ \nabla_+ {\Bar \nabla}_- 
\Psi ~-~ {\Bar \nabla}_+ {\nabla}_- {\Bar {\Psi }} ~~~,
\end{equation}
and by use of a {\it {Superspace}} technique we find,
\begin{equation}
\eqalign{
{\widehat {\cal F}}_{\pp , \, \mm}^{U(1)}\, |  ~=~ &{\widehat f_{\pp 
, \, \mm}}^{U(1)}  ~ - ~ i 2 {\psi}_{\mm}{}^- \nabla_+ \Psi \, |  ~ + ~  
i 2  {\Bar \psi}_{\pp}{}^+ {\Bar \nabla}_- {\Psi} \, | \cr
 & - ~ i 2 {\Bar {\psi}}_{\mm}{}^- {\Bar {\nabla}}_+ {\Bar {\Psi}} \, |  
~ + ~  i 2 {\psi}_{\pp}{}^+ {{\nabla}}_- {\Bar {\Psi}} \, | \cr
& + ~ 4 \, (\, {\Bar \psi}_{\pp}{}^+ {\psi}_{\mm}{}^-  ~-~ 
{\psi}_{\mm}{}^+ {\Bar \psi}_{\pp}{}^- \, ) \, {\Psi} \, | \cr
& - ~ 4 \, (\, {\psi}_{\pp}{}^+ {\Bar \psi}_{\mm}{}^-  ~-~ 
{\Bar \psi}_{\mm}{}^+ {\psi}_{\pp}{}^- \,) \, {\Bar {\Psi}} \, | 
~~~. }
\end{equation}
Thus, the following result is valid,
\begin{equation}
\int d^2 \s ~ {\rm e}^{-1} \, {\widehat f}_{\pp , \, \mm}^{U(1)} ~=~
\int d^2 \s ~ {\rm e}^{-1} \, [~ {\Tilde {\cal D}}{}^2 {\Psi} \, | ~+~ 
{\Bar {\Tilde {\cal D}}}{}^2 {\Bar {\Psi}}\, | ~]  ~~~,
\end{equation}
where ${\Tilde {\cal D}}^2$ is
\begin{equation}
{\Tilde {\cal D}}^2 ~\equiv ~ \nabla_+ {\Bar \nabla}_- ~+~ i 2 
{\psi}_{\mm}{}^- \nabla_+ ~-~ i 2 {\Bar \psi}_{\pp}{}^+ {\Bar \nabla}_-
~-~ 4\, (\, {\Bar \psi}_{\pp}{}^+ 
{\psi}_{\mm}{}^-  ~-~ {\Bar \psi}_{\mm}{}^+ {\psi}_{\pp}{}^- \,)
~~~.
\end{equation}
We thus find the following for the twisted chiral projector, 
$$
\int d \z^+ d {\bar \z}^{-} \,{\Tilde {\cal E}}^{-1} {\cal L}_{tc} ~=~ 
{\rm e}^{-1} [~ \nabla_+ {\Bar \nabla}_- ~+~ i 2 {\psi}_{\mm}{}^- \nabla_+ 
~-~ i 2 {\Bar \psi}_{\pp}{}^+ {\Bar \nabla}_-  {~~~~~~~~~~}  
$$
\begin{equation}
{~~~~~~~~~~~~~~~~~~~~~~~~~~~~}  ~-~ 4\, (~ {\Bar \psi}_{\pp}{}^+ 
{\psi}_{\mm}{}^-  ~-~ {\Bar \psi}_{\mm}{}^+ {\psi}_{\pp}{}^- ~) ~] \, 
{\cal L}_{tc} | ~~~, 
\end{equation}
where ${\Bar {\nabla}}_+ {\cal L}_{tc} \, = \, 
{{\nabla_-}} {\cal L}_{tc} \, = \, 0 $. Here ${\Tilde {\cal E}}^{-1}$ 
is used to denote the twisted chiral density measure.

The result of (63) raises new issues to be resolved regarding what is 
the complete list of density projectors for a given supergravity 
theory. As was first shown in \cite{GrWe}, both chiral and twisted 
chiral density projection formulae exist for 2D, $N$ = 2 superspace.  
It may well be the case that the number of such chiral-like projectors 
occur in theories whenever the are irreducible multiplets whose 
definition are totally expressed in term of first order derivative 
constraints acting on the superfields.

The case of 2D, $N$ = 2 supergravity also gives us a chance to present
another aspect of density projectors. Since the $U(1)$ theory is
distinct from the $U_A(1)$ theory, it possesses a very different
set of density projectors. Instead of repeating the step-by-step
derivation used in the $U_A(1)$ case, here we just summarize the 
results.

For the 2D, $N$ = 2 $U(1)$ supergravity theory, the chiral projector
is given (c.f. (53)) by
$$
\int d^2 \z \, { {\cal E}}^{-1} {\cal L}_{c} ~=~ 
{\rm e}^{-1} [~ \nabla_+ { \nabla}_- ~+~ i 2 {\Bar \psi}_{\mm}{}^- \nabla_+ 
~-~ i 2 {\Bar \psi}_{\pp}{}^+ {\nabla}_-  {~~~~~~~~~~}  
$$
\begin{equation}
 {~~~~~~~~~~~~~~~~~~~~~~~~~~~~}  ~-~ 4 \, (~ {\Bar \psi}_{\pp}{}^+ 
{\Bar \psi}_{\mm}{}^-  ~-~ {\Bar \psi}_{\mm}{}^+ {\Bar \psi}_{\pp}{}^- ~
) ~] \, {\cal L}_{c} | ~~~, 
\end{equation}
where ${\Bar {\nabla}}_+ {\cal L}_{c} \, = \, 
{\Bar \nabla}_- {\cal L}_{c} \, = \, 0 $
and for the twisted chiral density projector (c.f. (63))
$$
\int d \z^+ d {\bar \z}^{-} \, {\Tilde {\cal E}}^{-1} {\cal L}_{tc} ~=~ 
{\rm e}^{-1} [~ \nabla_+ {\Bar \nabla}_- 
~+~ i 2 {\psi}_{\mm}{}^- \nabla_+ ~-~ i 2 {\Bar \psi}_{\pp}{}^+ 
{\Bar \nabla}_- ~-~ 2{\Bar {\cal H}} 
$$
\begin{equation}
{~~~~~~~~~~~~~~~~~~~~~~~~~~~~~}  ~-~ 4 \, (~ {\Bar \psi}_{\pp}{}^+ 
{\psi}_{\mm} {}^- ~-~ {\Bar \psi}_{\mm}{}^+ {\psi}_{\pp}{}^- ~) ~] \,  
{\cal L}_{tc} | ~~~, 
\end{equation}
where ${\Bar {\nabla}}_+ {\cal L}_{tc} \, = \, {{\nabla}}_- 
{\cal L}_{tc} \,  = \, 0 $.
The importance of these last two equations is that they demonstrate
that for distinct superspace geometries, there correspond distinct
density projection operators.

\sect{IV. 3D, $N$ = 1 Local Ectoplasmic Integration}

Applying the same considerations to 3D, $N$ = 1 supergravity \cite{3d} begins
by once again knowing the form of the commutator algebra
$$[ \nabla_{\a} ~,~ \nabla_{\b} \}  = i 2 (\g^c)_{\a \b} \left [ \nabla_c 
- R {\cal M}_c \right ] ~~~, ~~~~~
~~~~~~~~~~~~~~~~~~~~~~~~~~~~~~~~~~ $$
$$~[ \nabla_\a ~,~ \nabla_b \}  =  i (\g_b)_{\a}{}^{\d} [~ \fracm 12 R
\nabla_{\d} ~+~ ( \S_{\d} {}^d ~+~ i \fracm 23 (\g^d)_{\d}{}^{\e}
( \nabla_{\e} R )) {\cal M}_d ~] ~~~~~~~~ $$
$$~~~~~~~~~+~ ( \nabla_{\a} R ) {\cal M}_b  ~~~~~, ~~~~~~~~~~~~~~~~~~~~
~~~~~~~~~~~~~ $$
$$[ \nabla_a ~,~ \nabla_b \}  =   - \fracm 12
\e_{a b c } [~ \S^{\a c} + i \fracm 23 (\g^c)^{\a \b} (\nabla_{\b}
R) ~] \nabla_{\a} ~~~~~~~~~~~~~~~~~~~~~~~~~~ $$
\begin{equation}
~~~~~~~~~~-~  \e_{a b c }[~  {\cal R}^{c d}  ~+~ \fracm 23 \eta^{c d} 
(\nabla^2 R ~-~ \fracm 32 R^2 )~] {\cal M}_d  ~~~, ~~~~~~
\end{equation}
where ${\cal R}^{a b} - {\cal R}^{b a} = \eta_{a b} {\cal R}^{a b} = 
(\g_d)^{\a \b} \S_{\b}{}^d = 0$ and 
\begin{equation}
 \nabla_{\a} \S_{\b}{}^c = i (\g_b)_{\a \b} {\cal R}^{b c} 
~-~ \fracm 23 [~ C_{\a \b} \eta^{c d} ~+~ i \fracm 12 (\g_b)_{\a \b} 
\e^{b c d} ~] \, (\nabla_d R)~~. 
\end{equation}
In writing these results, we have simplified their form by replacing
the usual Lorentz generator according to:
\begin{equation}
{\cal M}_{ b c} \to \e_{b c }{}^a {\cal M}_a ~~, 
\end{equation}
so that when acting on a spinor $\psi_{\a}$ or a vector $v_a$ we have
\begin{equation}
{\cal M}_a \psi_{\a} = i \fracm 12 (\g_a)_{\a}{}^{\b} \psi_{\b}
~~~,~~~{\cal M}_a v_b =  \e_{a b }{}^{c} v_{c}   ~~~~. 
\end{equation}

Three dimensions offer us a new possibility in the class of topological
invariants over the body of the supermanifold.  Here we may introduce
a two-form gauge field ${\cal B}$ whose three-form field strength $g
 = d {\cal B}$ can be used to construct a new class of topological 
invariants given by $\int g$. There exists a super 2-form that generalizes 
this component theory.  The field strength supertensor ${\cal G}_{A B C }$ 
satisfies the equations
\begin{eqnarray}
{\cal G}_{\a \b \g}  & = & 0 ~~~, \nonumber \\
{\cal G}_{\a  \b c} &=& i 2 (\g_{ c })_{ \a \b} { {\cal G}} ~~~, \nonumber \\
{\cal G}_{\a  b c} &=&  i \e_{a b c} (\g^{ a })_{ \a}{}^{ \b} \, ( \nabla_{\b} 
{\cal G} \,) ~~~, \nonumber \\
{\cal G}_{a  b c} & = & \e_{a b c} \, [\, \de^2 {\cal G} ~-~  R {\cal G} \, ] ~~~,
\end{eqnarray}
Now we may use the result of a general procedure to show that the
following equation must be true
\begin{equation}
{\cal G}_{a b c} | ~ = ~ g_{a b c} ~+~ \e_{a b c} \Big[ \, i \psi_a {}^{\a} 
(\g^a )_{\a}{}^{ \b} ( \, \nabla_{\b} {\cal G}| \, ) ~-~ i  \e^{ d e f} \psi_d 
{}^{\a} (\g_e )_{\a \b} \psi_f {}^{\b} (\, {\cal G} | \,) ~\Big]  ~~~~. 
\end{equation}
Combining this result with the last equation in (70) we obtain
\begin{equation}
\fracm 16 \e^{a b c} g_{a b c} = \de^2 {\cal G} |  ~-~ i \psi_a {}^{\a} (\g^a 
)_{\a} {}^{ \b} ( \nabla_{\b} {\cal G} | \,)  ~+~ i  \e^{ a b c} \psi_a {}^{\a} 
(\g_b )_{\a \b} \psi_c {}^{\b} ( {\cal G} | \,) ~-~ ( R {\cal G} | \, ) ~~~~,
\end{equation}
and introducing the notational device ${\cal D}^2$ we find
\begin{equation}
\eqalign{
\fracm 16 \e^{a b c} g_{a b c} &= ~ ( {\cal D}^2 {\cal G} | \,)   \cr
{\cal D}^2 & \equiv ~ \de^2  ~-~ i \psi_a {}^{\a} (\g^a )_{\a} {}^{ \b} \nabla_{\b}   
~+~ i  \e^{ a b c} \psi_a {}^{\a} (\g_b )_{\a \b} \psi_c {}^{\b} ~-~ R 
~~~.}
\end{equation}
Upon multiplying this by ${\rm e}^{-1}$ and integrating both sides, this becomes
\begin{equation}
{\Hat\D} ~\equiv~ 
\int d^3 x \, {\rm e}^{-1} \, \fracm 16 \e^{a b c} g_{a b c} ~=~
\int d^3 x \, {\rm e}^{-1} \, ( \, {\cal D}^2 {\cal G} | \,)  
~\equiv~ {\Tilde \D} ~~~,
\end{equation}
where the index on the far left is defined by the value of the integral
adjacent to it.  According to the Ethereal Conjecture we may define
the Ectoplasmic Integration theorem in 3D, $N$ = 1 superspace so that it reads
\begin{equation}
\int d^3 x d^2 \q \, {\rm E}^{-1} ~ {\cal L} ~ \equiv ~ 
\int d^3 x \, {\rm e}^{-1} \, (\,  {\cal D}^2 {\cal L} | \,)  ~~~,
\end{equation}
and thus
and the superfield topological invariant is given by,
\begin{equation}
{\Tilde \D} ~=~ 
\fracm 1{12} \int d^3 x d^2 \q {\rm E}^{-1} ~ [ \, i (\g^c)^{ \a \b} {\cal 
G}_{\a \b c } \,]  ~~~.
\end{equation}

The reader who has followed our arguments thus far, might offer a challenge
at this point, ``What independent arguments are there to support the 
conclusion that we have correctly identified the density projector?''  
An independent derivation of this density can be obtained via the normal 
coordinate expansion technique[2,4]. Another simple way to support this 
suggestion is to calculate the component version of the superfield expression 
$ \int d^3 x d^2 \q {\rm E}^{-1} R $.  This is known to be the 3D supergravity 
action. To obtain the correct component-level expression depends crucially 
on the form of the local density projector.   By solving the 3D, $N$ = 1 Bianchi
identities we find
\begin{equation}
{\rm R} | ~=~ {\rm B} ~~~, ~~~
\nabla_{\a} {\rm R} | ~=~ -  \e^{a b c} (\g_a)_{\a \b} \Psi_{b c}{}^{\b} 
~+~ i 2 {\rm B} (\g^b)_{\a \b} \psi_b {}^{\b} ~~,  {~~~~~~~~~~~~~~~}
\end{equation} 
\begin{equation}
\nabla^2 {\rm R} | ~=~ - \ha \e^{a b c} {\cal R}_{a b c} (\o) ~-~ i 2 \psi^a
{}^{\a} (\g^b)_{\a \b} \Psi_{a b} {}^{\b} ~+~ 2 {\rm B} \psi^a {}^{\a}
\psi_a {}_{\a} ~+~  i {\rm B} \e^{ a b c} \psi_a {}^{\a} (\g_b )_{\a \b} \psi_c ~~~, 
\end{equation}
where $ \Psi_{a b} {}^{\b}$ is the usual component level gravitino field
strength. This leads to
\begin{equation}
\int d^2 x d^2 \q \, {\rm E}^{-1}\,  R = \int d^3 x {\rm e}^{-1} \left [ \,- \ha 
\e^{a b c} \left ( {\cal R}_{a b c} (\o) ~+~  \psi_a {}_{\a} \Psi_{b c} {}^{\a} 
\right ) ~-~  B^2 \right ]   ~~~, 
\end{equation}
where the curvature tensor ${\cal R}_{a b c}$ is defined in terms of a
spin-connection, 
\begin{equation}
\o_a {}^b ~=~ \fracm 14   \e^{b c d} \left [ ~ C_{ c d a} ~-~ 2 C_{a c d}
~+~ i 4 \left (  \psi_c {}^{\a} (\g_a )_{\a \b} \psi_d {}^{\b} ~+~
 \psi_a {}^{\a} (\g_c )_{\a \b} \psi_d {}^{\b} \right ) ~  \right ]
~-~ \fracm 12 B \d_a {}^b ~~~.
\end{equation}
Again we find support for the Ethereal Conjecture.  We now turn our 
attention to four dimensional $N$ = 1 supersymmetric theories.

\sect{V. 4D, $N$ = 1 Local Ectoplasmic Integration}

Minimal 4D, $N$ = 1 supergravity is described by three fields strength
tensors denoted by $W_{\a \b \g}$, $G_{\a \dot \a}$, and $R$. The form of
the theory is completely described by
$$[ \de_{\a} ~,~ \de_{\b} \} ~=~ - 2 \,{\Bar R} {\cal M}_{\a \b} ~~~,
{~~~~~~~~~~~~~~~~~~~~~~~~~~~}{~~~~~~~~~~~~~~~~~~~~~~~~~~~~}$$
$$[ \de_{\a} ~,~ {\Bar \de}_{\dot \b} \} ~=~ i  \, \de_{\a \dot \b}~~~, 
{~~~~~~~~~~~~~~~~~~~~~~~~~~~~~~~~~~~}{~~~~~~~~~~~~~~~~~~~~~~~~~~~~}$$
$$[ \de_{\a} ~,~ \de_{\underline b} \} ~=~ -~ i C_{\a \b} \, [~ {\Bar R} \,
{\Bar \de}_{\dot \b} ~-~ G^{\g} {}_{\dot \b} \, \de_{\g} ~] 
~-~ i (\, {\Bar \de}_{\dot \b}
{\Bar R} \,) {\cal M}_{\a \b}
{~~~~~~~}{~~~~~~~} $$
$$ {~~~~~~~~~~~~~} +~ i  C_{\a \b} \, [~  {\Bar W}_{\dot \b \dot \g}
{}^{\dot \d}{\Bar {\cal M}}_{\dot \d }{}^{\dot \g} ~-~ (\de^{\g}
G_{\d \dot \b} ) {\cal M}_{\g }{}^{\d} ~] ~~~, {~~~~~~~~~~~}$$
$$[ \de_{\underline a} ~,~ \de_{\underline b} \} ~=~ \{ \, [~
C_{\dot \a \dot \b} {W}_{\a \b} {}^{\g} ~+~  C_{\a \b} (\, 
{\Bar \de}_{\dot \a}  G^{\g} {}_{\dot \b} \,)  
~-~ C_{\dot \a \dot \b} (\, \de_{\a} R  \,)\, \d_{\b} {}^{\g} ~] 
\de_{\g} {~~}\, \,\, \, \,$$
$${~~~~~~~~~~~~~~}\,\, \,  +~i  C_{\a \b} G^{\g} {}_{ \dot \b} \de_{\g 
\dot \a} ~-~ [\, C_{\a \b} (\, {\Bar \de}_{\dot \a} \de^{\d}  G_{\g 
\dot \b} \, )   {~~~~~~~~~~~~~~}$$ 
\begin{equation}
{~~~~~~~~~~~~~~~~}\, \,\, \, \, \,  ~-~ C_{\dot \a \dot \b} (\, \de_{\a}  
{W}_{\b \g} {}^{\d}  ~+~  (\, {\Bar \de}^2 {\Bar R} ~+~ R  {\Bar R} \,) 
C_{\g \b}  \d_{\a} {}^{\d} ) ~] {\cal M}_{\d} {}^{\g} \, \} $$
$$ ~+~ {\rm h}. ~{\rm c}.  {~~~~~~~~~~~~~~~~~~~~~~~~~~~~~~~~~~~~~~} \,
\end{equation}

We now apply our method by introducing a 3-form gauge field matter 
multiplet as first appeared very long ago \cite{pfm}.  In the context 
of 4D, $N$ = 1 superspace, such a supermultiplet is described by a 
super 4-form field strength $F_{\un A \, \un B \, \un C \, \un D}$ that 
is known to satisfy the constraints,
\begin{equation}
F_{\a \, \b \, \g ~ \un D} ~=~ F_{\ad ~ \b \, \g \, \un D} ~=~
F_{\ad \, \b \, \un c \, \un d} ~=~ 0 ~~~,~~~
F_{\a \, \b \, \un c \, \un d} ~=~  C_{\Dot \g \Dot \d} C_{\a (\g}
C_{\d ) \b} {\Bar {\cal F}}  ~~~. 
\end{equation}

After the imposition of the constraints and solving the Bianchi 
identities on $F_{\un A \, \un B \, \un C \, \un D}$ in the presence 
of the old minimal supergravity derivative commutator algebra
we find,
\begin{equation}
\eqalign{ {~~~~~~}
F_{\a \, \un b \, \un c \, \un d} &=~ - \e_{\un a \, \un b \, 
\un c \, \un d} {\Bar \nabla}^{\ad} {\Bar {\cal F}}  ~~~,
~~~ F_{\Dot \a \, \un b \, \un c \, \un d} ~=~  \e_{\un a \, \un b \, 
\un c \, \un d} {\nabla}^{\a} {{\cal F}}  ~~~,
~~~ {\Bar \nabla}_{\ad} \, { {\cal F}} ~=~ 0 ~~~,
\cr
F_{\un a \, \un b \, \un c \, \un d} &=~ i \e_{\un a \, \un b \, 
\un c \, \un d}\,  \Big\{ ~ [ \, (\, {\nabla}^2 \, + \,
3 {\Bar R} \, ) \,{\cal F}\, ]  ~-~ [ \, (\, {\Bar \nabla}^2 
 \, + \, 3 {R} \, ) \, {\Bar {\cal F}}\,]  ~ \Big\} ~~~, }
\end{equation}
\begin{equation}
\e_{\un a \, \un b \, \un c \, \un d} \equiv~  i \,  \frac 12 ~ 
[ ~ C_{\a \b} C_{\g \d} C_{ \Dot \a (\Dot \g } C_{\Dot \d ) \Dot \b}
- C_{\Dot \a \Dot \b} C_{\Dot \g \Dot \d} C_{ \a ( \g } C_{ \d )  
\b} ~] ~~~. 
\end{equation}

The $\q \to 0$ limit of {\it {all}} types of component level 
supercovariantized gauge field strengths is defined in an unambiguous 
manner as was discussed in {\it {Superspace}}. We need only slightly 
generalize the formulae given there to find (c.f. (9))
\begin{equation}
\eqalign{
\Big (  F_{\un a \, \un b \, \un c \, \un d} \,  | ~ \Big) =\, f_{
\un a \, \un b \, \un c \, \un d}\, &+~ [\, \frac 1{3!} \psi_{[ \un a |}
{}^{\a} ( F_{\a \,  | \un b \, \un c \, \un d] } \,| ~) ~+~ \frac 14 
\psi_{[ \un a |}{}^{\a} \psi_{| \un b |}{}^{\b} ( F_{\a \,\b \, | 
\un c \, \un d] } \, | ~)  ~+~ {\rm {h.\, c.}} ~] \cr
&+~ \frac 12 \psi_{[ \un a |}{}^{\a} {\Bar \psi}_{ | \un b |}{}^{
\dot \b} ( F_{\a \, {\dot \b} \, | \un c \, \un d] }  \, | ~) ~-~ [\, 
\frac 1{3!} \psi_{[ \un a |}{}^{\a} \psi_{| \un b |}{}^{\b} \psi_{| 
\un c |}{}^{\g} ( F_{\a \,\b \,  \g \, | \un d] }  \, | ~) ~+~ {\rm 
{h.\, c.}} ~] \cr
&-~  \frac 12 [\, \psi_{[ \un a |}{}^{\a} \psi_{| \un b |}{}^{\b} 
{\Bar \psi}_{| \un c |}{}^{\dot \g} ( F_{\a \, \b \,  {\dot \g} \, | 
\un d] }  \, | ~) ~+~ {\rm {h. \, c.}} ~] \cr 
&-~ [\, \psi_{ \un a }{}^{\a} \psi_{ \un b }{}^{\b} \psi_{\un c}
{}^{\g} \psi_{\un d}{}^{\d} (F_{\a \, \b \, \g \, \d }  \, | ~)  
~+~ {\rm {h.\, c.}} ~] \cr
&-~ [\, \frac 1{3!} \psi_{[ \un a |}{}^{\a} \psi_{| \un b |}{}^{\b} 
{\psi}_{| \un c |}{}^{\g} {\Bar \psi}_{| \un d |}{}^{\dot \d} ( F_{\a \, 
\b \, {\g} \, {\dot \d} }  \, | ~) ~+~ {\rm {h.\, c.}} ~]  \cr
&-~  \frac 14 \psi_{[ \un a |} {}^{\a} \psi_{| \un b |}{}^{\b} 
{\Bar \psi}_{| \un c |}{}^{\dot \g} {\Bar \psi}_{| \un d |}{}^{\dot 
\d} ( F_{\a \, \b \,  {\dot \g} \, {\dot \d} }  \, | ~)  ~~~. }
\end{equation}
Due to the constraints (83) on the 4-form supermultiplet, we can re-write 
this as 
\begin{equation}
f_{\un a \, \un b \, \un c \, \un d} ~=~  \Big( F_{\un a \, \un b \, 
\un c \, \un d} \,| ~\Big)  ~-~ [\,  \frac 1{3!} \psi_{[ \un a |}{}^{\a} 
(F_{\a \, | \un b \, \un c \, \un d] } \,| ~) ~+~ \frac 14 \psi_{[ \un a 
|}{}^{\a} \psi_{| \un b |}{}^{\b} ( F_{\a \, \b \, | \un c \, \un d] } 
\, | ~) ~+~ {\rm {h.\, c.}} ~]  ~~~.~~~~~
\end{equation}
Integrating both sides of this equation leads to,
\begin{equation}
\eqalign{ {~~~~~~}
{\Hat \D} ~=~ \int d^4 x ~  {\rm e}^{-1} \e^{\un a \, \un b \, \un c 
\, \un d} \, \Big[ ~ &\frac 1{4!} \, \Big( F_{\un a \, \un b \, \un c \, 
\un d}\,  | ~\Big)  \, - \, [ \, \frac 1{3!} \psi_{\un a }{}^{\a} 
(F_{\a \, \un b \, \un c \, \un d } \, | ~) \cr 
&+ \, \frac 14 \psi_{\un a}{}^{\a} \psi_{\un b }{}^{\b} 
( F_{\a \, \b \,  \un c \, \un d } \, | ~)  ~+~ {\rm {h.\, c.}} 
~ ] ~ \Big]  ~~~. }
\end{equation}
where ${\Hat \D}$ denotes the supersymmetric version of the index 
described by (11).  Next we use the solution to the Bianchi identities 
for $F_{\un a \, \un b \, \un c \, \un d}$ and $F_{\a \, \un b \, \un c \, 
\un d }$ (from (83)), which upon substitution into (87) yields,
\begin{equation}
\eqalign{ 
{\Hat \D} ~=~ \int d^4 x ~ {\rm e}^{-1} \, \Big[ ~  - i \, (
\, {\cal D}^2 {\cal F} \, | ) ~+~ {\rm {h. \, c.}} ~\Big] ~~~, }
\end{equation}
where the operator ${\cal D}^2$ is defined by
\begin{equation}
{\cal D}^2 ~\equiv ~ \nabla^2 ~+~ i  {\Bar \psi}{}^{\un a}{}_{\dot \a} 
\nabla_{\a} ~+~ 3 {\Bar R} ~+~  \fracm 12 C^{\a \b} {\Bar \psi}_{\un a }
{}^{( \dot \a} \, {\Bar \psi}_{\un b  }{}^{\dot \b )} 
~~~. 
\end{equation}

Now we see that a superdifferential operator appears acting on 
the superfield ${\cal F}$.  This superdifferential operator is,
in fact, the chiral density projector for old minimal supergravity. 
We continue by noting that the chiral density projector
$\int d^2 \q \, {\cal E}^{-1}$ may be defined by the equation
\begin{equation}
\int d^4 x \, d^2 \q  ~ {\cal E}^{-1} {\cal L}_c ~\equiv~ \int d^4 x ~  
{\rm e}^{-1} ~ (\, {\cal D}^2 \, {\cal L}_c \, {\bf |} ~) ~~~,
\end{equation}
for {\it {any}} chiral superfield, ${\cal L}_c$. Finally since any 
chiral superfield ${\cal L}_c$ in the presence of old minimal supergravity 
satisfies ${\cal L}_c = ({\Bar \nabla}{}^2 + R ){\cal L}$, where 
${\cal L}$ is a general superfield, we also have
\begin{equation}
\eqalign{
\int d^4 x \, d^2 \q \, d^2 {\bar \q} ~ {E}^{-1} {\cal L} &\equiv~ 
\fracm 12 \, 
\Big[ ~ \int d^4 x \, d^2 \q  ~ {\cal E}^{-1} ~( ~(\,{\Bar \nabla}{}^2 
~+~ R ) \, {\cal L} ~) ~+~ {\rm {h. \, c.}} ~\Big]  ~~~\cr
&=~   \fracm 12 \,\int d^4 x ~ {\rm e}^{-1} ~ \Big[ ~[ ~(\, {\cal D}^2 
\,(\,{\Bar \nabla}{}^2 ~+~ R ) \,  {\cal L} {\bf |} \,)  ~+~ {\rm {h. \, c.}}
 ~ ] ~\Big] ~~~.
} \end{equation}
On setting ${\cal L} = 1$, we derive the action for old minimal supergravity.  The 
fourth order differential operator $ {\cal D}^4 = \fracm 12 {\cal D}^2 \,(\,{\Bar \nabla}{}^2
 + R ) + {\rm {h.c.}}$ can be expanded in the form
\begin{equation}
\eqalign{
&\int d^4 x \, d^2 \q \, d^2 {\bar \q} ~ {E}^{-1} {\cal L} ~=~ 
\int d^4 x ~ {\rm e}^{-1} ~ \Big\{ ~ \Big[ ~ c^{(2,2) \, \a 
\b \dot \a \dot \b} {\nabla}_{\a} {\nabla}_{\b} {\Bar \nabla}_{
\dot \a} {\Bar \nabla}_{\dot \b} ~ + ~ c^{(1,2) \, \a \dot \a 
\dot \b} {\nabla}_{\a} {\Bar \nabla}_{\dot \a} {\Bar \nabla}_{\dot 
\b} \, \cr
&{~~~~~~~~~~~~~~~~~~~~~~~~~~~~~~~~~~~~~~~~~~~~~~~~~\,~~} + \, c^{(2,0) \, 
\a \b } {\nabla}_{\a} {\nabla}_{\b}  ~ + ~ c^{(0,2) \, \dot \a \dot \b} 
{\Bar \nabla}_{\dot \a} {\Bar \nabla}_{\dot \b} \,  \cr
&{~~~~~~~~~~~~~~~~~~~~~~~~~~~~~~~~~~~~~~~~~~~~~~~~~\,~~} + \,
 c^{(1,0) \, \a } {\nabla}_{\a}  ~ + ~ c^{(0,0)} ~
\Big] ~ {\cal L} | ~+~ {\rm {h.\, c.}} ~ \Big\} ~~~, {~~~~~~}
} \end{equation}
where the coefficients are given by
\begin{equation}
\eqalign{
c^{(2,2) \, \a \b \dot \a \dot \b} &=~ \fracm 14 C^{\a  \b } 
C^{\dot \a \dot \b}  ~~, ~~
c^{(1,2) \, \a  \dot \a \dot \b} ~=~  - i \fracm 12 {\Bar \psi}{}^{\a 
\dot \g}{}_{\dot \g}  C^{\dot \a \dot \b} ~~, ~~
\cr c^{(2,0) \, \a \b} &=~ - \fracm 12 C^{\a \b }  R  ~~~, 
~~~ c^{(0,2) \, \dot \a \dot \b} ~=~  - \fracm 12 C^{ \dot \a  
\dot \b } ( \, 3 {\Bar R} ~+~ \frac 12 C^{\a \b} {\Bar \psi}{}_{ \a
\dot \g}{}^{( \dot \g} \, {\Bar \psi}{}_{ \b \dot \d}{}^{\dot \d )} ~)   
~~~, \cr
c^{(1,0) \, \a } &=~ [~ (\nabla^{\a} R \,)
~+~ i {\Bar \psi}{}^{ \un a}{}_{\dot \a} R ~]  ~~~, \cr
c^{(0,0)} &=~ [~ (\nabla^2 R \,) ~+~ i {\Bar \psi}{}^{ \un a}
{}_{\dot \a} (\nabla_{\a} R\, ) ~+~  3 |R|^2  ~+~ \frac 12 C^{\a \b} 
{\Bar \psi}{}_{\un a}{}^{( \dot \a} \, {\Bar \psi}{}_{\un b}{}^{\dot 
\b )} R ~]  ~~~. \cr
} \end{equation}
This form of the density projector shows that we have achieved our goal
of calculating, from an a priori principle, the form of the coefficients
described abstractly in equation (4). We also see that the respective 
volumes of 4D, $N$ = 1 chiral and full superspaces for old minimal 
supergravity are
\begin{equation}
\eqalign{
\int d^4 x \, d^2 \q ~ {\cal E}^{-1} &=~ \int d^4 x \, {\rm e}^{-1} ~
[ ~ - C_{\dot \a \dot \b} \, c^{(0,2) \, \dot \a \dot \b}~ ] ~~~, \cr
\int d^4 x \, d^2 \q \, d^2 {\bar \q} ~ E^{-1} &=~ \int d^4 x \, 
{\rm e}^{-1} ~[ ~  c^{(0,0)} ~ + {\bar c}^{(0,0)}~ ] ~~~. \cr
} \end{equation}

The nonminimal 4D, $N$ = 1 supergravity formulation is the oldest ``off-shell''
form of the theory known, having first been introduced by Breitenlohner 
\cite{Breit} in a not quite irreducible form.  Nonminimal 4D, $N$ = 1 
supergravity is described by three fields strength tensors just as in the 
minimal case. Here these are denoted by $W_{\a \b \g}$, $G_{\a \dot \a}$, 
and $T_{\a}$. The derivation of the local density projector in the 
theory is much more complicated than in the case of the minimal case. 
So much so that to our knowledge, the explicit density projector for the 
nonminimal theory has never been presented in the literature previously.  
Using the ``Ethereal  Conjecture'', this calculation is rendered very 
simple.  We will next present its explicit derivation.

Our derivation begins by giving an explicit form of the nonminimal 4D, $N$ = 1
supergravity commutators. These may be written as
\begin{equation}
\eqalign{  {~~}
[ \de_{\a} ~,~ \de_{\b} \} ~=~ &\frac 12 T_{(\a} \de_{\b)} ~-~
2 \,{\Bar R} {\cal M}_{\a \b} ~~~, \cr
[ \de_{\a} ~,~ {\Bar \de}_{\Dot \b} \} ~=~ &i  \, \de_{\a \Dot \b}~~~, 
\cr
[ \de_{\a} ~,~ \de_{\underline b} \} ~=~ &\frac 12 \,T_{\b} \de_{\a 
\Dot \b} ~+~ i \,[~C_{\a \b} \, G^{\g} {}_{\Dot \b} ~+~ \frac 12 (\, 
{\Bar \de}_{\Dot \b} {T}_{\a}\, ) \d_{\b}{}^{\g}  ~] \de_{\g} \cr
&-~ i \,[~ C_{\a \b} \, (\de^{\g} G_{\d \Dot \b} ) {\cal M}_{\g}{}^{\d}
~+~  (\, {\Bar \de}_{\Dot \b} {\Bar R} \,) {\cal M}_{\a \b}~] \cr
&+~ i \, C_{\a \b} \, [~  {\Bar W}_{\Dot \b \Dot \g}{}^{\,\Dot \d} \,
{\Bar {\cal M}}_{\Dot \d }{}^{\Dot \g} ~+~  \frac 16 \, (  \,( \de^{\g} 
~+~ \fracm 12 T^{\g} \, ) \,  \de_{\g}\,  {\Bar  T}_{\Dot \g} \,) ~ 
{\Bar {\cal M}}_{\Dot \b}{}^{\Dot \g} ~] ~~~~. }
\end{equation}
The final commutator $[ \de_{\un a} ~,~ \de_{\un b} \}$ can be 
explicitly found from the equation 
\begin{equation}
[ \de_{\underline a} ~,~ \de_{\un b} \} ~=~  - i \, [ {\Bar \de}_{\Dot
\b} ~,~ [ \de_{\un a} ~,~ \de_{\b} \} \, \} 
- i \, [  \de_{\b} ~,~ [ \de_{\un a} ~,~ {\Bar \de}_{\Dot \b} \} \, \}
~~~~. \end{equation}
In the equations above we have also made a choice of the superconformal
symmetry parameter $n = 1$ \cite{SG}, so that we have 
\begin{equation}
{\Bar R} ~=~ - \frac 14 {\de}^{\a} {T}_{\a} ~~~~.
\end{equation}
Other consequences of the constraints (96) are that
\begin{equation}
{\Bar \de}_{\dot \a} [~ R ~-~ \fracm 14 \,  {\Bar T}{}^{\Dot \a} {\Bar 
T}_{\Dot \a} ~] ~=~ 0  ~~~,~~~ {\Bar \nabla}_{\Dot \d} \,
( \, {\Bar \de}{}^2 \, + \, \fracm 34 {\Bar T}^{\Dot \g} {\Bar \de}_{
\Dot \g} \, - \, {R} \, + \, \fracm 14  {\Bar T}^{\Dot \g}  {\Bar 
T}_{\Dot \g}  \, ) ~=~ 0
~~~.
\end{equation}

Now we once again simply use the constraints of the 4-form and solve
its Bianchi identities in the presence of the non-minimal supergravity
commutator algebra to find,
\begin{equation}
\eqalign{ {~~~~~~}
F_{\a \, \un b \, \un c \, \un d} &=~ - \e_{\un a \, \un b \, 
\un c \, \un d} ~[ \, ( \,{\Bar \nabla}^{\ad} ~-~ {\Bar T}^{\ad})\,
{\Bar {\cal F}} ~]   ~~~,~~~ (\,{\Bar \nabla}^{\ad} ~-~ {\Bar 
T}^{\ad}) {\cal F} ~=~ 0 
~~~, \cr
F_{\ad \, \un b \, \un c \, \un d} &=~  \e_{\un a \, \un b \, 
\un c \, \un d} ~[ \, ( \,{ \nabla}^{\a} ~-~ {T}^{\a})\,
{{\cal F}} ~]   ~~~, \cr
F_{\un a \, \un b \, \un c \, \un d} &=~ i \e_{\un a \, \un b \, 
\un c \, \un d}\,  \Big\{ ~\fracm 12 \, [ \,{\nabla}^{\e} \, (\, 
{\nabla}_{\e} \, + \, T_{\e} \, ) \,{\cal F}\, ]  ~-~ \fracm 12
\, [ \,{\Bar \nabla}^{\Dot \e} \,(\,  {\Bar \nabla}_{\Dot \e} \, + 
\, {\Bar T}_{\Dot \e} \, ) \, {{\cal F}}\,]  ~ \Big\} ~~~. }
\end{equation}
So that the quantity $\Hat \D$ here takes the form,
\begin{equation}
\eqalign{ 
{\Hat \D} ~=~ \int d^4 x ~ {\rm e}^{-1} \, \Big[ ~  - i \, (
\, {\cal D}^2 {\cal F} \, | )~+~ {\rm {h. \, c.}} ~\Big] ~~ }
\end{equation}
where now the operator ${\cal D}^2$ is defined by
\begin{equation}
{\cal D}^2 ~\equiv ~ \nabla^2 ~+~ [ \,i {\Bar \psi}{}^{\un a}{}_{\dot 
\a} ~-~ \fracm 12 T^{\a} \,] \nabla_{\a} ~+~ [\, 2 {\Bar R} ~-~ i
{\Bar \psi}{}^{\un a}{}_{\dot \a} T_{\a} ~+~ \fracm 12 C^{\a \b} {\Bar 
\psi}_{\un a }{}^{( \dot \a} \, {\Bar \psi}_{\un b  }{}^{\dot \b )} ~] 
~~~. 
\end{equation}
(This projector provides an explicit example of the comment that was 
made above equation (5) since $c_{(N_F - 1)}$ is not strictly
proportional to the gravitino.  This behavior generically occurs
in the presence of spinorial auxiliary fields.)

If we replace ${\cal F}$ by a chiral Lagrangian ${\Tilde {\cal L}}_c$ 
in (100) above, we seem to have reached our goal of finding a 
chiral-type density projector for the nonminimal theory
\begin{equation}
\int d^4 x \, d^2 \q ~{\cal E}^{-1} ~{\Tilde {\cal L}}_c \,=\, \int d^4 x 
~ {\rm e}^{-1} \,   ( \, {\cal D}^2 {\Tilde {\cal L}}_c \, 
| )   ~~,~~  (\,{\Bar \nabla}_{\ad} \,-\, {\Bar T}_{\ad}) \, {\Tilde 
{\cal L}}_c ~=~ 0 ~~~.
\end{equation}
However, the discerning reader will note a slight remaining problem.
Namely the modified definition of chirality satisfied by ${\Tilde {\cal 
L}}_c$ is not the usual one.  One way to remedy this is to note that 
there exists a superfield $T_{\a}$ satisfying $ \nabla_{\a} T_{\b} +  
\nabla_{\b} T_{\a} = 0$.  The solution of this implies that there exist 
another superfield $T$ such that $T_{\a} = \nabla_{\a} T$.  It therefore 
follows that
\begin{equation}
(\,{\Bar \nabla}_{\ad} \,-\, {\Bar T}_{\ad}) {\Tilde {\cal L}}_c 
\, =\, 0 ~~ \& ~~{\Tilde {\cal L}}_c \, \equiv \, e^{\Bar T} \, {\cal L}_c
~ \to ~ {\Bar \nabla}_{\ad} \,{\cal L}_c ~= ~ 0 ~~~.
\end{equation}
In order to use the superfield $T$ to obtain a component level expression,
we need to observe $T \, | = 0$ in a suitable Wess-Zumino gauge.
Assembling all of these pieces we find for the nonminimal supergravity
theory,
\begin{equation}
\eqalign{
\int d^4 x \, d^2 \q \, d^2 {\bar \q} ~ &{E}^{-1} {\cal L} ~\equiv~ 
\fracm 12 \, \Big[ ~ \int d^4 x \, d^2 \q  ~ {\cal E}^{-1} ~(\,
( \, {\Bar \de}{}^2 \, + \, \fracm 34 {\Bar T}{}^{\Dot \g} {\Bar \de}_{
\Dot \g} \, - \, {R} \, + \, \fracm 14  {\Bar T}{}^{\Dot \g}  {\Bar 
T}_{\Dot \g}  \, ) \, {\cal L} ~) \cr
&~~~~~~~~~~~~~~~~~~~+~ {\rm {h. \, c.}} ~\Big]  ~~~\cr
 &=~  \fracm 12 \,\int d^4 x ~ {\rm e}^{-1} ~ \Big[ ~[ ~(\, {\cal 
D}^2 \, e^{\Bar T} \, ( \, {\Bar \de}{}^2 \, + \, \fracm 34 
{\Bar T}^{\Dot \g} {\Bar \de}_{\Dot \g} \, - \, {R} \, + \, \fracm 14  
{\Bar T}^{\Dot \g}  {\Bar T}_{\Dot \g} \, ) \, {\cal L}\, ) \, 
{\bf |} ~ ] \cr
&~~~~~~~~~~~~~~~~~~~~~~~~+~ {\rm {h. \, c.}}  ~\Big] ~~~.
} \end{equation}
In this equation, the quantity ${\cal E}^{-1}$ is the usual density
for ordinary chiral superfields. In fact it is defined by
\begin{equation}
\int d^4 x \, d^2 \q \, d^2 {\bar \q} {E}^{-1} \, [\,  {\Bar R} ~-~ 
\fracm 14 \, {T}{}^{\a} { T}_{\a}  \, ]^{-1} ~\equiv~ \int d^4 x \, 
d^2 \q  ~ {\cal E}^{-1} ~~~.
\end{equation}
The appearance of ${\Bar T}$ still makes for an ungainly way to
proceed. So we may insert a factor of $1 = exp[- {\Bar T}]  exp[{\Bar T}]$
in front of ${\cal L}$ and ``push'' the factor of $exp[- {\Bar T}]$ through
the chiral projection operator until it annihilates the pre-factor of 
$exp[{\Bar T}]$. Now we can redefine ${\cal L}$ to absorb the other exponential.
The net effect of these operations is to ``change'' the chiral projection
operator according to
\begin{equation}
( \, {\Bar \de}^2 \, + \, \fracm 34 {\Bar T}^{\Dot \g} {\Bar \de}_{
\Dot \g} \, - \, {R} \, + \, \fracm 14  {\Bar T}{}^{\Dot \g}  {\Bar 
T}_{\Dot \g}  \, ) ~\to ~  {\Bar \de}{}^2 ~-~ \fracm 14 {\Bar T}{}^{\Dot\b} 
{\Bar \de}_{\Dot \b} ~+~ {R} ~=~ \fracm 12 {\Bar \de}{}^{\Dot \b} \, [ ~ 
{\Bar \de}_{\Dot \b} ~-~ \fracm 12 {\Bar T}_{\Dot \b} ~] ~~~,
\end{equation}
and the final form of the density projector for the nonminimal supergravity 
theory described by (103) can be cast into the form,
\begin{equation}
\int d^4 x \, d^2 \q \, d^2 {\bar \q} ~ {\rm E}^{-1} \, {\cal L} ~=~ 
\fracm 14 \Big\{ ~ \int d^4 x \,  {\rm e}^{-1}~ [ \, {\cal D}^2 \, 
{\Bar \nabla}{}^{\Dot \b} \, [ ~ {\Bar \nabla}_{\Dot \b} ~-~ \fracm 12 
{\Bar T}_{\Dot \b} ~] \, {\cal L} ~ ]  \, {\Big | } ~+~ {\rm {h. \, c.}}  
~ \Big\} ~~~.
\end{equation}
We believe that it is appropriate to note that equation (107) immediately 
above marks the first time, to our knowledge, that a density projection 
operator for nonminimal supergravity has appeared in the physics 
literature.

Some time ago we proposed \cite{GMOV} that the 4D, $N$ = 1 limit of heterotic 
string theory was most likely to be an unusual formulation of 4D, $N$ 
= 1 supergravity combined with a 4D, $N$ = 1 super gauge 2-form multiplet.
We presently call this the 4D, $N$ = 1 ``$\b$FFC supergeometry''\footnote{
The name ``$\b$FFC'' ($\equiv$ beta-function favored constraints) 
was suggested by H.Nishino.}.  The form of the commutator algebra for this
formulation is 
$$\eqalign{
[ \nabla_{\a }, \nabla_{\b} \}  &= 0   ~~, ~~ \cr
[ \nabla_{\a }, { \Bar {\nabla}}_{\dot \a} \}  &=~  
i {\nabla}_{ \un a} ~+~ H_{\b \dot \a} {\cal M}_{\a } {}^{\b} 
~-~ H_{\a \dot \b} {\Bar {\cal M}}_{\dot \a } {}^{\dot \b}~+~ 
H_{\un a} {\cal Y}  ~~~, \cr
[ \nabla_{\a }, \nabla_{\un b} \}  &=~  i (\nabla_{\b} H_{\g \dot \b}) 
\, [~{\cal M}_{\a} {}^{ \g} ~+~ \d_{\a} {}^{\g} \, {\cal Y} ~]  ~~ \cr
&~~~~~+~ i  [~ C_{\a \b} \, {\Bar W}_{\dot \b \dot \g} {}^{\dot \d}
~-~ \frac 13 \d_{\dot \b} {}^{\dot \d} (\, 2 \nabla_{\a} H_{\b \dot \g}
~+~ \nabla_{\b} H_{\a \dot \g} \, ) ~ ] {\Bar {\cal M}}_{\dot \d} 
{}^{\dot \g}  ~~~, {~~~~~~~~~~~~~~~~~~} \cr } $$
\begin{equation}
\eqalign{ 
[ { \nabla}_{\un a}, \nabla_{\un b} \} ~=~ \big \{ &~ \frac12  C_{\a 
\b } [~ i \, H^{\g} {}_{ ( \dot \a} \nabla_{ \g \dot \b)} ~-~ ( 
\nabla^{\g} {\Bar \nabla}_{ ( \dot \a } H_{\g \, \dot \b )} ) \, {\cal Y} 
~] \cr
&+~  [~ C_{\dot \a \dot \b} ( ~ W_{ \a \b} {}^{ \g} ~-~ \frac 16 
( {\Bar \nabla}{}^{\dot \g} H_{ ( \a \dot \g}) \d_{ \b )} {}^{\g}
\,) ~-~ \frac 12 C_{\a  \b} ( {\Bar \nabla}_{ ( \dot \a } H^{\g} 
{}_{\dot \b ) })~] \nabla_{\g} \cr
&-~  C_{\dot \a \dot \b} \, [~ W_{ \a \b \g \d} ~+~ i \frac 14 
C_{\g (\a | } ({\nabla}_{| \b ) }{}^{ \dot \e} H_{ \d \dot \e}) ~+~ 
\frac 1{12} C_{\g (\a }  C_{\b ) \d } ({\nabla}^{\e} {\Bar 
\nabla}{}^{\dot \e } H_{ \e \dot \e})~] {\cal M}^{\g \d} \cr
&+ \frac 12 C_{\a \b } [~{\nabla}_{ \g} {\Bar \nabla}_{ ( \dot \a } 
H_{\d} {}_{\dot \b ) } ~] {\cal M}^{\g \d} ~+~ {\rm h.c.} ~~~ \big \} ~~.  
\cr }   \end{equation}
The self-consistency of the Bianchi identities of the commutator
algebra above requires that the following differential equation
must also be satisfied.
$$ \nabla^{\un a} H_{\un a} ~=~0 ~~,~~ {\Bar \nabla}_{\dot \b} 
W_{\a \b \g} ~=~ 0 ~~,~~  \nabla^{\b}\nabla_{\b} H_{\un a} 
~=~ 0~~~, $$
\begin{equation}
\nabla^{\a} W_{\a \b \g} ~=~ - \frac16 \nabla_{( \b} 
{ \Bar \nabla}{}^{\dot \g} H_{\g ) \dot \g} ~-~ \frac 12 
{ \Bar \nabla}{}^{\dot \g} \nabla_{( \b} H_{\g ) \dot \g}~~~~.~~
\end{equation}
Finally, since the theory contains a gauge 2-form, there occurs a
super 3-form field strength whose various components are given
by
$$ 
H_{\a \b \g} ~=~ H_{\a \b \dot \g} ~=~ H_{\a \b \g} ~=~ H_{\a \b 
\un c} ~=~ H_{\a \dot \b \un c} ~-~ i \frac 12 C_{\a \g} C_{\dot \b 
\dot \g} ~=~ 0  ~~~, $$
\begin{equation}
{H}_{\a \un b \un c} ~=~ 0 ~~,~~ {H}_{\un a \un b \un c} 
~=~ i \frac 14  [~ C_{\b \g} C_{ \dot \a ( \dot \b } {H}_{\a \dot 
\g )} ~-~ C_{ \dot \b \dot \g } C_{ \a ( \b } {H}_{ \g ) \dot \a
} ~ ]  ~~. 
\end{equation}

Repeating the step of solving the Bianchi identity in the presence
of the constraints leads to
\begin{equation}
\eqalign{ {~~~~~~}
F_{\a \, \un b \, \un c \, \un d} &=~ - \e_{\un a \, \un b \, 
\un c \, \un d} {\Bar \nabla}{}^{\ad} {\Bar {\cal F}}  ~~~,
~~~ F_{\Dot \a \, \un b \, \un c \, \un d} ~=~  \e_{\un a \, \un b \, 
\un c \, \un d} {\nabla}^{\a} {{\cal F}}  ~~~,
~~~ {\Bar \nabla}_{\ad} \, { {\cal F}} ~=~ 0 ~~~,
\cr
F_{\un a \, \un b \, \un c \, \un d} &=~ i \e_{\un a \, \un b \, 
\un c \, \un d}\,   ~ [ \, {\nabla}^2 \,{\cal F}  ~-~ 
 {\Bar \nabla}^2  {\Bar {\cal F}} \,]   ~~~. }
\end{equation}
These are substituted into (87) and as previously, we see that a 
superdifferential operator appears acting on the superfield ${\cal F}$.  
This time the superdifferential operator is the chiral density projector 
for $\b$FFC supergravity.  We continue by noting that the chiral density 
projector $\int d^2 \q \, {\cal E}^{-1}$ may be defined by the equation
\begin{equation}
\eqalign{
\int d^4 x d^2 \q  ~ {\cal E}^{-1} {\cal L}_c ~\equiv& ~\int d^4 x ~  
e^{-1} ~ (\, {\cal D}^2 \, {\cal L}_c \, | ~) ~~~, \cr
{\cal D}^2 ~\equiv& ~\nabla^2 ~+~ i  {\Bar \psi}{}^{\un a }{}_{
\dot \a} \nabla_{\a} ~+~ \fracm 12 C^{\a \b} {\Bar \psi}_{\un a }
{}^{( \dot \a} \, {\Bar \psi}_{\un b  }{}^{\dot \b )} 
 ~~~. }
\end{equation}
for {\it {any}} chiral superfield, ${\cal L}_c$. Finally since any 
chiral superfield ${\cal L}_c$ in the presence of $\b$FFC supergravity 
satisfies ${\cal L}_c = {\Bar \nabla}{}^2 {\cal L}$, where 
${\cal L}$ is a general superfield, we also have
\begin{equation}
\int d^4 x d^2 \q d^2 {\bar \q} ~ {E}^{-1} {\cal L} ~\equiv~ \fracm 12 
\Big[ ~
\int d^4 x d^2 \q  ~ {\cal E}^{-1} ~( \,{\Bar \nabla}{}^2 {\cal L}
~) ~+~ {\rm {h. \, c.}} ~\Big]  ~~~. 
\end{equation}
On setting ${\cal L} = 1$, we derive 
\begin{equation}
V_{\b FFC} ~=~ \int \, d^4 x \, d^2 \q \, d^2 {\bar \q} ~ {\rm E}^{-1}
~=~ 0 ~~~,
\end{equation}
since any purely derivative operator acting on a constant vanishes, i.e. 
the volume of the full $\b$FFC superspace vanishes.  Interestingly enough, 
however, the volume of the 4D, $N$ = 1 $\b$FFC chiral superspace is 
non-vanishing.  If we introduce a complex parameter $\m_c $ we find
\begin{equation}
\eqalign{
V_{\b FFC}^c &=~ \int d^4 x d^2 \q  ~ {\cal E}^{-1} ~ \m_c 
~=~ \int d^4 x ~ 
e^{-1} [\, {\cal D}^2_{\b} ( \, \m_c \, | \,)~  ] \cr
&=~ \fracm 12  {\m_c} \, \int d^4 x ~ e^{-1} ~ [ \, {\Bar 
\psi}_{\a ( \dot \a | } {}^{\dot \a} {\Bar \psi}{}^{\, \a}
{}_{| \dot \b ) }{}^{\dot \b} \,  \, ] ~~~.}
\end{equation}
This last expression (taken together with its conjugate) is recognizable 
as a mass term for the gravitino.  Thus we find the very elegant result
that the mass of the gravitino in 4D, $N$ = 1 $\b$FFC supergravity is
proportional to the volume of chiral superspace.  Since we have proposed that
$\b$FFC supergeometry is the limit of the supergravity theory associated
with heterotic and superstring theory, the results in (114) and (115) must
have interesting ``string'' implications.

We simply close this section by noting that we have presented a number of 
results that have not appeared previously to our knowledge.  These were
given in (107), (114) and (115). With this we end our discussion of examples
of how to derive density projectors. The methods that we have described in the 
last two sections may be extended to a number of other supergravity theories
with no impediments.

\sect{VI. Stage-II Density Projection Operators}

In the previous sections, we have presented evidence from a wide number 
of examples which show that density projection operators can often be 
{\it {derived}} by looking at topological indices. One feature of all 
of the examples is that the construction of the density projection operators 
were obtained by essentially algebraic means.  We shall call these Stage-I 
density projection operators. There are cases, however, where this is 
not the case and the derivation of density projection operators seem 
to require solving for some ``prepotential{\footnote {By prepotential, we 
mean in the original sense for which the word was coined \cite{gspot}, not 
in the \newline ${~~~~\,}$ more recent usage as widely appears in $N$ = 2 
SUSY YM theory.}.''   The density projection operators of this type
will be called Stage-II density projection operators.  Derivations for 
Stage-II projectors are more complicated than for Stage-I projectors. As 
an illustration of such a theory, we shall describe the treatment
of the local 2D, $N$ = 4 theory.

A minimal off-shell 2D, $N$ = 4 supergravity theory \cite{GMOV} consists of 
the component fields $ (e_a{}^m , ~ \psi_a {}^{\a i}, ~ A_{ai} {}^j , ~ B,  
~ G, ~H ) $.  These are the components of that remain after imposing the 
following constraints on the 2D, $N$ = 4 superspace supergravity covariant 
derivative, (with $ \phi_{\a \, \b} \equiv - i [  C_{\a \b} G + i (\g^3
)_{\a \b} H ] $)
\begin{equation}
\begin{array}{lll}
~[\de \sb{\a i} , \de \sb{\b j} \} & = & 2 \bar B [C \sb {\a \b}
C \sb{ ij} \cm ~-~ ( \g \sp 3 ) \sb{ \a \b} \cy \sb{i j}] ~~,  \\ 
~[\de \sb{\a i} , \bar \de \sb{\b} \sp j \}   &=& 2  [~i  \d \sb i \sp j 
(\g \sp{c}) \sb{\a \b} \de \sb{c} ~+~  \d \sb i \sp j \phi \sb{\a}{}^{\g}
(\g^3)_{\g \b}  \cm  ~-~ i \phi \sb{\a \b} \cy \sb i \sp j  ~] ~~,   \\
~[\de \sb{\a i} , \de \sb{b} \}   &=& i {1 \over 2} \phi \sb {\a} \sp{\g}
(\g \sb b) \sb \g \sp {\b} \de \sb{\b i} ~+~ i {1 \over 2} (\g \sp 3 \g \sb b) 
\sb {\a} \sp{ \b} \bar B C \sb{i j} \bar \de \sb \b \sp j   \\     
&~&~~ -~ i(\g \sp 3 \g \sb b) \sb {\a \b}
\bar \S \sp \b \sb i \cm ~+~ i (\g \sb b) \sb{\a \b} \bar \S \sp \b \sb j
\cy \sb i \sp j ~~,  \\  
~[\de \sb{a} , \de \sb{b} \}   &=& - \ha \e \sb{a b} 
[ (\g \sp 3) \sb \a \sp \b \S \sp{ \a i} \de \sb{ \b i}
~+~ (\g \sp 3) \sb \a \sp \b \bar \S \sp \a \sb i \bar \de \sb \b \sp i
~+~ \car \cm ~+~ i \cf \sb i \sp j \cy \sb j \sp i ]~~ .   
\end{array}       
\end{equation}
The consistency of the Bianchi identities constructed from the commutator
algebra above required the conditions, 
\begin{equation}
\begin{array}{lll}
\bar \de \sb \a \sp i B &=&0~~~~~~,~~~
\de \sb{\a i} B = -2 C \sb{i j} ( \g \sp 3 ) \sb{ \a \b} \S \sp{ \b j}~~,
 \\
\de \sb{\a i} G &=& \bar \S \sb{ \a i}~~~,~~~  
\de \sb{\a i} H = i( \g \sp 3 ) \sb{ \a} \sp{ \b} \bar \S \sb{ \b i},
~~~, \\
\bar \de \sb \a \sp i \S \sp{\b j} &=& i  C \sp{i j}
(\g \sp 3 \g \sp a) \sb {\a } \sp{ \b} \de \sb a B  ~~,  \\
\de \sb{\a i} \S \sp{\b j} &=& {1 \over 2} \d \sb \a \sp \b
\d \sb i \sp j [ \car ~-~ 2 G \sp 2 ~-~ 2 H \sp 2 ~-~ 2 B \bar B ]
~+~ i (\g \sp 3) \sb \a \sp \b \cf \sb i \sp j  \\
& &+~ i {1 \over 2} \d \sb i \sp j (\g \sp a ) \sb \a \sp \b (\de \sb a G) 
-~  {1 \over 2} \d \sb i \sp j (\g \sp 3 \g \sp a ) \sb \a \sp \b 
(\de \sb a H) ~~~.     
\end{array}     
\end{equation}
The component gauge fields occur in the above supertensors in the 
following manner.  
\begin{equation}
\begin{array}{lll} 
\car {\big |} &=& \e^{a b} {\large \{ } ~ {\car}_{a b} (\hat \omega) ~+~  
[ ~ i 2 (\g^3 \g_a) _{\a \b} \psi_b {}^{\a i} \bar \S^\b {}_i ~+~ {\rm 
{h.c.}} ~  ]  \\ 
& &~~~~~~ +~ 4 \phi_\a {}^\g (\g^3)_{\g \b} \psi_a {}^{\a i} {\bar \psi}_b 
{}^\b {}_i  ~-~2  [~ C_{i j} \bar B \psi_a {}^{\a i}  \psi_{b \a}{}^{j}
~+~ {\rm {h.c.}} ~  ]~ {\large \} } ~~, \\
&   & \\
\S ^{\a i} {\big |} & = & \e^{ab} {\large \{ } ~ \psi_{ab} {}^{\b i}
(\g^3)_\b {}^\a ~-~ i \psi_a {}^{\b i} {\phi}_\b {}^\g 
( \g^3 \g_b)_\g {}^\a   ~+~ i C^{ij} B \bar \psi_a {}^\b {}_j 
(\g_b)_\b {}^\a  ~ {\large \} } ~~, \\ 
&   & \\
\cf_i {}^j {\big |} &=& \e^{ab} {\large \{ } ~ {\rm F}_{a b}(A)_{i}{}^j 
~-~ i 2  (\g_a)_{\a \b}  [~ \psi_b {}^{\a j} \bar \S^\b {}_i ~+~ \bar \psi_b 
{}^\a {}_i \S^{\b j} \\ 
& &~~~~~~~~~~~~~~~~~~-~ \frac 12 \d_i^j (\psi_b {}^{\a k} \bar \S^\b {}_k 
~+~  \bar \psi _b {}^\a {}_k \S^{\b k}) ~] \\
& & ~~~~~~~~~~~~~~~~~~-~ 4  \phi_{\a \b} [\psi_a {}^{\a j} \bar
\psi_b{}^\b{}_i ~-~ \frac 12 \d_i^j \psi_a {}^{\a k} \bar
\psi_b{}^\b{}_k ] \\             
& & ~~~~~~~~~~~~~~~~~~-~ 2 (\g^3)_{\a \b} [~\bar B ( C_{i k} \psi_a {}^{\a k} 
\psi_b {}^{\b}{}^{k} ~-~ \frac 12 \d_i^j C_{k l} \psi_a {}^{\a k} \psi_b 
{}^{\b l})
\\ & & ~~~~~~~~~~~~~~~~~~+~ B ( C^{j k} \bar \psi_a {}^{\a}{}_{i}
 \bar \psi_{b}{}^{ \b}{}_{k} ~-~ \frac 12 \d_i^j  C^{k l} \bar \psi_a{}^{\a}
{}_{k} \bar \psi_{b}{}^{\b}{}_{l})~ ] ~{\large \} } ~~ ,   \end{array}        
\end{equation}
where $r(\hat \omega)$ is the usual two-dimensional curvature in
terms of ${\rm e}_a {}^m$ and $\hat \omega_m $. 

Now let us repeat the by now familiar steps which follow from our presentation
thus far. This begins by totally contracting the indices on the last equation in 
(117) yielding
\begin{equation}
\fracm 12 \de \sb{\a i} \S \sp{\a i} ~=~ [ \car ~-~ 2 G \sp 2 ~-~ 2 
H \sp 2 ~-~ 2 B \bar B ] ~~~.
\end{equation}
We next use the first result of (118) to write
\begin{equation} 
\begin{array}{lll}
\e^{a b} \, {\car}_{a b} (\hat \omega) &=& \Big[ ~ \fracm 12 \de 
\sb{\a i} \S \sp{\a i} ~+~ 2 G \sp 2 ~+~ 2 H \sp 2 ~+~ 2 B \bar B \\
&{~}& {~~~~} -\,  \e^{a b} \, [ ~ i 2 (\g^3 \g_a) _{\a \b} \psi_b {}^{\a i} 
\bar \S^\b {}_i ~+~ {\rm {h.c.}} ~  ]  \\ 
& &~~~~ -~ 4\, \e^{a b} \,  \phi_\a {}^\g (\g^3)_{\g \b} \psi_a {}^{\a i} 
{\bar \psi}_b {}^\b {}_i  \\ & &~~~
~+~2  \e^{a b} \, [~ C_{i j} \bar B \psi_a {}^{\a i}  \psi_{b \a}{}^{j}
~+~ {\rm {h.c.}} ~  ]~~ \Big] \,  {\Big |} ~~~.
\end{array} 
\end{equation}
Upon multiplying both sides of this equation by $-\frac 12$, integrating
over the 2D manifold and using (11), we find
\begin{equation}
\begin{array}{lll}
{\Hat \D}   &=& \int d^2 \s \,{\rm e}^{-1} ~ \Big\{ 
\Big[ ~ \fracm 12 \de \sp{\a i} {\Bar \de} \sb{\a i} G ~+~  \e^{a b} 
\, [ ~ i  (\g^3 \g_a) _{\a \b} \psi_b {}^{\a i} \de^\b {}_i G~+~ {\rm {h.c.
}} ~  ] {~~~~~~~~~~~~} \\
&{~}& {~~~~~~~~~~~~~~~~~~} -~  G \sp 2 ~-~  H \sp 2 ~-~  B \bar B 
~+~ 2\, \e^{a b} \, \psi_a {}^{\a i} {\bar \psi}_b {}^\b {}_i 
 \, \phi_\a {}^\g (\g^3)_{\g \b}  \\ 
& &{~~~~~~~~~~~~~~~~~~} -~  \e^{a b} \, [~ C_{i j} \psi_a {}^{\a i}  
\psi_{b \a}{}^{j} \bar B  ~+~ {\rm {h.c.}} ~  ]~~ \Big] ~ {\Big |} ~ \Big\} ~~~.
\end{array} 
\end{equation}

The fact that this superspace is substantially different from our previous
cases can be seen by noting that this last result is {\it {not}} of the
form of some differential operator acting on a single superfield which 
is characteristic of a Stage-II density projector.  The basic problem
is that $G$, $H$ and $B$ {\it {all}} simultaneously appear in this
expression and there are no algebraic relations among them.

At first this seems to be an insurmountable problem. In fact, we are
presently aware of three different ways to find an explicit expression
for a 2D, $N$ = 4 density projector. Although we will not pursue these
to their logical conclusion\footnote{This will be the topic of a future 
work.}, we deem it useful to discuss how these approaches work at 
least in principle. Let us call these three methods;
 
(a.) the 2D, $N$ = 4 SG variational method,

(b.) the 2D, $N$ = 4 VM-I variational method,

(c.) the 2D, $N$ = 4 VM-II variational method.

All three of these methods can be used and remarkably enough, {\it {none}}
of them require the process of solving the constraints in terms of 
prepotentials.  This was the reason why we used the word ``seem'' in the 
definition of Stage-II projectors.  They each rely instead on an observation 
first noted by Wess and Zumino \cite{wz}.  Namely, in a constrained supersymmetric 
gauge theory, the complete set of variations which preserve the constraints 
may be derived by a consistency method applied to the constraints.  It was 
by use of this observation that the first proof of the correct superspace
supergravity action was demonstrated in the literature.

\noindent
(a.)  2D,~ $N$ \,=\, 4 ~ SG 

Applying the approach of Wess and Zumino, one is ultimately led to
a set of equations of the forms
\begin{equation}
\begin{array}{lll}
\d G &=& {\cal D}_1 {\cal V} ~~~, \\
\d H &=& {\cal D}_2 {\cal V} ~~~, \\
\d B &=& {\cal D}_3 {\cal V} ~~~, 
\end{array} 
\end{equation}
where ${\cal D}_i$ are certain differential operators that are
derived simultaneously with the derivations of the unconstrained
variations here denoted by ${\cal V}$.  These differential operators
are then substituted into (122) appropriately 
\begin{equation}
\begin{array}{lll}
{\Hat \D}   &=& \int d^2 \s \,{\rm e}^{-1} ~ \Big\{ 
\Big[ ~ \fracm 12 \de \sp{\a i} {\Bar \de} \sb{\a i} {\cal D}_1 ~+~  \e^{a b} 
\, [ ~ i  (\g^3 \g_a) _{\a \b} \psi_b {}^{\a i} \de^\b {}_i 
{\cal D}_1 ~+~ {\rm {h.c.
}} ~  ] {~~~~~~~~~~~~} \\
&{~}& {~~~~~~~~~~~~~~~~~~} -~  G {\cal D}_1 ~-~  H {\cal D}_2 ~-~  
\fracm 12 B {\Bar {\cal D}_3} ~-~  \fracm 12 {\Bar B} {{\cal D}_3} \\
&{~}& {~~~~~~~~~~~~~~~~~}~+~ 2\, \e^{a b} \, \psi_a {}^{\a i} 
{\bar \psi}_b {}^\b {}_i  \,[ \, C_{\a \b} {\cal D}_2 \, -\, i \,
(\g^3)_{\a \b} {\cal D}_1 \,]  \\ 
& &{~~~~~~~~~~~~~~~~~~} -~  \e^{a b} \, [~ C_{i j} \psi_a {}^{\a i}  
\psi_{b \a}{}^{j} {\Bar {\cal D}_3}  ~+~ {\rm {h.c.}} ~  ]~~ \Big] ~ 
{\cal V} ~{\Big |} ~ \Big\} ~~~.
\end{array} 
\end{equation}
and lead to a density projector as described by (11). 

\noindent
(b.) and (c.)  2D,~ $N$ \,=\, 4 ~ VM-II, \, VM-II 

The methods utilizing the 2D, $N$ = 4 vector multiplets essentially work 
in the same manner. One first calculates ${\Hat \D}$ starting from the 
rigid results (see appendix D) then covariantizes with respect to 
supergravity. After this is done, the variational approach of Wess and 
Zumino is applied {\it {only}} to the portions of the commutator 
algebra describing the matter multiplets. In these cases, this leads 
ultimately to a set of equations analogous to (122) which are then 
substituted in the expressions for the respective ${\Hat \D}$'s to 
yield an explicit expression for the projector.

\sect{VII. The Ethereal Conjecture and Superspace Constraints}

So far by the survey of numbers of examples, we hope that the reader
has found our arguments (which support the Ethereal Conjecture),
so convincing that it may be taken as a working hypothesis to study 
aspects of supersymmetry theories that have been mysterious for many 
years.  One such aspect is the matter of the constraints themselves.  
In the introduction we alluded to our belief that the fundamental 
reason that constraints are imposed in supersymmetrical field theories 
may also be due to the Ethereal Conjecture. In this section we wish to
display some evidence for how such a vague belief can be supported
by explicit calculations. In order to illustrate this, we shall 
discuss some of its implications within the context of 2D, (1,0) 
theory where all calculations are easily carried out.

Let us begin by introducing a (1,0) supervielbein denoted
by ${\Bar {\rm E}}_A {}^M$ to distinguish it from the (1,0)
supervielbein ${\rm E}_A {}^M$ of equation (13) via
\begin{equation}
{\Bar {\rm E}}_A {}^M ~=~ {\cal A}_A {}^B {\rm E}_B {}^M  ~~~. 
\end{equation}
Since the superfield quantities ${\cal A}_A {}^B$ are completely
arbitrary, the supervielbein ${\Bar {\rm E}}_A {}^M$ correspondingly is 
completely arbitrary satisfying no constraints. Let us note that
the decomposition in (124) is quite general and may be applied
to more complicated supergravity theories.  For example, the
fiducial vielbein (above denoted by ${\rm E}_B {}^M$) can be chosen
to describe an on-shell supergravity theory.

Our strategy is quite simple.  Even in a (1,0) superspace geometry 
defined by ${\Bar {\rm E}}_A {}^M$, supergravity covariant derivatives 
\begin{equation}
{\Bar \nabla}_A ~\equiv ~ {\Bar {\rm E}}_A ~+~ {\Bar \o}_A \, {\cal M} ~~~,
\end{equation}
can be defined.  After the introduction of the connections, in turn 
these unconstrained supergravity covariant derivatives lead to 
unconstrained superspace torsion and curvature superfields via the 
equations,
\begin{equation}
\eqalign{
[ {\Bar \nabla}_+ , {\Bar \nabla}_+ \} &=~  ( ~ {\Bar T}_{+ \, , \, +}{}^B 
{\Bar \nabla}_B + {\Bar {\cal R }}_{+ \, , \, +} \, \cm ) ~~,~~   \cr
[ {\Bar \nabla}_+ , {\Bar \nabla}_{\pp} \}  &=~ ( ~ {\Bar T}_{+ \, , \, \pp}{}^B 
{\Bar \nabla}_B + {\Bar {\cal R}}_{+ \, ,\, \pp} \, \cm ) ~~ , ~~  \cr
[ {\Bar \nabla}_{+} , {\Bar \nabla}_{\mm} \} &=~  ( ~ {\Bar T}_{+\, , \, \mm}
{}^B {\Bar \nabla}_B + {\Bar {\cal R}}_{+ \, ,\, \mm} \, \cm )  ~~ , ~~ \cr
[ {\Bar \nabla}_{\pp} , {\Bar \nabla}_{\mm} \}  &=~ - ~ ( ~ {\Bar T}^B {\Bar 
\nabla}_B + {\Bar {\cal R}} \, \cm ) ~ \quad . }
\end{equation}
In this modified theory, we have imposed {\it {no}} {\it a} {\it {priori}}
constraints. An important question to ask is, ``How are the topological 
indices of the modified theory related to the indices defined previously?''

Let us attempt to answer this by investigating a brief calculation.  On the 
basis of dimensional analysis and Lorentz covariance, the
quantity\footnote{The rules of ``superspace conjugation \cite{ggrs}'' 
are such that ${\Hat {\cal X}}$ as 
defined is a real quantity.}  $({\Hat {\cal X}})$ defined by
\begin{equation}
{\Hat {\cal X}} ~=~ i \, \fracm 12 \, \int d^2 \s \, d \zeta^- \, {\Bar 
{\rm E}}{}^{-1} ~ {\Bar {\cal R}}_{+ \, ,\, \mm} ~~~,
\end{equation}
is a candidate  to describe an index in the (1,0) superspace.  By use of 
the normal coordinate technique as espoused in \cite{ectonorcor,norcor}, 
this becomes
\begin{equation}
{\Hat {\cal X}} ~=~  i \,  \fracm 12 \,\int d^2 \s ~\Big\{ ~ {\rm e}^{-1} ~ 
\Big[ \, {\Bar \de}_+ ~+~ {\Bar T}_{a\, , \, +} {}^a ~+~ \fracm 12 \, {\Bar 
T}_{+\, , \, +}{}^+ ~-~ \fracm 12 \, {\Bar \psi}{}_{a}{}^+ \, {\Bar T}_{+\, , \, +}{}^a  
~\Big] ~   {\Bar {\cal R}}_{+ \, , \, \mm} \, {\Big |} ~\Big\}  ~~~.
\end{equation}
Next we observe that one of the Bianchi identities takes the form
\begin{equation}
{\Bar \de}_{( +} \, {\Bar {\cal R}}_{+ ) \, ,\, \mm} ~+~ {\Bar \de}_{\mm} 
{\Bar {\cal R}}_{+ \, , \, +} ~-~ {\Bar T}_{+ \, , \, +}{}^D \,
{\Bar {\cal R}}_{ D \, ,\,  \mm} ~-~ {\Bar T}_{\mm \, , \, ( +}{}^D \,
{\Bar {\cal R}}_{ D \, , \, +) } ~=~ 0  ~~~,
\end{equation}
which may be rewritten in the form
\begin{equation}
\eqalign{ {~~~~}
{\Bar \de}_{+} {\Bar {\cal R}}_{+ \, , \, \mm} ~=~ &{~} \fracm 12 ~ \Big[ \,
{\Bar T}_{+ \, , \, +}{}^{\pp} \,{\Bar {\cal R}}_{ \pp \, ,\,  \mm} ~+~ 
{\Bar T}_{+ \, , \, +}{}^+ \,{\Bar {\cal R}}_{ + \, ,\, \mm} \, \Big] 
- \, \fracm 12 ~ \Big[ \, {\Bar \de}_{\mm} {\Bar {\cal R}}_{ + \,  , 
\, +} \, \Big]  \cr
&+ \, \, \Big[ \, {\Bar T}_{\mm \, , \, +}{}^{\pp} \,{\Bar {\cal R}}_{ \pp
\, , \, +} ~+~ {\Bar T}_{\mm \, , \, +}{}^{\mm} \,{\Bar {\cal R}}_{ \mm \, ,
\, +} ~+~ {\Bar T}_{\mm \, , \, +}{}^+ \,{\Bar {\cal R}}_{ +\, , \, +} \, 
\Big] ~~~.
}
\end{equation}
Substitution of this result into the leading term in the expression for 
${\Hat {\cal X}}$ yields,
\begin{equation}
\eqalign{
{\Hat {\cal X}} ~=~ i \, \fracm 12 \, \int d^2 \s ~ {\rm e}^{-1} ~\Big\{  
&{~} \fracm 12 ~ \Big[ \, {\Bar T}_{+ \, , \, +}{}^{\pp} \,{\Bar {\cal 
R}}_{ \pp \, ,\,  \mm} {\Big |}  ~+~ {\Bar T}_{+ \, , \, +}{}^+ \,{\Bar 
{\cal R}}_{ + \, ,\, \mm} {\Big |} \, \Big] - \, \fracm 12 ~ \Big[ \, 
{\Bar \de}_{\mm} {\Bar {\cal R}}_{ + \,  , \, +}{\Big |}  \, \Big]  \cr
&+ \, \, \Big[ \, {\Bar T}_{\mm \, , \, +}{}^{\pp} \,{\Bar {\cal R}}_{ \pp
\, , \, +}{\Big |}  ~+~ {\Bar T}_{\mm \, , \, +}{}^{\mm} \,{\Bar {\cal 
R}}_{\mm \, ,\, +} {\Big |}  ~+~ {\Bar T}_{\mm \, , \, +}{}^+ \,{\Bar {\cal 
R}}_{+\, , \, +} {\Big |} \cr
&+~ \Big[ \, {\Bar T}_{a\, , \, +} {}^a ~+~ \fracm 12 \, {\Bar T}_{+\, , 
\, +}{}^+ ~-~ \fracm 12 \, {\Bar \psi}{}_{a}{}^+ \, {\Bar T}_{+\, , \, +}{}^a  
~\Big] ~   {\Bar {\cal R}}_{+ \, , \, \mm} \, {\Big |} ~\Big\} 
\,  ~~~. }
\end{equation}

Next the leading term in (131) contains a supercovariantized curvature ${\Bar {\cal 
R}}_{ \pp \, ,\,  \mm} { |}$ that possesses an expansion in terms
of the gravitino (see C.4 in an appendix),
\begin{equation}
{\Bar {\cal R}}_{ \pp \, ,\,  \mm} {\Big |} ~=~ {\bar r}_{\pp \, , \, \mm}
~+~  {\Bar \psi}{}_{\pp}{}^+ \, {\Bar {\cal R}}_{ +\, , \, \mm}  {\Big |}
~-~  {\Bar \psi}{}_{\mm}{}^+ \, {\Bar {\cal R}}_{ +\, , \, \pp}  {\Big |}
~+~ {\Bar \psi}{}_{\pp}{}^+ \, {\Bar \psi}{}_{\mm}{}^+ \,  {\Bar {\cal R
}}_{ + \, , \, +}{\Big |}
\end{equation}
The first term in ${\Bar {\cal R}}_{ \pp \, ,\,  \mm} { |} $ allows
the definition of the quantity $ {\Tilde {\cal X}} $ via the
definition
\begin{equation}
{\Tilde {\cal X}} ~\equiv~  - \frac 12 \int d^2 \s ~ {\rm e}^{-1} ~ 
{\bar r}_{\pp\, , \, \mm}  ~~~.
\end{equation}
Using these facts, we finally arrive at the equation ${\Hat {\cal X}} \,-\, 
{\Tilde {\cal X}} \, =\, {\cal O}_T$ where
\begin{equation}
\eqalign{ {~~~}
{\cal O}_T ~=~  i \, \fracm 12 \, \int d^2 \s ~ {\rm e}^{-1} ~\Big\{  
&{~} \fracm 12 ~ (\, {\Bar T}_{+ \, , \, +}{}^{\pp} \, -\,  i 2 \, )  
\,{\Bar {\cal R}}_{ \pp \, ,\, \mm} ~+~ \Big[ \,\, {\Bar T}_{+ \, , 
\, +}{}^+ ~-~ {\Bar T}_{+ \, , \, \pp}{}^{\pp} \,  \cr
&-~ \frac 12 \, {\Bar \psi}{}_{\mm}{}^+ \,{\Bar T}_{+ \, , \, +}{}^{\mm}~
- ~ \frac 12 \, {\Bar \psi}{}_{\pp}{}^+ \,  (\, {\Bar T}_{+ \, , 
\, +}{}^{\pp} \, -\,  i 2 \, )  ~ \Big] \, {\Bar {\cal R}}_{ + \, , 
\, \mm} \,  \, \cr
&-~ \frac 12 {\Bar \de}_{\mm} {\Bar {\cal R}}_{ + \,  , \, +} ~+~ 
\Big[ \,  {\Bar T}_{\mm \, , \, +}{}^{+} ~+~ i \, {\Bar \psi}{}_{
\pp}{}^+ \, {\Bar \psi}{}_{\mm}{}^+ \, \Big] \,  {\Bar {\cal R}}_{ 
+ \,  , \, +} \cr
&-~ \, \Big[ \, \, {\Bar T}_{\mm \, , \, +}{}^{\pp} ~+~ i \,
{\Bar \psi}{}_{\mm}{}^+ \,\, \Big] \,  {\Bar {\cal R}}_{ +\, , \, \pp}\,
\, \Big\} \, {\Big |} ~ ~~~.
}
\end{equation}
It can be seen that when the usual (1,0) superspace constraints are
imposed,
\begin{equation}
{\Bar T}_{+ \, , \, +}{}^{\pp} ~=~  i 2 ~~~,~~~{\Bar T}_{+ \, , \, +}{}^+ 
~=~ {\Bar T}_{+ \, , \, \pp}{}^{\pp}  ~=~ {\Bar T}_{+ \, , \, +}{}^{\mm} 
~=~ {\Bar {\cal R}}_{ + \,  , \, +} ~=~{\Bar {\cal R}}_{ +\, , \, 
\pp} ~=~ 0
\end{equation}
all of the terms of ${\cal O}_T$, which we regard as the obstruction 
to the topological triviality of the ectoplasm (or ``ectoplasmic
obstruction''), vanish up to total derivatives.  In our formulation of 
the Ethereal Conjecture, ${\cal O}_T$ must be trivial in order for 
the conjecture to be valid.  So this equation shows that within the 
context of (1,0) supergravity, the E.C. is {\it {not}} satisfied 
without the imposition of the supergravity constraints.   This strongly 
suggests that the reasons for the constraints in {\it {all}} supergravity 
theories have their origins in topology!  However, additional study of 
this matter is needed in order to construct a rigorous proof of this 
more generally. The Ethereal Conjecture is, we believe, equivalent 
to the triviality of ${\cal O}_T$.

The method of carrying out the calculation of ${\cal O}_T$ above can be 
graphically illustrated. If we imagine that superspace is a sphere, the 
quantity${\Hat {\cal X}}$ corresponds to an ANZ based calculation of an 
index throughout the bulk of superspace (i.e.~the interior of the sphere). 
Since purely bosonic $p$-forms have their support only on the boundary
of the sphere, the quantity ${\Tilde {\cal X}}$ corresponds to the index 
calculated on the purely bosonic sub-manifold of superspace (i.e.~the
surface of the sphere). The quantity ${\cal O}_T$ measures the difference
of these two definitions.

\sect{VIII. Future Prospectives}

~~~~We have seen that there is evidence from a number of supergravity 
theories that topology is at the heart of the process of defining the 
integration of Grassmann variables over local supersymmetry manifolds.  
If our conjecture is taken as a working assumption, future efforts may 
have a basis for adding to a new level of understanding of off-shell 
field representations of supersymmetric theories. In particular,
we must begin to understand how to extend the argument of the last 
section to the cases of more interesting supergravity theories.

Since perhaps the most important theories which would prove of the 
widest interest are 10D theories, it is appropriate to review 
exactly what topological invariants might be available to test the 
Ethereal Conjecture.  For the case of the coupled 10D, $N$ = 1 
supergravity-Yang-Mills system, the topological invariants may be 
denoted by 
\begin{equation}
\D_{p \, q\, r} (SG \,+\, YM) ~=~ \int Tr \Big[ ~ F^{r} \, \wedge \, 
R^{p} \, \wedge \, H^{q} \, \wedge \, (d \Phi)^{10 - 2(p + r) - 3 q} 
~ \Big]  ~~~.
\end{equation}
The integers $p$, $q$ and $r$ take on the values as indicated in the
following table
\begin{center}
\renewcommand\arraystretch{1.2}
\begin{tabular}{|c|c| }\hline
$(p, \, r) $ & $q$  \\ \hline \hline
$~(0,\,0) ~$ & $q \,=\,0, \, 1, \, 2, \, 3 ~~~\,\,$  \\ \hline
$  ~(1,\,0)  ~(0,\,1) ~$ & $q \,=\,0, \, 1, \, 2 ~~~~~~~~$  \\ \hline
$ ~(2,\,0) ~(1,\,1)  ~(0,\,2 ) ~$ & $q \,=\,0, \, 1, \, 2 ~~~~~~~~$  
\\ \hline
$ ~(3,\,0) ~(2,\,1) ~(1,\,2) ~(0,\,3) ~ $ & $\,q \,=\,0, \, 1 
~~~~~~~~~~~~$  \\ \hline
$~(4,\,0) ~(3,\,1) ~(2, \, 2) ~(1,\,3) ~(0,\,4) ~$ & $\,q \,=\,0  
~~~~~~~~~~~~~~~$  \\ \hline
$~(5,\,0) ~(4,\,1) ~(3, \, 2) ~(2,\,3) ~(1,\,4) ~(0,\,5) ~$ & 
$\,q \,=\,0  ~~~~~~~~~~~~~~~$  \\ \hline
\end{tabular}
\end{center}
\centerline{{\bf Table I}}
The trace here is taken  with respect to the matrix representation of 
the Lorentz generator and the Yang-Mills gauge group\footnote{There
may occur multiple ways to define these traces. For example, in 4D, the 
Pontrjagin and \newline ${~~~~~}$ Euler indices arise as two distinct 
ways of evaluating the trace over the Lorentz generators.}. The list of 
38 {\it a} {\it {priori}} indices in Table I includes the special cases 
of the decoupled theories.  Not all of the invariants in Table I are 
non-trivial. Any $(p,r)$-vector that contains a 1 as a component vanishes 
due to the tracing operation. This leaves only 21 non-trivial invariants.
For 10D supersymmetric Yang-Mills (i.e. $\D_{0 \, 0\, 5}$) the single 
invariant is
\begin{equation}
\D (YM)  ~=~ \int  Tr \Big[ ~F \, \wedge \, F \, \wedge \, F 
\, \wedge \, F \, \wedge \, F ~\Big]  ~\equiv~ \int \, F^5 ~~~,
\end{equation}
and for decoupled 10D supergravity  $\D_{p \, q\, 0}$ the invariants are
\begin{equation}
\D_{p \, q} (SG) ~=~ \int Tr \Big[ ~ R^{p} \, \wedge \, H^{q} \, \wedge 
\, (d \Phi)^{10 - 2p - 3 q} ~ \Big] 
~~~.~~~
\end{equation}
Here the integers $p$ and $q$ take on the values as indicated in the
following table
\begin{center}
\renewcommand\arraystretch{1.2}
\begin{tabular}{|c|c| }\hline
$p$ & $q$  \\ \hline \hline
$~p \,=\, 0 ~$ & $q \,=\,0, \, 1, \, 2, \, 3 ~~~\,\,$  \\ \hline
$p \,=\, 1 $ & $q \,=\,0, \, 1, \, 2 ~~~~~~~~$  \\ \hline
$p \,=\, 2 $ & $q \,=\,0, \, 1, \, 2 ~~~~~~~~$  \\ \hline
$p \,=\, 3 $ & $\,q \,=\,0, \, 1 ~~~~~~~~~~~~$  \\ \hline
$p \,=\, 4 $ & $\,q \,=\,0  ~~~~~~~~~~~~~~~$  \\ \hline
$p \,=\, 5 $ & $\,q \,=\,0  ~~~~~~~~~~~~~~~$  \\ \hline
\end{tabular}
\end{center}
\centerline{{\bf Table II}}
As found by counting, there are only eleven such invariants.  Although 
there are many additional topological invariants in the coupled case, we 
would expect that once the constraints are found in the decoupled cases 
to realize the Ethereal Conjecture separately, the additional ones would 
follow as consequences.

We can also foresee the possibility to use such an argument for
type-II theories also.  As is well-known the purely bosonic spectrum
of 10D, $N$ = 1 supergravity consists of $e_a {}^m$, $b_{a \, b}$
and $\Phi$. In the case of the type-IIA theory, there is a supplementary
purely bosonic spectrum given by ${\Hat A}_a$ and ${\Hat A}_{a \, b \, c }$
and hence field strengths  ${\Hat F}_{a \, b}$ and ${\Hat L}_{a \, b \, c 
\,d}$.  In the case of the type-IIB theory, there is a supplementary
purely bosonic spectrum given by $\Hat \Phi$, ${\Hat A}_{a \, b}$ 
and ${\Hat A}_{a \, b \, c \,d }$ and hence field strengths $\pa_a
{\Hat \Phi}$, ${\Hat N}_{a \, b\, c}$ and ${\Hat K}_{a \, b \, c \, d
\, e}$.

In the type-IIA case the candidates for the topological invariants
are 
\begin{equation}
\D_{p \, q\, r \, s}^{IIA} ~=~ \int Tr \Big[ ~ {\Hat F}^{r} \, \wedge \,
{\Hat L}^{s} \, \wedge \,  
R^{p} \, \wedge \, H^{q} \, \wedge \, (d \Phi)^{10 - 2(p + r) - 3 q - 4 s} 
~ \Big]  ~~~.
\end{equation}
Here $p$, $q$, $r$ and $s$ denote integers similar to those discussed in
the $N$ = 1 case.  For the type-IIB case we have
\begin{equation}
\D_{p \, q\, r \, s\, t}^{IIB} ~=~ \int Tr \Big[ ~ (d {\Hat \Phi})^{r} 
\, \wedge \, {\Hat N}^{s} \, \wedge \,  {\Hat K}^{t} \, \wedge \, R^{p} \, 
\wedge \, H^{q} \, \wedge \, (d \Phi)^{10 - r - 2 p - 3 (s + q) - 5 t} 
~ \Big]  ~~~.
\end{equation}
Here $p$, $q$, $r$, $s$ and $t$ once again denote an appropriate
set of integers\footnote{In the enumeration above, we have not taken 
into account any redundancy that might result\newline ${~\,~~~}$ from 
the existence of possible distinct definitions of performing the 
tracing operation.}.

We wish to observe that the ultimate formulation of covariant
string field theory must ultimately confront many of the issues
that we discussed in our introduction. In particular, in covariant
string field theory there must also be developed a theory of local
integrations. The zero-modes for covariant string field theory 
appear to play the role of the spacetime coordinates of superspace
and the oscillator modes play the role of the Grassmann coordinates.
We believe that it will be the case that the concept of stringy
p-forms should play an important role. 

The super forms in (18) obviously belong to the general class of 
super forms given by
\begin{equation}
f_{{\m}_1 \, \cdots \, {\m}_{p}{\un m}_1 \, \cdots \, {\un 
m}_{q}} ~=~ (-1)^{[\fracm{N_B}{2}]}  \, {\rm E}_{{\m}_{q}}
{}^{{\un C}_{q} } \, \cdots \,{\rm E}_{{\m}_1}{}^{{\un C}_1}  
\, {\rm E}_{{\un m}_{p}}{}^{ {\un A}_{p} }  \, \cdots \,{\rm 
E}_{{\un m}_1}{}^{{\un A}_1} \, F_{{\un A}_1 \, \cdots  \, 
{\un A}_{p}{\un C}_1 \, \cdots  \, {\un C}_{q}}  ~~~.
\end{equation}
It is our contention that the Ethereal Conjecture likely is
equivalent to the following statement 

${~~~~~}$ {\it {The topology of super manifolds with $N_F$ 
and $N_B$ fermionic and bosonic \newline ${~~~~~}$ coordinates, 
respectively, arises solely from its purely bosonic 
$p$-forms.}}  

\noindent
The mixed and purely fermionic super $p$-forms although topologically
insignificant, are useful for other aspects.  For example, the case
of $p = 1, \, q = 0$ and $p = 0, \, q = 1$ has been used previously
\cite{pfm,pfm2} to define supersymmetric gauge phase factors.

Finally, the coefficients given in (9) are such that ${\Hat \D}$ defined 
by (11) corresponds to the integration of the super $N_B$-form $f_{{\un 
m}_1 ... {\un m}_{N_B}}$ over a hypersurface in superspace. The hypersurface 
corresponds to the bosonic spacetime manifold.  From this vantage point, it 
should be clear that the theory of ectoplasmic integration that we have 
described is based on the use of superdifferential forms and follows the 
path that is standard for ordinary bosonic differential forms.  This 
observation dramatically emphasizes that the definition of ectoplasmic 
integration defined in the present work is {\it {logically}} {\it {independent}} 
of ANZ local superspace integration theory [9]. Dramatically, our new approach 
to local superspace integration is {\it {totally}} {\it {independent}} of 
the superdeterminant.  More remarkably however, the local ectoplasmic 
integration operator ${\cal D}^{N_F}$ derived on the basis of super p-forms 
(11-13) agrees {\it {exactly}} with the local ANZ integration 
operator\footnote{Whimsically, we might call this the ANZIO-i.e.~the ANZ 
integration operator.} ${\cal D}^{N_F}$ derived on the basis of a normal 
coordinate expansion of the superdeterminant (1). Whether this statement
is necessarily true for {\it {all}} supergravity theories is an interesting 
question to pursue.
$${~~~}$$
``{\it {I have no special talents. I am only passionately curious.}}'' -- 
Einstein
$${~~~}$$

%%%%%%%%%%%%%%%%%%%%%%%%%%%%%%%%%%%%%%%%%%%%%%%%%%%%%%%%%%%%%%%%%%%%
\noindent {\bf{Acknowledgment} }
%%%%%%%%%%%%%%%%%%%%%%%%%%%%%%%%%%%%%%%%%%%%%%%%%%%%%%%%%%%%%%%%%%%%
\indent \newline
${~~~~}$The author wishes to acknowledge the hospitality of the 
Mathematical Sciences Research Institute where this work was initiated
and which inspired him to re-consider the problem of a local superspace 
integration theory and the possible role of topology. He has also 
benefitted greatly from numerous critical comments and inputs from 
M.~T.~Grisaru, W.~Siegel and especially M.~E.~Knutt-Wehlau for her 
assistance in section seven.
\newpage
%%%%%%%%%%%%%%%%%%%%%%%%%%%%%%%%%%%%%%%%%%%%%%%%%%%%%%%%%%%%%%%%%%%%
\noindent{{\bf {Appendix A: Conventions for 2D and 3D Spinors }}}
%%%%%%%%%%%%%%%%%%%%%%%%%%%%%%%%%%%%%%%%%%%%%%%%%%%%%%%%%%%%%%%%%%%%
 
For two dimensional superspaces and using covariant notation, we use 
the following conventions for the quantities associated with spinors.

$$ \eta_{ a b} ~=~ (1 , -1)  ~~~,~~~ \e_{a b} \e^{c d} ~=~  - \d_{[a}{}^c
  \d_{b]}{}^d  ,~~~ \e^{0 1} ~=~  +1 ~~~, $$
$$  (\g^a)_{\a}{}^{ \g} (\g^b)_{\g}{}^{ \b} =   \eta^{ a b} \, \d_{\a}{}^{\b}
- \e^{ a b}  (\g^3)_{\a}{}^{\b} ~~~, $$
$$ (\g^3)_{\a}{}^{ \g} (\g^a)_{\g}{}^{ \b} ~=~ - \, \e^{a b} 
(\g_b)_{\a}{}^{ \b}   ~~~.  \eqno(A.1)  $$

Some useful Fierz identities are:
$$ C_{\a \b}  C^{\g \d} ~=~  \d_{[\a}{}^\g  \d_{\b]}{}^\d ~~, $$
$$ (\g^a)_{\a \b} (\g_a)^{\g \d} ~+~ (\g^3)_{\a \b} (\g^3)^{\g \d}
~=~ - \d_{(\a}{}^\g  \d_{\b)}{}^\d  ~~, $$
$$ (\g^a)_{(\a}{}^\g (\g_a)_{\b)}{}^\d ~+~ (\g^3)_{(\a}{}^\g 
(\g^3)_{\b)}{}^\d ~=~ \d_{(\a}{}^\g  \d_{\b)}{}^\d    ~~ ,  $$
$$ (\g^a)_{(\a}{}^\g (\g_a)_{\b)}{}^\d ~=~ -2 (\g^3)_{\a \b} 
(\g^3)^{\g \d} ~~, $$
$$ 2 (\g^a)_{\a \b} (\g_a)^{\g \d} ~+~ (\g^3)_{(\a}{}^\g (\g^3)_{\b)}
{}^\d   ~=~ - \d_{(\a}{}^\g  \d_{\b)}{}^\d ~~,$$
$$ (\g_a)_\a {}^\d \d_\b ^\g ~+~ (\g^3 \g_a)_\a {}^\g (\g^3)_\b {}^\d
~=~ (\g^3 \g_a)_{\a \b} (\g^3)^{\g \d} ~~.    \eqno(A.2) $$

For three dimensional superspaces, we use the following conventions for the
quantities associated with spinors.

$$ \eta_{ a b} = (1 , -1 , -1)  ~~~,~~~ \e_{a b c} \e^{d e f} ~=~ 
\d_{[a}{}^d \d_{b}{}^e \d_{c]}{}^f ,~~~ \e^{0 1 2} =  +1 ~~~, $$
$$  (\g^a)_{\a}{}^{ \g} (\g^b)_{\g}{}^{ \b} =   \eta^{ a b} \d_{\a}{}^{\b}
+ i \e^{ a b c}  (\g_c)_{\a}{}^{\b} ~~.   \eqno(A.3)    $$

Some useful Fierz identities are:
$$ C_{\a \b} C^{\g \d} ~=~  \d_{[\a}{}^\g  \d_{\b]}{}^\d ~~, ~~~~
~~~~~~~~~~~~~~~~$$
$$ (\g^a)_{\a \b} (\g_a)^{\g \d} ~=~ - \d_{(\a}{}^\g  \d_{\b)}{}^\d  ~~, 
~~~~~~~~~~~~~~~~~~~~~~~ $$
$$ \e^{a b c } (\g_b)_{\a \b} (\g_c)_\g {}^\d ~=~ - i C_{\a \g} (\g^a)_\b 
{}^\d - i (\g^a)_{\a \g} \d_{\b} {}^\d ~~~.~~~~~
   \eqno(A.4)   $$

\newpage 
%%%%%%%%%%%%%%%%%%%%%%%%%%%%%%%%%%%%%%%%%%%%%%%%%%%%%%%%%%%%%%%%%%%%%%%%%%
\noindent{{\bf {APPENDIX B: Supercovariantized Field Strengths  }}}
%%%%%%%%%%%%%%%%%%%%%%%%%%%%%%%%%%%%%%%%%%%%%%%%%%%%%%%%%%%%%%%%%%%%%%%%%%

In this appendix, we present some samples of component-level field
strength tensors.  We concentrate on the gauge (D - 1)-form multiplet 
in D dimensions. The reason for looking at this particular multiplet
is that the D-form gauge field strength is a topological invariant and
provides candidates for the quantities $\tilde I$ in equation (2).  We 
present these results without derivation\footnote{The interested reader 
can find the method of derivation of this general class of results 
\newline ${~~~~~}$ by reviewing \cite{ggrs}}.  In each of the following  
cases, the equation 
is to be understood to be valid only at $\q = 0$ order.  Additionally, 
the first term on the rhs of each equation represents the usual 
component-level field strength that is present without the presence of 
supersymmetry.

~~~~(A.) The 2D, $N$ = 1, 2, 4 Yang-Mills supercovariantized field strength
takes \newline \indent ${~~~~~~~~~~}$ the form,
$$ F_{a b} | ~=~ f_{a b} ~+~ \psi_{[ a|} {}^{\a} F_{\a ~| b ] } ~+~ 
\psi_{a} {}^{\a} \psi_{b} {}^{\b} F_{\a ~\b  } ~~~.  
\eqno(B.1)$$

~~~~(B.) The 3D, $N$ = 1 supercovariantized field antisymmetric tensor gauge
field  \newline \indent ${~~~~~~~~~~}$ strength takes the form,
$$\begin{array}{lll} 
G_{a b c} | &=& g_{a b c } ~+~ \fracm 12 \psi_{[ a|} {}^{\a} G_{\a ~ |b  c] }
~+~  \fracm 12 \psi_{[ a|} {}^{\a} \psi_{ |b|} {}^{\b} G_{\a \b | c] } \\
& & ~-~  \psi_{ a} {}^{\a} \psi_{ b} {}^{\b} \psi_{ c} {}^{\g} G_{\a \b \g } ~~~.
\end{array}         \eqno(B.2)    $$

~~~~(C.) The 4D, $N$ = 1 supercovariantized field antisymmetric rank three gauge
 \newline \indent ${~~~~~~~~~~}$ field strength takes the form,
$$\begin{array}{lll} 
~~~~~~~~~~{F}_{a b c d} | &=& f_{a b c d} ~+~ \fracm 1{3!} (\psi_{[ a|} 
{}^{\a} {F}_{\a ~|b c d] } ~+~ {\bar \psi}_{[ a|} {}^{\dot \a}{\cal 
F}_{\dot \a ~|b c d]} ) \\ 
& &    \\
&+&  \fracm 12 \psi_{[ a|} {}^{\a}{\bar \psi}_{ |b|} {}^{\dot \b} 
{F}_{\a \dot \b | c d] } ~+~\fracm 14 \psi_{[ a|} {}^{\a} \psi_{ |b|}
{}^{\b}  {F}_{\a \b | c d] } ~+~\fracm 14 {\bar \psi}_{[ a|} {}^{\dot \a} 
{\bar \psi}_{ |b|} {}^{\dot \b} {F}_{\dot \a \dot \b | c d] } \\
& &    \\
&-& \fracm 1{3!} \psi_{[ a|} {}^{\a} \psi_{ |b|} {}^{\b} \psi_{ |c|} {}^{\g}
{F}_{\a \b \g | d] } ~-~\fracm 1{3!} {\bar \psi}_{[ a|} {}^{\dot \a} 
{\bar \psi}_{ |b|} {}^{\dot \b} {\bar \psi}_{ |c|} {}^{\dot \g}
{F}_{\dot \a \dot \b \dot \g  | d] } \\
& &    \\
&-& \fracm 12 \psi_{[ a|} {}^{\a} \psi_{ |b|} {}^{\b} {\bar \psi}_{ |c|} 
{}^{\dot \g} {F}_{\a \b \dot \g | d] } ~-~\fracm 12 {\bar \psi}_{[ a|}
{}^{\dot \a}  {\bar \psi}_{ |b|} {}^{\dot \b} {\psi}_{ |c|} {}^{\g}
{F}_{\dot \a \dot \b \g  | d] } \\
& &    \\
&-& \fracm 1{3!} \psi_{[ a|} {}^{\a} \psi_{ |b|} {}^{\b} {\psi}_{ |c|}
{}^{\g} {\bar \psi}_{ |d]} {}^{\dot \d} {F}_{\a \b \g \dot \d }
~-~ \fracm 1{3!} {\bar \psi}_{[ a|} {}^{\dot \a}{\bar \psi}_{ |b|} {}^{\dot 
\b}{\bar \psi}_{ |c|} {}^{\dot \g} {\psi}_{ |d]} {}^{\d} {F}_{\dot \a \dot 
\b \dot \g  \d } \\
& &    \\
&-& \psi_{a} {}^{\a} \psi_{b} {}^{\b} {\psi}_{c}{}^{\g}
 {\psi}_{d} {}^{\d} {F}_{\a \b \g \d } ~-~
 {\bar \psi}_{a} {}^{\dot \a} {\bar \psi}_{ b} {}^{\dot \b} {\bar 
\psi}_{ c} {}^{\dot \g} {\bar \psi}_{ d} {}^{\dot \d}  {F}_{\dot \a
\dot\b \dot \g  \dot \d } \\
& & \\
& &~~-~ \fracm 14 {\psi}_{[ a|} {}^{\a} \psi_{ |b|} {}^{\b} {\bar \psi}_{ |c|} 
{}^{\dot \g} {\bar \psi}_{ |d]} {}^{\dot \d}  {F}_{\a \b \dot \g \dot \d }
~~~. \end{array}         \eqno(B.3)    $$

~~~~The analogs of these identities for superspace supergravity were first given
a long time ago \cite{3d} and take the form (all quantities on the rhs are
to be understood as first being evaluated at $\q = 0$)
$$\begin{array}{lll} 
T \sb{a b} \sp{ C} \de \sb{ C} | ~+~ R \sb{ a b}  \sp{ \G} {\cal M} \sb{ \G} |
&=& [ \de \sb{a} ~,~ \de \sb{b}  ] ~+~ {\psi}_{ [a|} 
{}^{\dot \g} {\bar \psi}_{ |b]} {}^{\dot \d} [ \de \sb{\g} ~,~ {\bar \de} \sb{
\dot \d}  \} \\
&+& \left [ ~ {\psi}_{ [a|} {}^{\g} [ \de \sb{\g} ~,~ {\de} \sb{ |b]} \} ~+~
{ \psi}_{ a} {}^{\g} {\psi}_{b} {}^{\d} [ \de \sb{\g} ~,~ {\de} \sb{\d}  \}
~+~ {\rm {h.c.}} \right ] ~~~. \end{array}         \eqno(B.4)    $$
Here the first term on the rhs is the usual set of field strengths in a
non-supersymmetric theory and ${\cal M} \sb{ \G}$ denotes the generators of
the tangent space.

All of the field strengths discussed in this section can be seen as
special cases of the formula,
$$
f_{{\un m}_1 \, \cdots \, {\un m}_{p}} | ~=~ (-1)^{[\fracm{N_p}{2}]}    
\, {\rm E}_{{\un m}_{p}}{}^{ {\un A}_{p} }  \, \cdots \,{\rm 
E}_{{\un m}_1}{}^{{\un A}_1} \, F_{{\un A}_1 \, \cdots  \, 
{\un A}_{p}} |  ~~~,
\eqno(B.5)    $$
where the leading term $f_{{\un m}_1 \, \cdots \, {\un m}_{p}}$ corresponds
the field strength of some component level gauge field, $F_{{\un A}_1 \, 
\cdots  \, {\un A}_{p}} $ corresponds to the appropriate superspace field 
strength and ${\rm E}_{{\un m}}{}^{{\un A}}$ is identified with the component
level gravitino and vielbein as described in (19).

\newpage 
%%%%%%%%%%%%%%%%%%%%%%%%%%%%%%%%%%%%%%%%%%%%%%%%%%%%%%%%%%%%%%%%%%%%%%%%%%
\noindent{{\bf {APPENDIX C: Some Facts About 4D, $N$ = 1 Super p-forms  }}}
%%%%%%%%%%%%%%%%%%%%%%%%%%%%%%%%%%%%%%%%%%%%%%%%%%%%%%%%%%%%%%%%%%%%%%%%%%

Although the Bianchi identities (B.~I.'s) for the 4D, $N$ = 1 super 4-form
have been explicitly given before \cite{pfm}, for the convenience of the
reader we will here provide explicit expressions.  The equation which
we schematically write as
$$0 ~=~ 
\nabla_{\, [{\un A}_1 |} J_{ | \, {\un A}_2 ... {\un A}_5 )} ~-~
T_{\, [ {\un A}_1 \, {\un A}_2 |} {}^L   J_{ L \, | \, {\un A}_3 ... 
{\un A}_5 )}
\eqno(C.1)
$$
is explicitly of the form,
$$
\begin{array}{lll} 
0 &=& { {\fracm 1{4!} }} \nabla_{( \a } {J}_{\b \g\, \d \,
\e ) } ~-~ { {\fracm 1{2\times 3!} }}\, T_{ ( \a \, \b | }{}^L
{J}_{L | \g \,  \d \,  \e )}  ~~, \\
&{~}& \\
0 &=& { {\fracm 1{3!} }} \nabla_{( \a } {J}_{\b \g\, \d )
\dot \e  } ~+~ {\Bar \nabla}_{\dot \e } {J}_{\a \b  \g \, \d  }
~-~ { {\fracm 1{4} }}\, T_{ ( \a \, \b | }{}^L
{J}_{L | \g \,  \d )  \dot \e } \\
&{~}& \\
0 &=& { {\fracm 1{2} }} \nabla_{( \a } {J}_{\b \g \,) \dot \d
\, \dot \e  } ~+~ {\Bar \nabla}_{( \dot \e } {J}_{\dot \d )}{}_{ \a \, \b
\, \g  } ~-~ { {\fracm 1{2} }}\, T_{ ( \a \, \b | }{}^L
{J}_{L | \g \,) \, \dot \d \, \dot \e } ~-~  T_{  \dot \d \, \dot \e }{}^L
{J}_{L \, \a \, \b \, \g \,} \\
&{~}&~-~ { {\fracm 1{2} }}\, T_{ ( \a \,| \, ( \dot \d }{}^L {J}_{L |
\dot \e \, ) \, |  \b \, \g \,)   } ~~,  \\
&{~}& \\
0 &=& { {\fracm 1{3!} }} \nabla_{( \a } {J}_{\b \g\, \d )
\un e  } ~+~ {\nabla}_{\un e} {J}_{\a \b  \g \, \d  } ~-~
{ {\fracm 1{4} }}\, T_{ ( \a \, \b | }{}^L
{J}_{L | \g \,  \d )  \un e } ~-~ { {\fracm 1{3!} }}\, T_{   
\un e \, ( \a \,| }{}^L {J}_{L |\b \,  \g \,  \d )} ~~, \\
&{~}& \\
0 &=& { {\fracm 1{2} }} \nabla_{( \a } {J}_{\b \g \, ) \dot \d
\un e  } ~+~ {\nabla}_{\un e} {J}_{\a \b  \g \, \dot \d  } ~-~
{ {\fracm 1{2} }}\, T_{ ( \a \, \b | }{}^L
{J}_{L | \g \, )\, \dot \d  \un e } ~-~ { {\fracm 1{2} }}\, T_{   
\un e \, ( \a \,| }{}^L {J}_{L |\b \,  \g \,)\, \dot \d } \\
&{~}& ~+~  T_{ \dot \d \,  \un e}{}^L
{J}_{L \, \a \, \b \,  \g  }  ~~, \\
&{~}& \\
0 &=& { {\fracm 12}} \nabla_{( \a } {J}_{\b \g) \un c \, \un d  } ~ + ~
{\nabla}_{ [ \un c } {J}_{\un d ] \a  \b \g}
~-~ T_{ \un c \, \un d  }{}^L {J}_{L \a \b \g}
~-~  { {\fracm 1 2} } T_{( \a \b |} {}^L {J}_{L | \g ) \un c \un d}\\
&{~}& ~-~ T_{ ( \a | [ \un c |} {}^L {J}_{L | \un d ] | \b  \g ) }
 ~~, \\
&{~}& \\
0 &=& \nabla_{ ( \a } {J}_{\b ) \dot \g \un c \, \un d  } ~ + ~
{\Bar \nabla}_{\dot \g } {J}_{\a \b  \un c \, \un d  } ~ + ~
{\nabla}_{ [ \un c } {J}_{\un d ] \a  \b \dot \g}
~-~ T_{ \un c \, \un d  }{}^L {J}_{L \a \b \dot \g} 
~-~ T_{ \a \b } {}^L {J}_{L \dot \g  \un c \un d}\\
&{~}& 
~-~ T_{ (\a |\dot \g } {}^L {J}_{L | \b) \un c \un d}
~-~ T_{ ( \a | [ \un c |} {}^L {J}_{L | \un d ] | \b )  \dot \g  } 
~-~ T_{ \dot \g [ \un c |} {}^L {J}_{L | \un d ] \a \b }
 ~~, ~~\\
&{~}& \\
0 &=& \nabla_{ ( \a } {J}_{\b ) \un c \, \un d \, \un e } ~ + ~
{\nabla}_{ [ \un c } {J}_{\un d \, \un e ] \a  \b} ~ - ~
T_{\a \b } {}^L {J}_{L \un c \, \un d \, \un e} {~}
~ - ~ { { \fracm 1 2}} T_{ [ \un c \, \un d | }{}^L
{J}_{L | \un e ] \a \b } \\
&{~}& ~ + ~{ {\fracm 1 2}} T_{ ( \a | [ \un c | }{}^L
{J}_{L | \un d \, \un e ]  |\b ) }
~~, ~~ \\
&{~}& \\
0 &=& \nabla_{ \a } {J}_{\dot \b \, \un c \, \un d \, \un e} ~+~
{\Bar \nabla}_{ \dot \b } {J}_{\a  \, \un c \, \un d \, \un e}
~+~ { {\fracm 1 2}} \nabla_{ [ \un c | } {J}_{| \un d \,
\un e ] \a \dot \b} ~ - ~ T_{ \a \dot \b }{}^L {J}_{L \un c \, \un d
\, \un e} \\
&{~}& ~ + ~ { {\fracm 1 2} } T_{ \a [ \un c |} {}^L {J}_{L |
\un d \, \un e ] \dot \b }  ~ + ~ { {\fracm 1 2} } T_{ \dot \b
[ \un c |} {}^L {J}_{L | \un d \, \un e ] \a } ~ - ~ { {\fracm
1 2} } T_{ [ \un c \, \un d |} {}^L {J}_{L | \un e ]   \a \dot \b }
~~, ~~ \\
&{~}& \\
0 &=& \nabla_{ \a } {J}_{\un b \, \un c \, \un d \, \un e}
~-~ { {\fracm 1 6}} \nabla_{[ \un b } {J}_{\a \, \un c \, \un d
\, \un e ]} ~-~ { {\fracm 1 6} } T_{ \a [ b |} {}^L {J}_{L \a |
\un c \, \un d \, \un e ] } ~-~  { {\fracm 1 4}} T_{ [ \un b \,
\un c |} {}^L {J}_{L | \un d \, \un e ] \a }~
 ~~,  ~~ \\
&{~}& \\
0 &=& { {\fracm 1{4!} }} \nabla_{[ \un a }
{J}_{\un b \un c \, \un  d \, \un e ] } ~-~ { {\fracm 1{2\times
3!} }}\, T_{ [ \un a \, \un b | }{}^L {J}_{L | \un c \,  \un d \,
\un e ]}  ~~. ~~  
\end{array} \eqno(C.2)
$$

In a similar manner we find that a super 4-form field strength
$F_{\un A \un B \un C \un D}$ can be expressed as the super
exterior derivative of a super 3-form gauge field $J_{\un A \un 
B \un C}$ schematically via the equation
$$
F_{{\un A}_1 ... {\un A}_4} ~=~ 
\nabla_{\, [{\un A}_1 |} J_{ | \, {\un A}_2 ... {\un A}_4 )} ~-~
T_{\, [ {\un A}_1 \, {\un A}_2 |} {}^L   J_{ L \, | \, {\un A}_3  
{\un A}_4 )}  ~~~,
\eqno(C.3) $$
has the explicit representation given by,
$$
\begin{array}{lll} 
{F}_{\a \, \b \, \g \, \d}  &=& {\scriptstyle {\fracm 1{3!} }} \nabla_{(\a} 
{J}_{\b \g\, \d ) } ~-~ {\scriptstyle {\fracm 1{4} }}\, T_{ ( \a \, \b | }{}^L 
{J}_{L | \g \,  \d )}  ~~, \\
&{~}&  \\ 
{F}_{\a \, \b \, \g \, \dot \d}  &=& {\scriptstyle {\fracm 1{2} }} \nabla_{( 
\a } {J}_{\b \g\, ) \, \dot \d  } ~+~ {\Bar \nabla}_{\dot \d } {J}_{\a \b  
\g } ~-~ {\scriptstyle {\fracm 1{2} }}\, T_{ ( \a \, \b | }{}^L {J}_{L 
| \g \, )\, \dot \d  }  ~-~ {\scriptstyle {\fracm 1{2} }}\, T_{ ( \a  |\, 
\dot \d }{}^L {J}_{L |\b \, \g \, )  } ~~, \\
&{~}&  \\ 
{F}_{\a \, \b \, \dot \g \, \dot \d}  &=&  \nabla_{( \a } {J}_{\b \, ) \,
\g \, \, \dot \d  } ~+~ {\Bar \nabla}_{( \dot \g } {J}_{\dot \d ) \,\a 
\, \b} ~-~  T_{  \a \, \b  }{}^L {J}_{L \, \dot \g \,  \dot \d  } ~-~  
T_{ \dot \g \, \dot \d }{}^L {J}_{L \, \a \, \b  } \\ 
&{~}& ~-~ T_{ ( \a |\, (\, \dot \g |}{}^L 
{J}_{L |\b \, ) \, |  \dot \d )  } ~~, \\
&{~}&  \\ 
{F}_{\a \, \b \, \g \, \un d}  &=& {\scriptstyle {\fracm 1{2} }} \nabla_{( 
\a } {J}_{\b \g ) \un d  } ~-~ {\nabla}_{\un d} {J}_{\a \, \b \, \g } ~-~ 
{\scriptstyle {\fracm 1{2} }}\, T_{ ( \a \, \b | }{}^L 
{J}_{L | \g )  \un d }  ~+~ {\scriptstyle {\fracm 1{2} }}\, T_{   \un d \, 
( \a \,| }{}^L {J}_{L |\b \,  \g  )}  ~~, \\
&{~}&  \\ 
{F}_{\a \, \b \, \dot \g \, \un d}  &=& \nabla_{( \a } {J}_{\b \,) \dot \g \,
\un d  } ~+~ {\Bar \nabla}_{\dot \g } {J}_{\a \, \b \, \un d }~-~ {\nabla}_{
\un d} {J}_{\a \, \b \, \dot \g \, \un d  } ~-~ T_{  \a \, \b  }{}^L 
{J}_{L \, \dot \g \,  \un d } \\ 
&{~}&  ~-~ T_{ \dot \g ( \a \, | }{}^L {J}_{L | \b \,) \,  \un d }~+~ T_{\un 
d \, ( \a \,| }{}^L {J}_{L \, |\b \,)\, \dot \g } ~~, \\
&{~}&  \\ 
{F}_{\a \, \b \, \un c \, \un d}  &=& \nabla_{( \a } {J}_{\b \,) \un c \,
\un d  } ~+~ {\nabla}_{[ \un c |} {J}_{\a \, \b \, | \un d ]} ~-~ T_{\a \, 
\b  }{}^L {J}_{L \, \un c \,  \un d } ~-~ T_{ [\un c\, | ( \a \, | }{}^L 
{J}_{L | \b \,) \, | \un d ] }  \\
&{~}& ~-~ T_{ \un c \, \un d }{}^L {J}_{L \, \a \, \b  } ~~, \\
&{~}&  \\ 
{F}_{\a \, \dot \b \, \un c \, \un d}  &=& \nabla_{ \a } {J}_{\dot \b \, 
\un c \, \un d} ~+~ {\Bar \nabla}_{ \dot \b } {J}_{\a  \, \un c \, \un d} 
~+~ {\scriptstyle {\fracm 1 2}} \nabla_{ [ \un c | } {J}_{| \un d ] \a 
\dot \b} ~ - ~ T_{ \a \dot \b }{}^L {J}_{L \un c \, \un d} \\
 &{~}&  ~ + ~  T_{ \a [ \un c |} {}^L {J}_{L | \un d  ] \dot \b } ~ + ~  
T_{ \dot \b [ \un c |} {}^L {J}_{L | \un d  ] \a } ~ - ~  T_{  \un c \, 
\un d } {}^L {J}_{L   \a \dot \b } ~~, ~~ \\
 &{~}&  \\ 
{F}_{\a \, \un b \, \un c \, \un d}  &=& \nabla_{ \a } {J}_{\un b \, 
\un c \, \un d } 
~-~ {\scriptstyle {\fracm 1 2}} \nabla_{[ \un b } {J}_{\a \, \un c \, 
\un d ]} ~-~ {\scriptstyle {\fracm 1 2} } T_{ \a [ b |} {}^L {J}_{L \a | 
\un c \, \un d] }  ~-~  {\scriptstyle {\fracm 1 2}} T_{ [ \un b \, \un c |} 
{}^L {J}_{L | \un d \,] \a }~  ~~,  ~~  \\
 &{~}& \\
{F}_{\un a \, \un b \, \un c \, \un d}  &=& {\scriptstyle {\fracm 1{3!} }} \,
\nabla_{[ \un a } {J}_{\un b \un c \, \un  d ] } ~-~ {\scriptstyle {\fracm 
1{4} }}\, T_{ [ \un a \, \un b | }{}^L {J}_{L | \un c \,  \un d 
]}  ~~. 
\end{array} \eqno(C.4)
$$

\newpage
%%%%%%%%%%%%%%%%%%%%%%%%%%%%%%%%%%%%%%%%%%%%%%%%%%%%%%%%%%%%%%%%%%%


\begin{thebibliography}{66}
\bibitem{c3}To our knowledge B.Zumino first raised such questions in the
early eighties.
\end{thebibliography}
\end{document}